\documentclass[a4paper,11pt]{article}

\RequirePackage[T1]{fontenc}
\RequirePackage{fix-cm}
\RequirePackage{stix}

\usepackage{enumitem}
\usepackage{graphicx}
\usepackage{float}
\usepackage{mathtools}
\usepackage{amsthm}
\usepackage{color}
\usepackage{mathrsfs}
\usepackage{subcaption}

\newcommand{\red}{\color{black}}
\newcommand{\blue}{\color{black}}

\newcommand{\ket}[1]{|#1\rangle}
\newcommand{\bra}[1]{\langle #1|}

\newtheorem{Result}{Result}

\newtheorem{Claim}{Claim}
\newtheorem{theorem}{Theorem}

\newtheorem{lemma}[theorem]{Lemma}
\newtheorem{corollary}[theorem]{Corollary}
\newtheorem{proposition}[theorem]{Proposition}

\allowdisplaybreaks[4]

\title{Existence of universal resource and uselessness of too entangled states {\red for quantum metrology}}
\author{%
Rina Miyajima\thanks{NTT Communication Science Laboratories, NTT Corporation,
  3--1 Morinosato Wakamiya,
  Atsugi,
  Kanagawa
  243-0198,
  Japan},\
Yuki Takeuchi\footnotemark[1]
\thanks{The current affiliation is Information Technology R\&D Center, Mitsubishi Electric Corporation.}
\ and\ 
Seiseki Akibue\footnotemark[1]
}
\date{\vspace{-6ex}}

\begin{document}

\setlength{\abovedisplayskip}{3pt} % 上部のマージン
\setlength{\belowdisplayskip}{3pt} % 下部のマージン

\maketitle

%########### body
\section{{\red Introduction}}
Quantum mechanics provides prominent capabilities for information processing
beyond those of classical information processing~\cite{nielsenchung}.
However, it is still not completely clear how these quantum advantages arise.
In particular, it would be intriguing to see how the mechanism of quantum advantage depends on quantum information processing tasks.
In this {\red paper}, we investigate the connection between the two primary quantum information processing tasks, quantum metrology and computation.
To this end, we consider the following two fundamental questions: 
(i) Whether the full quantum advantage can be obtained from a universal resource?
(ii) How much entanglement {\red is} sufficient/necessary for quantum advantage? 

These two questions have been extensively investigated for quantum computation.
Measurement-based quantum computation (MBQC)~\cite{raussendorf2001one,
raussendorf2001computational,
walther2005experimental,
anders2009computational} is a universal quantum computing model, which proceeds by measuring each qubit of a universal resource state (e.g., the cluster state~{\red\cite{briegel2001persistent}}) one by one.
On the other hand, it has also been shown that too entangled quantum states cannot be universal resource states for MBQC~\cite{gross09,bremner2009random}.
This would be a remarkable difference from quantum communication, given a fact that Alice can send any single-qubit state to a distant party, Bob, without any error by consuming one maximally entangled state~\cite{bennett93,boschi98}.

We obtain MBQC-like properties for quantum metrology (for details, see Results 1 and 2).
More concretely, we show (i) the existence of universal resource states for a certain class of linear Hamiltonians and (ii) the uselessness of highly entangled states for quantum metrology of linear Hamiltonians.
We also show that random pure states are basically not useful even if we consider more general Hamiltonians(, which is a corollary of Results 3 and 4).
Since random pure states have high entanglement~\cite{epsilon_net,aspects,lubkin1978entropy,page1993average,foong1994proof}, this result strengthens the uselessness of highly entangled states for quantum metrology.

\section{{\red Summary} of our contribution}
Quantum metrology~\cite{degen17,paris2009quantum,metrology_rev}
{enables us} to measure unknown physical {parameters, such as gravitational waves~\cite{ligo2011gravitational,aasi2013enhanced,abbott16}, 
magnetic fields~\cite{wasi10, sewell12}, and 
temperature~\cite{correa15,de2019quantum,mehboudi2019thermometry},} beyond the {precision achieved by classical metrology}.
Classically,
when using $n$ probes which is 
a realization of $n$ independently and 
{identically} distributed (i.i.d.) random variables $X$,
the achievable mean-squared error is $1/n$.
This {limit} is called the standard quantum limit (SQL)~\cite{glm04}.
However, {
by using entangled $n$ probes,
the achievable error can be more negligible than {the SQL}. }
The ultimate limit of {the} precision attainable through quantum mechanics
is {$1/n^2$, which is} known as {the} Heisenberg limit {(}HL)~\cite{glm06}.
In this {\red paper},
we focus on phase estimation~\cite{glm04,glm06,glm11,Toth14,hb93}.
Phase estimation is a concrete quantum sensing protocol {that can potentially achieve {the HL}} and {proceeds} as follows:
(1)~{Prepare} a quantum state $\rho$ as a probe.
(2)~{Interact $\rho$} with {an object subject to sensing.
As a result,} unknown parameter $\theta$ is encoded {into the quantum state through the time evolution with the Hamiltonian $H$ corresponding to the object, and}
$\rho_\theta= e^{-iH\theta}\rho e^{iH\theta}$
{is obtained.}
(3)~{Estimate the} phase $\theta$ of {the} quantum state $\rho_\theta$ by measuring $\rho_\theta$.

A historically important question
has been what quantum state
should we prepare as a probe.
Hereafter,
we consider
$n$-qubit quantum {systems}.
{When $H$ is a linear Hamiltonian, i.e.,}
\begin{align}
H_{0,1}&=h \otimes I \otimes \cdots \otimes I
+ I \otimes h \otimes I \otimes \cdots \otimes I
+\cdots 
+ I \otimes \cdots \otimes I \otimes h,\label{LH}
\end{align}
where $h=\lambda_0 \ket{0}\bra{0}+\lambda_1 \ket{1}\bra{1}$ is a single-qubit non-degenerate Hermitian operator with $\lambda_0 \neq \lambda_1$, 
the phase estimation can achieve {the SQL} and HL when $\rho$ is 
an optimal separable state
{and the}
{\red Greenberger-Horne-Zeilinger (GHZ)} state $\ket{\phi_\mathrm{GHZ}}=(\ket{0^n}+ \ket{1^n})/\sqrt{2}$, respectively.
The fluctuation of the estimated value is given by the inverse of
the quantum Fisher information (QFI)
\cite{braunstein94,hayashi05}.
In phase estimation, 
for a pure state $\rho=\ket{\psi}\bra{\psi}$,
the QFI can be computed as 
\begin{equation}\label{qfi}
F_Q(\ket{\psi}, H)
=4(\langle \psi |H^2| \psi \rangle-
\langle \psi |H| \psi \rangle^2).
\end{equation}
The GHZ state $\ket{\phi_\mathrm{GHZ}}$ takes 
the maximal value of QFI,
and its value is
$F_Q(\ket{\phi_\mathrm{GHZ}},H_{0,1})
=n^2(\lambda_1-\lambda_0)^2$, which corresponds to the HL.
However, 
when the Hamiltonian $H$ is just locally rotated, i.e., it becomes
\begin{align}
H_{+,-}&=h' \otimes I \otimes \cdots \otimes I
+ I \otimes h' \otimes I \otimes \cdots \otimes I
+\cdots 
+ I \otimes \cdots \otimes I \otimes h',\label{LH'}
\end{align}
where $h'=\lambda_0 \ket{+}\bra{+}+\lambda_1 \ket{-}\bra{-}$ is a single-qubit non-degenerate Hermitian operator with $\lambda_0 \neq \lambda_1$, 
the QFI of $\ket{\phi_\mathrm{GHZ}}$ is
$F_Q(\ket{\phi_\mathrm{GHZ}},H_{+,-})
=n(\lambda_1-\lambda_0)^2$ {\red for $n\ge3$}, which corresponds to the SQL.
This means that the performance of $\ket{\phi_\mathrm{GHZ}}$
is the same as that of the product states for $H_{+,-}$.
Thus, 
preparing the GHZ state with respect to the computational basis is not always the best.
On the other hand,
the QFI of 
$\ket{\phi_\mathrm{superposition}}=(\ket{0}^{\otimes n} +\ket{1}^{\otimes n}
+\ket{+}^{\otimes n} 
+\ket{-}^{\otimes n})/\|\ket{0}^{\otimes n} +\ket{1}^{\otimes n}
+\ket{+}^{\otimes n} 
+\ket{-}^{\otimes n}\|_2$
is {\red$F_Q(\ket{\phi_\mathrm{superposition}},H_{0,1})
=F_Q(\ket{\phi_\mathrm{superposition}},H_{+,-})
=\Theta(n^2)$}.
In this sense,
the quantum state
$\ket{\phi_\mathrm{superposition}}$
is better than $\ket{\phi_\mathrm{GHZ}}$.

Then, the natural question arises:
is there a quantum state which
is suitable for quantum metrology of
any Hamiltonian?
Our result partially answers to this question.
We show that
there are symmetric states whose values of QFI are $\Theta(n^2)$ (HL) for
any linear Hamiltonian of the following form:
\begin{align}
H_{L}=h_1 \otimes I \otimes \cdots \otimes I
+ I \otimes h_2 \otimes I \otimes \cdots \otimes I
+\cdots 
+ I \otimes \cdots \otimes I \otimes h_n,\label{LH''}
\end{align}
for some single-qudit Hermitian operator
$h_i=\sum_{j=1}^{d} \lambda_{i,j} \ket{\phi_j}\bra{\phi_j}$
and there exists $j \neq j'$ such that
$\sum_{i=1}^n \lambda_{i,j} - \sum_{i=1}^n \lambda_{i,j'}=\Theta(n)$.
Note that 
$\{\ket{\phi_1},\ket{\phi_2},\cdots,\ket{\phi_d}\}$
is a fixed orthonormal basis.
Our first result is summarized as follows (see also Fig.~1):
\setcounter{Result}{0}
\begin{Result}
Let $S_{L}$ be 
a set of
linear $n$-qudit Hamiltonians
such as {\red(\ref{LH''})}.
Set $d>13$.
Denote by $\ket{\psi} \leftarrow
Sym^n(\mathbb{C}^d)$,
a quantum state
sampled uniformly {at} random
from all $n$-qudit pure symmetric states.
For any positive constant $c$,
an upper bound on
\begin{equation}
\underset{\ket{\psi} \leftarrow
Sym^n(\mathbb{C}^d)}{\mathrm{Pr}}\left(\sup_{H_L \in S_{L}} \Big(
\Theta(n^2) -
F_Q(\ket{\psi},H_L)
\Big)> c\right),
\end{equation}
which is a probability 
that there exists an element of 
$S_{L}$ such
the quantum Fisher information of
$\ket{\psi}$ is lower than
$\Theta(n^2)$,
converges to $0$ in the limit of $n \rightarrow \infty$.
Here,
$F_Q(\ket{\psi},H)$ is
the quantum Fisher information, defined in 
(\ref{qfi}).
\end{Result}
\noindent Result~1 means that
random symmetric states are useful
even if the most unsuitable Hamiltonian is chosen from 
a set of linear Hamiltonians
for each sampled probe.
We call such symmetric states
as ``universal resource states''.

As a potential application of Result~1,
we give the delegation 
of quantum metrology~\cite{shettell2022quantum}.
We consider the following situation:
{\red there are a server and a client.
The client has a quantum register, and his/her magnetic field includes the confidential information which the client wishes to conceal from the server.
The client can perform phase estimation of $U=e^{-iH\theta}$ securely as follows:
(1)~The server prepares a ``{universal resource} state'' and sends it to the client.
(2)~The client interacts the state in (1)
with his/her magnetic field and then measures it.
(3)~The client obtains an estimated value by repeating (1) and (2).
This protocol is a quantum-metrology analogue of \cite{morimae2013blind} and improves \cite{takeuchi2019quantum}.
A thorough analysis is beyond the scope of this paper.}
%there are the server and 3 clients, $A$, $B$, and $C$.
%Each client has the quantum registers, and their magnetic fields $h_A$, $h_B$, and $h_c$ have the confidential information which the clients wish to conceal from the server.
%The clients can perform phase estimation of $U=e^{-iH\theta}$ where $H=h_A \otimes I \otimes  I+I \otimes h_B \otimes I+I \otimes I \otimes h_c$ as follows:
%(1)~The server prepares a ``{universal resource} state''.
%(2)~Each client intaracts the state in (1) with his/her magnetic field.
%(3)~The clients calculate the estimated value by LOCC between every client. 
%From \cite{hayashi2004asymptotic}, the highest accuracy of the estimated value of a phase $\theta$, computed in (3) can be attained by LOCC.
%But explicit construction is still open.

\setcounter{figure}{0}
\begin{figure*}[t]
  \centering
\includegraphics[width=0.85\columnwidth]{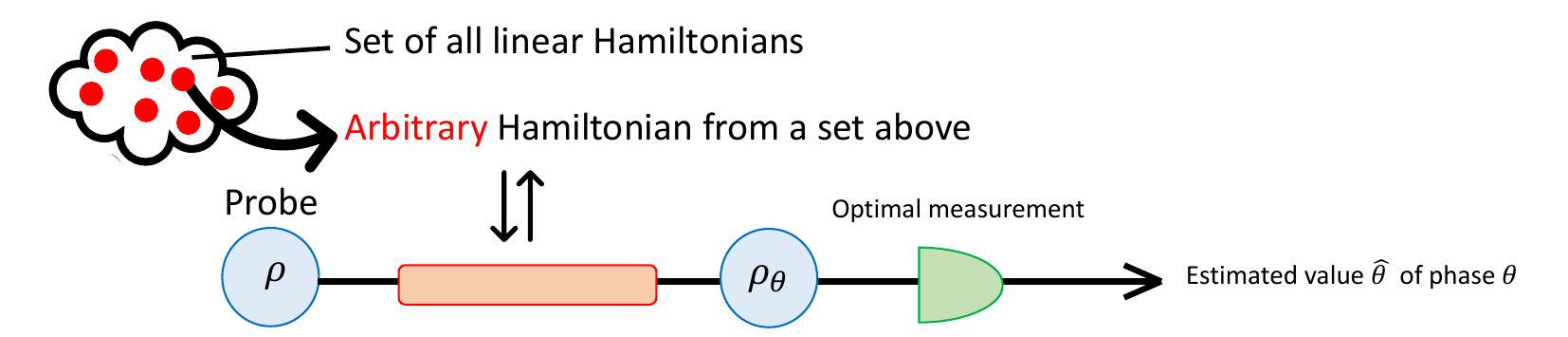} 
\vspace{-10pt}
  \caption{Schematic of Result~1.
Result~1 means that
random symmetric states are useful
even if the most unsuitable Hamiltonian is chosen from 
a set of linear Hamiltonians
for each sampled probe.}
\end{figure*}

Then, we consider the second question (ii).
To this end, we focus on
{\red the} geometric measure of entanglement (GME) defined as 
$E_g(\ket{\Psi})=-\log_2 \sup_{\ket{\alpha}:\mathrm{product}}|\langle \alpha|\Psi\rangle|^2$~\cite{shimony1995degree, wei2003geometric, gross09, bremner2009random}.
We show that
very high GME
leads to low values in the QFI (i.e., the uselessness)
for linear Hamiltonians.
Our second result is summarized as follows:
\begin{Result}
Let $1<c<2$ and $n^{c-1}>\log_e n$.
If a geometric measure of entanglement $E_g(\ket{\Psi})$ is 
larger than $n-\{2(n^{c-1}-\log_e n)+c\log_e n\}/\log_e 2$,
then the QFI 
$F_Q(\ket{\Psi},H_S)$ is less than $n^c$
for linear $n$-qubit Hamiltonians $H_S$.
Here, a geometric measure of entanglement
is defined as $E_g(\ket{\Psi})=-\log_2 \sup_{\ket{\alpha}:\mathrm{product}}|\langle \alpha|\Psi\rangle|^2$,
the QFI 
is defined as (\ref{qfi}), and 
$H_S=h_S \otimes I \otimes \cdots \otimes I
+ I \otimes h_S \otimes I \otimes \cdots \otimes I
+\cdots 
+ I \otimes \cdots \otimes I \otimes h_S$,
where $h_S=\lambda_0 \ket{\phi_{0}}\bra{\phi_{0}}+\lambda_1 \ket{\phi_{1}}\bra{\phi_{1}}$ ($\lambda_0 \neq \lambda_1$)
is some non-degenerate single-qubit Hermitian operator.
\end{Result}

From Result 2, high GME is not useful
in quantum metrology
of linear Hamiltonians.
However, 
there is still a possibility that 
highly entangled states may be useful for other kinds of Hamiltonians
\cite{boixo2007generalized,boixo2008quantum,zwierz10,zwierzx12}.
It seems to be challenging
to directly analyze a quantum state with high GME
for more general Hamiltonians.
Alternatively, we focus on random states
which {\red have} high GME.
For random states,
GME is larger than {\red or equal to}
$n-{\red 2}\log_2 n-3$
with {\red probability at least} $1-e^{-n^2}$~\cite{gross09}.
We show that the achievable precision of random states
is the same
as that of product states
for {\red several} locally diagonalizable Hamiltonians.
Our result is a generalization of
\cite{random_qfi} and summarized as follows:
\begin{Result}
Let $S_{LD}$ be 
a set of
locally diagonalizable $n$-qudit Hamiltonians
prameterized by
at most $d^{o(n)}$ parameters,
where locally diagonalizable Hamiltonians
are
Hamiltonians that can be diagonalized by {a} product basis such as 
$\{\ket{i_1 i_2 {\red\cdots} i_n}
:i_j \in \{1,2,\cdots,d\}\}$.
Denote by $\ket{\psi} \leftarrow
(\mathbb{C}^d)^{\otimes n}$,
a quantum state
sampled uniformly {at} random
from all $n$-qudit pure states.
For any positive constant $c$,
an upper bound {on}
\begin{equation}\label{intro_bound}
\underset{\ket{\psi} \leftarrow
(\mathbb{C}^{\red d})^{\otimes n}}{\mathrm{Pr}}\left(\sup_{H_{LD} \in S_{LD}} \Big(F_Q(\ket{\psi},H_{LD})
-
\max_{\ket{\Phi}: \mathrm{separable}}F_Q(\ket{\Phi},H_{LD})\Big) > c\right){,}
\end{equation}
{which is a} probability 
that there exists an element of 
$S_{LD}$
 such that
 {the} QFI of
 $\ket{\psi}$ is higher than
the nearby value of that of 
an optimal separable state,
converges to $0$ {in the limit of $n \rightarrow \infty$}.
Here,
$F_Q(\ket{\psi},H)$ is
{the} QFI defined in (\ref{qfi}).
\end{Result}
\noindent

Combining with the fact that the optimal entangled state, such as GHZ or our universal state, provides quadratic better QFI compared to the optimal separable state in the case of linear Hamiltonian, we conclude that the generic states are useless for linear {\red Hamiltonians}.
However, it was not known whether such QFI gap between optimal separable states and optimal entangled states exists in the case of general Hamiltonian.
We identify
the set of Hamiltonians
in which 
the accuracy attained by the generic states(,
which is almost the same as that of a particular symmetric product state)
 is much lower than
that of an optimal state in all quantum states.
For convenience, we define the following claim:
\setcounter{Claim}{0}
\begin{Claim}
The scaling of
the maximal QFI of all {symmetric product states}
with respect to $n$
is {{\red different} from that of the optimal entangled state}.
That is,
the accuracy attained by a {symmetric product state}
 is much lower than
that of an optimal state in all quantum states.
\end{Claim}

\begin{figure*}[t]
  \centering
\includegraphics[width=0.8\columnwidth]{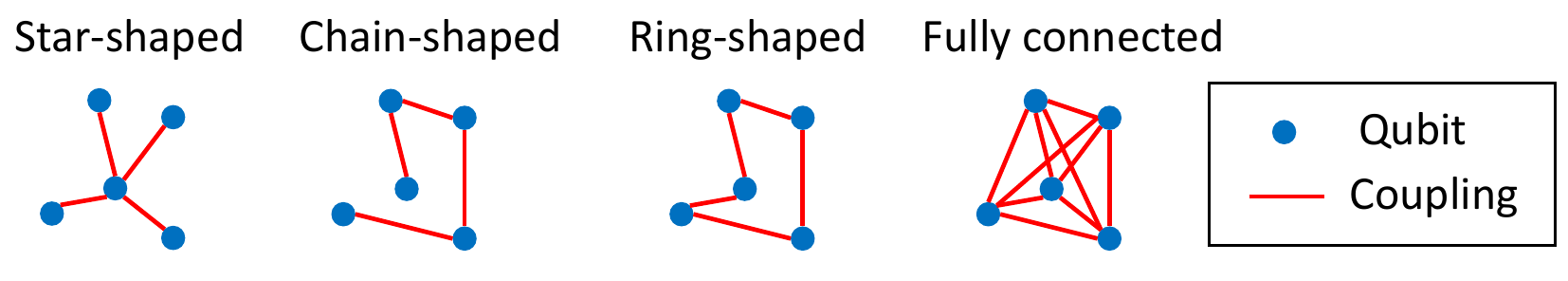} 
\vspace{-10pt}
  \caption{
{Examples} of {2-body} Hamiltonians {with 5 qubits. From left to right, the graphs represent star-shaped, {\red chain}-shaped, {\red ring}-shaped, and fully connected Hamiltonians}.
}
\end{figure*}

{Hereafter}, 
we focus on 2-{body} 
locally diagonalizable Hamiltonians
$H_{LD}=\sum_{(i,j) \in S} h_i \otimes h_j$,
{where} $h_i=\lambda_0 \ket{\phi_{0,i}}\bra{\phi_{0,i}}+\lambda_1 \ket{\phi_{1,i}}\bra{\phi_{1,i}}$ ($0 <\lambda_0 < \lambda_1$)
{is some non-degenerate single-qubit Hermitian operator} {\red applied on the $i$th qubit}
for any $i=1,\cdots,n$, and $S$ is some set of pairs of qubits.
{Several typical physical models, such as the Ising model, are described as 2-body Hamiltonians.
Furthermore, as} the larger the size of interactions is, the more difficult the implementation is, 
and thus the limitation {to 2-body Hamiltonian would be} reasonable from {{a} practical perspective}.
{We treat a 2-body} Hamiltonian as a graph 
by regarding {qubits and couplings as vertices and edges, respectively (see also Fig.}~2).
We show the following result:

\begin{Result}
Consider a 2-body locally diagonalizable
$n$-qubit Hamiltonian $H_{LD}$, as a graph 
by regarding qubits and couplings as vertices and edges, respectively (see also Fig.~2).
Let $d_k$ be the degree of a vertex $v_k$
which corresponds to 
{the $k$th qubit}
(i.e., the number of edges connected 
to the vertex $v_k$) and 
$d=(d_1,d_2,\cdots,d_n)$.
The maximal QFI of all quantum states 
{and} the maximal QFI of 
all {symmetric product states}
are {respectively} given
as follows:
\begin{align}
\label{max_arbitrary_intro}
\max_{{\ket{\Psi}}:\mathrm{arbitrary}}
F_Q \left(\ket{\Psi},H_{LD}\right)
&=\Theta \left((\|d\|_1 )^2\right)
=\Theta \left((|d_1|+\cdots+|d_n|)^2\right),\\
\label{max_sep_intro}
\max_{{\ket{\phi}^{\otimes n}}: \mathrm{product}}F_Q\left({\ket{\phi}^{\otimes n}},H_{LD}\right) 
&=\Theta \left((\|d\|_2 )^2\right)
=\Theta \left(|d_1|^2+\cdots+|d_n|^2\right).    
\end{align}
This means that
for arbitrary 
2-body locally diagonalizable
$n$-qubit Hamiltonians $H_{LD}$ 
such that
{the scaling of a 2-norm $\|d\|_2=\sqrt{|d_1|^2+\cdots+|d_n|^2}$ with respect to $n$
is  different
from that of a 1-norm $\|d\|_1=|d_1|+\cdots+|d_n|$,
the scaling of
the maximal QFI of all {symmetric product states}
with respect to $n$
is different
from that of all quantum states.}
That is,
the accuracy attained by a {symmetric product state}
 is much lower than
that of an optimal state in all quantum states.
\end{Result}

The {values} inside the parentheses
in (\ref{max_arbitrary_intro}) {and} (\ref{max_sep_intro}) are the same for star-shaped Hamiltonians,
and hence {\red Result~4} implies that
Claim {1} does not hold for these Hamiltonians.
In contrast, for ring-shaped{,} chain-shaped{,} and {\red fully connected} Hamiltonians, 
the {scalings} of 
the {values} inside the parentheses
in (\ref{max_arbitrary_intro}) {and} (\ref{max_sep_intro}) are different{,}
and hence
Claim~1 holds.

Thus, Claim~1 is expected to hold
for 2-body Hamiltonians on
regular graphs
(i.e., graphs where each vertex has the same degree.)
This expectation indeed holds from Result 4.
This is because a set of
Hamiltonians stated in Result~4
includes
a set of 
Hamiltonians on all regular graphs
and Hamiltonians on 
the graphs that are similar to regular graphs.

In conclusion, 
we demonstrate the existence of { universal resource} states for quantum metrology 
for a certain class of linear Hamiltonians.
In addition, we show that too entangled states are not useful in quantum metrology for a wider class 
of Hamiltonians including
linear Hamiltonians.
Since we analyze a wider class of 
Hamiltonians than \cite{random_qfi},
experimenters
will be one step closer to the implementation of quantum metrology.

The outline of this paper is as follows:
In Section~4, we present the details of Result~1 and Result~3.
We adopt the notion of $\epsilon$-net \cite{epsilon_net,aspects} and
show that for an arbitrary linear Hamiltonian
such as (\ref{LH''}),
the QFI of random symmetric states
is $\Theta(n^2)$ (HL)
with high probability.
By the same discussion as the proof of Result~1,
we show that for an arbitrary Hamiltonian
in a set of locally diagonalizable Hamiltonians
parameterized by at most $d^{o(n)}$
parameters,
the QFI of random pure states
is at most almost the same as
that of the {\blue optimal separable state} with high probability.
{\red In} Section~5, we present the details of Result~2.
We show that
very high GME
leads to low values in QFI (not useful)
for linear Hamiltonians.
In Section~6, we present the details of Result~4.
%We prove that in a locally diagonalizable 2-body Hamiltonian on a regular graph, the difference between the maximal QFI of all quantum states and the maximal QFI of all symmetric product states is the largest.
{\red We} clarify
the class of locally diagonalizable 
2-body $n$-qubit Hamiltonians
in which
these two values above have different {\red scalings}
with respect to $n$.
Finally, we conclude our paper in Section~7.

\section{Preliminary}
Throughout this paper, we focus on phase estimation and 
consider $n$-qudit systems.
Let $H$ be an $n$-{\red qudit} Hamiltonian.
As we described in Section~{\red 2},
phase estimation is a concrete quantum sensing protocol and  proceeds as follows:
\begin{enumerate}[label=(\arabic*)]
\item Prepare a quantum state $\rho$ as a probe.
\item Interact $\rho$ with an object subject to sensing.
As a result, unknown parameter $\theta$ is encoded into the quantum state through the time evolution with the Hamiltonian $H$ corresponding to the object, and
\begin{equation}\label{rho_theta}
\rho_\theta= e^{-iH\theta}\rho e^{iH\theta}
\end{equation}
is obtained.
\item Estimate the phase $\theta$ of the quantum state $\rho_\theta$ by measuring $\rho_\theta$.
\end{enumerate}

Actually,
the fluctuation of the estimated value 
$\Delta^2 \hat{\theta}$ is given by the inverse of
the quantum Fisher information (QFI) $F_Q(\rho_\theta)$
as follows \cite{braunstein94,hayashi05}:
\begin{equation}\label{CRB}
\Delta^2 \hat{\theta} \geq 1/F_Q(\rho_\theta).
\end{equation}
In our case such as
a single-parameter estimation,
the equality in (\ref{CRB}) holds
when an optimal measurement is performed.
In phase estimation, 
for a pure state $\rho=\ket{\psi}\bra{\psi}$,
the QFI can be computed as follows:
\begin{equation}\tag{\ref{qfi}}
F_Q(\ket{\psi},H)
\coloneqq F_Q(e^{-iH\theta}\rho e^{iH\theta})
=4(\langle \psi |H^2| \psi \rangle-
\langle \psi |H| \psi \rangle^2).
\end{equation}
Define $f(\psi)$ as 
the following
function of a pure state $\rho=\ket{\psi}\bra{\psi}$:
\begin{equation*}
f(\psi)
=\frac{1}{4}F_Q(e^{-iH\theta}\rho e^{iH\theta})
=\langle \psi |H^2| \psi \rangle-
\langle \psi |H| \psi \rangle^2.
\end{equation*}
Let $H_{\mathrm{L}}$ be a linear Hamiltonian
such as
\begin{equation}\tag{\ref{LH''}}
H_{L}=h_1 \otimes I \otimes \cdots \otimes I
+ I \otimes h_2 \otimes I \otimes \cdots \otimes I
+\cdots 
+ I \otimes \cdots \otimes I \otimes h_n,
\end{equation}
for some single-qudit Hermitian operator
$h_i=\sum_{j=1}^{d} \lambda_{i,j} \ket{\phi_j}\bra{\phi_j}$.
Let $H_{\mathrm{S}}$ be a linear Hamiltonian
such as
\begin{equation}\label{LH''_S}
H_{S}=h_S \otimes I \otimes \cdots \otimes I
+ I \otimes h_S \otimes I \otimes \cdots \otimes I
+\cdots 
+ I \otimes \cdots \otimes I \otimes h_S,
\end{equation}
for some single-qudit non-degenerate Hermitian operator
$h_S=\sum_{j=1}^{d} \lambda_{j} \ket{\phi_j}\bra{\phi_j}$
and there exists $j \neq j'$ such that
$\lambda_j \neq \lambda_{j'}$.
Note that 
$\{\ket{\phi_1},\ket{\phi_2},\cdots,\ket{\phi_d}\}$
is a fixed orthonormal basis.
When $h_1=h_2=\cdots=h_n$ holds,
a Hamiltonian $H_L$ has the form of $H_S$.
That is, a Hamiltonian $H_S$ is a special case of
$H_L$.

Given $\pi \in S_n$ an element of the symmetric group $S_n$,
let $V_d(\pi)$ be the permutation {\red matrix},
namely the unitary matrix that satisfies
\[
V_d(\pi)\ket{\psi_1}\otimes \cdots \otimes \ket{\psi_n}
=\ket{\psi_{\pi^{-1}(1)}}{\red\otimes}\cdots \otimes \ket{\psi_{\pi^{-1}(n)}},
\]
for all $\ket{\psi_1}, \cdots, \ket{\psi_n}
\in \mathbb{C}^d$.

For a finite-dimensional complex linear space $\mathcal{H}$, let $Sym^k(\mathcal{H})$ be
a symmetric subspace of $\mathcal{H}^{\red k}$,
i.e.,
\[
Sym^k(\mathcal{H})=\{\ket{\psi} \in \mathcal{H}^{\otimes k}
: V_d(\pi)\ket{\psi}=\ket{\psi},\forall \pi \in S_k\}.
\]
Let $Asym^k(\mathcal{H})$ be
an anti-symmetric subspace of
$\mathcal{H}^k=(\mathbb{C}^{d^n})^{\otimes k}$,
i.e.,
\[
Asym^k(\mathcal{H})=\{\ket{\psi} \in \mathcal{H}^{\otimes k}
: V_d(\pi)\ket{\psi}=sgn(\pi)\ket{\psi},\forall \pi \in S_k\}.
\]

Denote by $\ket{\psi} \leftarrow
Sym^n(\mathbb{C}^d)$,
a quantum state
sampled uniformly at random
from all symmetric $n$-qudit pure states.
Denote by $\ket{\psi} \leftarrow
(\mathbb{C}^{\red d})^{\otimes n}$,
a quantum state
sampled uniformly at random
from all $n$-qudit pure states.
This $\ket{\psi} \leftarrow
(\mathbb{C}^d)^{\otimes n}$
is called Haar random states \cite{haar_random}.

For a vector
$v =(v_1,\cdots,v_{\red d})
\in \mathbb{C}^d$
and $p\in [1,\infty]$,
the {\red $p$}-norm of $v$ is denoted by
$\|v\|_p$ {\red and} is
defined as 
$
\|v\|_p \coloneqq
 \left(\sum_{i=1}^{d} |v_i|^p\right)^{1/p}.
 $
The Schatten {\red $p$}-norm of a  
$d \times d$ matrix A
is denoted by
$\|A\|_p$ {\red and} is
defined as 
$
\|A\|_p \coloneqq
 \mathrm{Tr}
 ((\sqrt{A^{\dag} A})^p)^{1/p}.
$
The infinite norm, denoted as
$\|\cdot\|_\infty$, of a matrix is defined as its largest singular value.

\section{Details of Result~1 and Result~3}
In Section~4, we present the details of Result~1
and Result~3.
For proofs, see Appendix.

\subsection{The expectation of the quantum Fisher information (QFI) of random pure states and
that of random symmetric states~\cite{random_qfi}}

For an arbitrary (fixed) Hamiltonian,
the expectation of QFI of 
random pure states and that of random symmetric states {\red are} given 
as follows~\cite[Appendix~C]{random_qfi}:

\begin{lemma}\label{expectation}
(The expectation of the QFI of 
random pure states and 
that of random symmetric states)
The expectation of QFI $f(\psi)$
of Haar random states is
\begin{equation}\label{expectation_random}
\underset{\ket{\psi} \leftarrow
(\mathbb{C}^d)^{\otimes n}}{E}[f(\psi)]
=\frac{\mathrm{Tr}[H^2]}{d^n+1}
-\frac{\mathrm{Tr}[H]^2}{d^n(d^n+1)}.
\end{equation}
That of random symmetric states is
\begin{align}\label{expectation_symmetric}
\underset{\ket{\psi} \leftarrow Sym^n(\mathbb{C}^d)}{E}[f(\psi)]
&=\frac{\mathrm{Tr}[\Pi_{Sym^n(\mathbb{C}^d)}H^2\Pi_{Sym^n(\mathbb{C}^d)}]}{|D|+1}-\frac{\mathrm{Tr}[\Pi_{Sym^n(\mathbb{C}^d)}H\Pi_{Sym^n(\mathbb{C}^d)}]^2}{|D|(|D|+1)},
\end{align}
where $\Pi_{Sym^n(\mathbb{C}^d)}:
(\mathbb{C}^d)^{\otimes n} \rightarrow Sym^n(\mathbb{C}^d)$
is a projection:
\[
\Pi_{Sym^n(\mathbb{C}^d)}=\frac{1}{n!}
\sum_{\pi \in S_n} V_d(\pi)
\]
and $|D|=\dim Sym^n(\mathbb{C}^d)=_{n+d-1}C_n$.
\hfill $\blacksquare$
\end{lemma}

Lemma~\ref{expectation}
follows from (C12) of 
\cite[Appendix~C]{random_qfi},
but we give a proof in 
Appendix in our paper.

\subsection{The concentration of the QFI of random pure states and that of random symmetric states~\cite{random_qfi}}

From Levy's lemma \cite{haar_random,aspects},
the values of
the QFI of random pure states
concentrate on the expectation
given in Lemma~\ref{expectation}.
In other words,
the QFI of random pure states
is almost the same as
the expectation
given in Lemma~\ref{expectation},
with high probability.

\begin{lemma}\label{concentration}
(Concentration of the QFI of random pure states and 
that of random symmetric states)
Let $\epsilon>0$.
Then,
\begin{align*}
 \underset{\ket{\psi} \leftarrow
(\mathbb{C}^d)^{\otimes n}}{\mathrm{Prob}}\left(
\left|
f(\psi)-
\underset{\ket{\psi} \leftarrow
(\mathbb{C}^d)^{\otimes n}}{E}[f(\psi)]
\right|
\geq \epsilon
\right)
\leq &2 \exp
\left(
-\frac{2d^{n} \epsilon^2}{
9\pi^3(2\|H^2\|_{\infty}+
2\sqrt{2} \|H\|^2_{\infty})^2
}
\right),\\
\underset{\ket{\psi} \leftarrow
(\mathbb{C}^d)^{\otimes n}}{\mathrm{Prob}}\left(
f(\psi)-
\underset{\ket{\psi} \leftarrow
(\mathbb{C}^d)^{\otimes n}}{E}[f(\psi)]
> \epsilon
\right)
\leq &2 \exp
\left(
-\frac{2 d^{n} \epsilon^2}{
9\pi^3 \log_e 2 (2\|H^2\|_{\infty}+
2\sqrt{2} \|H\|^2_{\infty})^2
}
\right).
\end{align*}
Furthermore,
\begin{align*}
 \underset{\ket{\psi} \leftarrow
Sym^n(\mathbb{C}^d)}{\mathrm{Prob}}\left(
\left|
f(\psi)-
\underset{\ket{\psi} \leftarrow
Sym^n(\mathbb{C}^d)}{E}[f(\psi)]
\right|
\geq \epsilon
\right)
\leq &2 \exp
\left(
-\frac{2 _{n+d-1}C_n\epsilon^2}{
9\pi^3(2\|H^2\|_{\infty}+
2\sqrt{2} \|H\|^2_{\infty})^2
}
\right),\\
\underset{\ket{\psi} \leftarrow
Sym^n(\mathbb{C}^d)}{\mathrm{Prob}}\left(
f(\psi)-
\underset{\ket{\psi} \leftarrow
Sym^n(\mathbb{C}^d)}{E}[f(\psi)]
< -\epsilon
\right)
\leq &2 \exp
\left(
-\frac{2 _{n+d-1}C_n \epsilon^2}{
9\pi^3 \log_e 2 (2\|H^2\|_{\infty}+
2\sqrt{2} \|H\|^2_{\infty})^2
}
\right).
\end{align*}
\hfill $\blacksquare$
\end{lemma}

\cite{random_qfi} analyzed 
linear Hamiltonians such as (\ref{LH})
and 
\cite[Theorem 1]{random_qfi}
is a special case of the first half of Lemma~\ref{concentration} above.
The second half of Lemma~\ref{concentration} above
is given in
\cite[Theorem 2]{random_qfi}.

\subsection{The class of Hamiltonians for our analysis}

Hereafter,
we analyze locally diagonalizable Hamiltonians which 
are represented by Hermitian operators
diagonalized by a product basis.

When a considered quantum system is
$n$-qubit system,
locally diagonalizable Hamiltonians
are described as follows:
\begin{align}\label{localhamiltonian_qubit}
H_{LD}
=\sum_{(i_1,\cdots, i_n)} \lambda_{(i_1,\cdots, i_n)}
\left(
\bigotimes_{j ={\red 1}}^{n}\{\ket{\phi_j}\bra{\phi_j}\ \mathrm{if} \ i_j=0 \ \mathrm{or}\
\ket{\phi_j^{\perp}}\bra{\phi_j^{\perp}}\ \mathrm{if} \ i_j=1\}
\right),
\end{align}
where $(i_1,\cdots, i_n) \in 
\{0,1\} \times \cdots \times \{0,1\}$ and  
$\{\ket{\phi_j},\ket{\phi_j^{\perp}}\}$ 
is an orthonormal basis (ONB) of 
$\mathbb{C}^2$.
When a considered quantum system is
$n$-qudit system,
locally diagonalizable Hamiltonians
are described as follows:
\begin{align}\label{localhamiltonian}
H_{LD}
=\sum_{(i_1,\cdots, i_n)} \lambda_{(i_1,\cdots, i_n)}
\left(
\bigotimes_{j ={\red 1}}^{n}\{\ket{\phi_k}_{j}\bra{\phi_k}_{j}\ \mathrm{if} \ i_j=k \}
\right),
\end{align}
where $(i_1,\cdots, i_n) \in  \{1,2,\cdots,d\} \times \cdots \times \{1,2,\cdots,d\}$ and  
$\{\ket{\phi_1}_j, \cdots, \ket{\phi_d}_j\}$ 
is an ONB of $\mathbb{C}^d$
for all $j=1,\cdots,n$.
Note that an underscript $j$ ($j=1,\cdots,n$) means 
being a quantum state in the $j$-th quantum system
$\mathbb{C}^d$
(the whole quantum system is $(\mathbb{C}^d)^{\otimes n}$).
Note that for all $j,j'=1,\cdots,n$, 
$\ket{\phi_k}_j$ and $\ket{\phi_k}_{j'}$
{\red are} not necessarily the same.

Linear Hamiltonians are
a special case of locally diagonalizable Hamiltonians.
When $H$ is a linear Hamiltonian 
which has the following form:
\begin{equation*}
H_{\mathrm{S}}=h_S \otimes I \otimes \cdots \otimes I
+ I \otimes h_S \otimes I \cdots \otimes I
+\cdots 
+ I \otimes \cdots \otimes I \otimes h_S,
\end{equation*}
for some single-qubit non-degenerate Hermitian operator $h_S=\lambda_0 \ket{\phi_0}\bra{\phi_0}
+\lambda_1 \ket{\phi_1}\bra{\phi_1}$,
the coefficients which appear in (\ref{localhamiltonian_qubit}) are
\[
\lambda_{(i_1,\cdots,i_n)}=(n-k)\lambda_0+k \lambda_1
\]
where $k=i_1+i_2+\cdots+i_n$.

\subsection{Evaluation of the expectation of QFI of 
random pure states and that of random symmetric states}

\cite{random_qfi} 
computes
the expectation of QFI of 
random pure states and
that of symmetric random states
for linear Hamiltonians which has 
the following form:
\begin{equation}\tag{\ref{LH''_S}}
H_{S}=h_S \otimes I \otimes \cdots \otimes I
+ I \otimes h_S \otimes I \otimes \cdots \otimes I
+\cdots 
+ I \otimes \cdots \otimes I \otimes h_S,
\end{equation}
for some single-qudit non-degenerate Hermitian operator
$h_S=\sum_{j=1}^{d} \lambda_{j} \ket{\phi_j}\bra{\phi_j}$
and there exists $j \neq j'$ such that
$\lambda_j \neq \lambda_{j'}$.

The expectation of QFI of 
random pure states and
that of symmetric random states
{\red are} given as follows
\cite[Appendix~C]{random_qfi}:

\begin{lemma}\label{gutaitekinaatai}
(The expectation of QFI of 
random pure states and
that of symmetric random states
for linear Hamiltonians
such as (\ref{LH''_S}))
For linear Hamiltonians
such as (\ref{LH''_S}),
\begin{align*}
\underset{\ket{\psi} \leftarrow  (\mathbb{C}^d)^{\otimes n}}{E}[f(\psi)] 
&=\frac{1}{4}
\underset{\ket{\psi} \leftarrow  (\mathbb{C}^d)^{\otimes n}}{E}[F_Q(\ket{\psi},H_S)] 
=n\frac{d^n}{d^n+1}
  \left(\frac{\mathrm{Tr}(h_S^2)}{d}
  -\frac{\mathrm{Tr}(h_S)^2}{d^2}
  \right).\\
\underset{\ket{\psi} \leftarrow Sym^n(\mathbb{C}^d)}{E}[f(\psi)] 
&=\frac{1}{4}\underset{\ket{\psi} \leftarrow  Sym^n(\mathbb{C}^d)}{E}[F_Q(\ket{\psi},H_S)]
=\frac{n(n+d)}{d+1}\frac{_{n+d-1}C_n}{_{n+d-1}C_n +1}
\left(\frac{\mathrm{Tr}(h_S^2)}{d}
  -\frac{\mathrm{Tr}(h_S)^2}{d^2}
  \right).
\end{align*}
\hfill $\blacksquare$
\end{lemma}

Lemma~\ref{gutaitekinaatai}
follows from (C27) and (C28)
of \cite[Appendix~C]{random_qfi}.

The expectation of QFI of 
random pure states is $\Theta(n)$ and 
is almost the same as 
QFI of an optimal separable state.
On the other hand,
the expectation of QFI of 
random symmetric states is $\Theta(n^2)$ and 
is almost the same as 
QFI of a truly optimal state.

\cite{random_qfi} evaluated
the expectation of QFI
of random pure states and
that of random symmetric states
only for linear Hamiltonians
which has the form of (\ref{LH''_S}).
Unlike \cite{random_qfi},
we analyze a wider class of Hamiltonians.
Concretely,
we give the evaluation of the following values:
\begin{itemize}
\item (in Proposition~\ref{lessthansep})
the expectation of QFI
of random pure states for 
locally diagonalizable Hamiltonians
which has the following form:
\begin{align}\tag{\ref{localhamiltonian}}
H_{LD}
=\sum_{(i_1,\cdots, i_n)} \lambda_{(i_1,\cdots, i_n)}
\left(
\bigotimes_{j ={\red 1}}^{n}\{\ket{\phi_k}_{j}\bra{\phi_k}_{j}\ \mathrm{if} \ i_j=k \}
\right),
\end{align}
where $(i_1,\cdots, i_n) \in  \{1,2,\cdots,d\} \times \cdots \times \{1,2,\cdots,d\}$ and  
$\{\ket{\phi_1}_j, \cdots, \ket{\phi_d}_j\}$ 
is an orthonormal basis (ONB) of $\mathbb{C}^d$.
\item (in Proposition~\ref{largerthanHS})
the expectation of QFI
of random symmetic states for 
linear Hamiltonians which {\red have} the following form:
\begin{equation}\tag{\ref{LH''}}
H_{L}=h_1 \otimes I \otimes \cdots \otimes I
+ I \otimes h_2 \otimes I \otimes \cdots \otimes I
+\cdots 
+ I \otimes \cdots \otimes I \otimes h_n,
\end{equation}
for some single-qudit Hermitian operator
$h_i=\sum_{j=1}^{d} \lambda_{i,j} \ket{\phi_j}\bra{\phi_j}$
and there exists $j \neq j'$ such that
$\sum_{i=1}^n \lambda_{i,j} {\red -} \sum_{i=1}^n \lambda_{i,j'}{\red =\Theta(n)}$.
\end{itemize}

For any locally diagonalizable Hamiltonians
such as (\ref{localhamiltonian}),
we first prove that
the expectation of QFI of random pure states
is less than or equal to 
the QFI of 
an {\blue optimal separable state}.

\begin{proposition}\label{lessthansep}
(For any locally diagonalizable Hamiltonians
such as (\ref{localhamiltonian}),
the QFI of random pure states
is less than or equal to
the QFI of an {\blue optimal separable state}.)
For any locally diagonalizable Hamiltonians $H_{LD}$
such as (\ref{localhamiltonian}),
\[
4\underset{\ket{\psi} \leftarrow (\mathbb{C}^d)^{\otimes n}}{E}[f(\psi)] 
=\underset{\ket{\psi} \leftarrow (\mathbb{C}^d)^{\otimes n}}{E}[F_Q(\ket\psi,H_{LD})] 
\leq \max_{\ket\phi: \mathrm{separable}}F_Q(\ket\phi,H_{LD}),
\]
where $\ket\phi^{\otimes n}$ is 
a symmetric product state with respect to
product basis which {\red diagonalizes} a Hamiltonian $H_{LD}$.
\hfill $\blacksquare$
\end{proposition}

From Lemma~{\ref{concentration}} and Proposition~{\ref{lessthansep}},
it can be said that
random pure states can only achieve at most the same accuracy as that of an {\blue optimal separable state} with high probability,
in quantum metrology for 
locally diagonalizable Hamiltonians.

For an arbitrary linear Hamiltonian $H_L$
such as (\ref{LH''}),
we prove that
there exists a linear Hamiltonian $H_S$
such that 
which has the form of (\ref{LH''_S}) and
the expectation of QFI of random symmetric states for $H_L$
is larger than or equal to 
the expectation of QFI of random symmetric states for $H_S$.
This means that 
the expectation of QFI of random symmetric states for linear Hamiltonians such as 
(\ref{LH''}) is $\Theta(n^2)$.

\begin{proposition}\label{largerthanHS}
(The expectation of QFI of random symmetric states for linear Hamiltonians such as 
(\ref{LH''}) is $\Theta(n^2)$.)
Let $H_L$ be 
an arbitrary linear Hamiltonian
which has the following form:
\begin{equation}\tag{\ref{LH''}}
H_{L}=h_1 \otimes I \otimes \cdots \otimes I
+ I \otimes h_2 \otimes I \otimes \cdots \otimes I
+\cdots 
+ I \otimes \cdots \otimes I \otimes h_n,
\end{equation}
for some single-qudit Hermitian operator
$h_i=\sum_{j=1}^{d} \lambda_{i,j} \ket{\phi_j}\bra{\phi_j}$
and there exists $j \neq j'$ such that
$\sum_{i=1}^n \lambda_{i,j}- \sum_{i=1}^n \lambda_{i,j'}=\Theta(n)$.
Define
a linear Hamiltonian $H_S'$ as follows:
\begin{equation}\label{H_S}
H_S'=\frac{1}{|S_n|} \sum_{\pi \in S_n} V_d(\pi) \ H_L \ V_d(\pi),
\end{equation}
where $|S_n|=n!$ is the number of
elements in the symmetric group $S_n$.
Then, the following {\red inequality} holds:
\begin{align*}
\underset{\ket{\psi} \leftarrow Sym^n(\mathbb{C}^d)}{E}[F_Q(\ket{\psi},H_L)] 
&\geq \underset{\ket{\psi} \leftarrow Sym^n(\mathbb{C}^d)}{E}[F_Q(\ket{\psi},H'_S)]\\
&=\frac{4 n(n+d)}{d+1}\frac{_{n+d-1}C_n}{_{n+d-1}C_n +1}
\left(\frac{\mathrm{Tr}({h_S'}^2)}{d}
  -\frac{\mathrm{Tr}(h_S')^2}{d^2}
  \right).   
\end{align*}
\hfill $\blacksquare$
\end{proposition}

\noindent
Note that the Hamiltonian $H_S'$
defined in (\ref{H_S}) is represented as follows:
\begin{equation}
H_S'=h_S' \otimes I \otimes \cdots \otimes I
+ I \otimes h_S' \otimes I \otimes \cdots \otimes I
+\cdots 
+ I \otimes \cdots \otimes I \otimes h_S'
\end{equation}
for the following single-qudit non-degenerate Hermitian operator
\[
h_S'=\sum_{j=1}^{d} 
\frac{\sum_{i=1}^n \lambda_{i,j}}{n} 
\ket{\phi_j}\bra{\phi_j}.
\]
Note that
\[
\frac{\mathrm{Tr}({h_S'}^2)}{d}
  -\frac{\mathrm{Tr}(h_S')^2}{d^2}> 0.
\]
It follows from 
the assumption that there exists $j \neq j'$ such that
$\sum_{i=1}^n \lambda_{i,j}- \sum_{i=1}^n \lambda_{i,j'}=\Theta(n)$.

From Lemma~{\ref{concentration}} and Proposition~{\ref{largerthanHS}},
the QFI of random symmetric states
for linar Hamiltonians such as (\ref{LH''})
concentrates on $\Theta(n^2)$.
This means that
random symmetric states can achieve
almost the same accuracy as that of 
an optimal state with high probability.

\subsection{Definition of a set of local diagonalizable Hamiltonians}

In this section,
we define a set of local diagonalizable Hamiltonians.
First, we define 
a set $S_{LD}^{\mathrm{Result\ 3}}$ of local diagonalizable Hamiltonians,
which appears in Result~3.
Then, as a special case of $S_{LD}^{\mathrm{Result\ 3}}$,
we define 
a set $S_{LD}^{\mathrm{Result\ 1}}$ of local diagonalizable Hamiltonians,
which appears in Result~1.
Finally, 
we define a set {\red $S_{L}$} of linear Hamiltonians, as a subset of $S_{LD}^{\mathrm{Result\ 1}}$.
The set  $S_{L}$ appears in Result~1. 

\subsubsection{Definition of a set $S_{LD}^{\mathrm{Result\ 3}}$ of local diagonalizable Hamiltonians}

First, we define the following set:
\begin{align*}
I_{\mathrm{coff}}&=\{1,2,\cdots,s_{\mathrm{coff}}\},\\
I_{\mathrm{basis}}
&=\{i_1,i_2,\cdots,i_{s_{\mathrm{basis}}}\}
\subset \{1,\cdots, n\},\\
I_{\mathrm{basis}}^c
&= \{1,\cdots, n\}-I_{\mathrm{basis}}.
\end{align*}
Let
$\{\ket{1}_j, \cdots, \ket{d}_j\}$
be an orthonormal basis (ONB) of $\mathbb{C}^d$.
Also, let
$\{\ket{\phi_1}_j, \cdots, \ket{\phi_d}_j\}$
be an ONB of $\mathbb{C}^d$.
For all $m \in I_{\mathrm{coff}}\cup \{0\}$,
we define
\begin{align*}
A_m&=\sum_{(i_1,\cdots, i_n)}
a_{(i_1,\cdots, i_n),m}
\left(
\bigotimes_{j \in I_{\mathrm{basis}}}\{\ket{\phi_k}_j\bra{\phi_k}_j\ \mathrm{if} \ i_j=k\}
\right) \otimes \left(
\bigotimes_{j \in I_{\mathrm{basis}}^c}\{\ket{k}_j\bra{k}_j\ \mathrm{if} \ i_j=k\}
\right),
\end{align*}
where $(i_1,\cdots, i_n) \in  \{1,2,\cdots,d\} \times \cdots \times \{1,2,\cdots,d\}$and  
$\{\ket{\phi_1}_j, \cdots, \ket{\phi_d}_j\}$ 
is an ONB of $\mathbb{C}^d$
for all $j=1,\cdots,n$.
Note that an underscript $j$ ($j=1,\cdots,n$) means 
being a quantum state in the $j$-th quantum system
$\mathbb{C}^d$
(the whole quantum system is $(\mathbb{C}^d)^{\otimes n}$).
Note that for all $j,j'=1,\cdots,n$, 
$\ket{\phi_k}_j$ and $\ket{\phi_k}_{j'}$
{\red are} not necessarily the same,
and $\ket{k}_j$ and $\ket{k}_{j'}$
{\red are} not necessarily the same.
By using this, we define
a set $S_{LD}^{\mathrm{Result\ 3}}$ of Hamiltonians as follows:
\begin{align*}
S_{LD}^{\mathrm{Result\ 3}}
=\Biggl\{
H
=A_0+\sum_{m \in I_{\mathrm{coff}}} \mu_m A_m
: \mu_m \in [-B,-A]&\cup [A,B]\ \ \mathrm{and}\ \ 
\{\ket{\phi_1}, \cdots, \ket{\phi_d}\}\ 
\mathrm{is} \ \ \mathrm{ONB}\ \ \mathrm{of} \ \ 
\mathbb{C}^d
\Biggr\}
\end{align*}
where $B>A>0$ and 
$A_0$ is a fixed Hamiltonian.

\subsubsection{Definition of a set $S_L$ of linear Hamiltonians in Result~1}

We define a set of coefficient as:
\[
I_{\mathrm{coff}}=\{1,2,\cdots,n\}\times \{1,2,\cdots,d\}{\red,}
\]
a Hamiltonian $A_{j,k}$ for $(j,k) \in I_{\mathrm{coff}}$ as:
\begin{align*}
A_{1,k} =&\ket{\phi_k}\bra{\phi_k} \otimes I \otimes I \otimes \cdots \otimes I \otimes I,\\
A_{2,k} =&I \otimes \ket{\phi_k}\bra{\phi_k} \otimes I \otimes \cdots \otimes I \otimes I,\\
&\vdots \\
A_{n,k}=&I \otimes I \otimes I \otimes \cdots \otimes I \otimes \ket{\phi_k}\bra{\phi_k}{\red ,}
\end{align*}
and a fixed Hamiltonian $A_0$ as $A_0=0$.
Then, 
\begin{align*}
S_{LD}^{\mathrm{Result\ 1}}&=\Biggl\{
H
=\sum_{(j,k) \in I_{\mathrm{coff}}} \mu_{j,k} A_{j,k}
:\mu_{j,k} \in [-B,-A] \cup [A,B],
\ \ \mathrm{and}\ \ 
\{\ket{\phi_1}, \cdots, \ket{\phi_d}\}\ 
\mathrm{is} \ \ \mathrm{ONB}\ \ \mathrm{of} \ \ 
\mathbb{C}^d
\Biggr\}\\
&{\red=}\Biggl\{
H
=h_1 \otimes I \otimes \cdots \otimes I+ \cdots+ I \otimes \cdots \otimes I \otimes h_n
{\red:}\mu_{j,k} \in [-B,-A] \cup [A,B],
\{\ket{\phi_1}, \cdots, \ket{\phi_d}\}\ 
\mathrm{is} \ \ \mathrm{ONB}\ \ \mathrm{of} \ \ 
\mathbb{C}^d
\Biggr\},
\end{align*}
for some single-qudit Hermitian operator
$h_j=\sum_{k=1}^{d} \mu_{j,k} \ket{\phi_k}\bra{\phi_k}$.
We define $S_L$ as follows:
\begin{align*}
S_{L}
=\Biggl\{
H
=h_1 \otimes I \otimes \cdots \otimes I+ \cdots+ I \otimes \cdots \otimes I \otimes h_n
&{\red:}\mu_{j,k} \in [-B,-A] \cup [A,B],\\
&\ \ \{\ket{\phi_1}, \cdots, \ket{\phi_d}\}\ 
\mathrm{is} \ \ \mathrm{ONB}\ \ \mathrm{of} \ \ 
\mathbb{C}^d,\\
&\ \ \exists j \neq j' \ \mathrm{s.t.}
\sum_{i=1}^n \lambda_{i,j} - \sum_{i=1}^n \lambda_{i,j'}=\Theta(n)
\Biggr\}.
\end{align*}

\subsection{Construction of an $\epsilon$-net $\mathcal{N}_S$
for {\red sets} of Hamiltonians $S_{LD}^{\mathrm{Result\ 1}},S_{LD}^{\mathrm{Result\ 3}}$}

In this section,
we construct an $\epsilon$-net $\mathcal{N}_S$
for {\red sets} of Hamiltonians $S_{LD}^{\mathrm{Result\ 1}},S_{LD}^{\mathrm{Result\ 3}}$.
First, we define a set $\mathcal{N}_S$ and
prove that $\mathcal{N}_S$ is 
an $\epsilon$-net
for {\red sets} of Hamiltonians $S_{LD}^{\mathrm{Result\ 1}},S_{LD}^{\mathrm{Result\ 3}}$,
that is,
\begin{enumerate}[label=(\Roman*)]
\item A set $\mathcal{N}_S$ is finite.
\item For an arbitrary $H \in S_{LD}^{\mathrm{Result\ 1}}(S_{LD}^{\mathrm{Result\ 3}})$,
there exists $H_{\mathrm{rep}} \in \mathcal{N}_S$ such that
\[
\|H- H_{\mathrm{rep}}\|_\infty \leq \epsilon.
\]
\end{enumerate}

First, we define an $\epsilon$-net
of pure states.
Let $\mathcal{H}$ be a $\mathbb{C}$-linear vector space of dimension $D$.
By \cite[Lemma II.4]{epsilon_net},
for $0< \epsilon_p<1$,
there exists a set $\mathcal{N}_{T,\mathcal{H}}$ of pure {\red states}($\in \mathcal{H}$) such that
\[
|\mathcal{N}_{T,\mathcal{H}}| \leq\left(\frac{5}{\epsilon_p}\right)^{2D}.
\]
That is,
for an arbitrary $\ket{\phi} \in \mathcal{H}$,
there exists 
$\ket{\tilde{\phi}} \in \mathcal{N}_{T,\mathcal{H}}$ such that 
\[
\|\ket{\phi}\bra{\phi}-\ket{\tilde{\phi}}\bra{\tilde{\phi}}\|_1 \leq 
2\|\ket{\phi}-\ket{\tilde{\phi}}\|_2 \leq \epsilon_p.
\]

Let $0< \epsilon_c, \epsilon_p<1$.
We define a set $\mathcal{N}_S$ as follows:
\begin{align*}
\mathcal{N}_S
=\Biggl\{
H
=B_0+\sum_{m \in I_{\mathrm{coff}}} ({\blue {\blue \pm B}} {\blue \mp 2\epsilon_c} k_m) B_m&:k_m \in \left\{0,1,2, \cdots, \left\lceil \frac{B-A}{2 \epsilon_c}\right\rceil \right\},\\
&\ \  \ket{\Phi_1} \in \mathcal{N}_{T,\mathbb{C}^d},\ 
\ket{\Phi_2} \in \mathcal{N}_{T,\mathrm{span}\{\ket{\Phi_1}\}^{\perp}},\ 
\cdots,\\
&\ \ \ket{\Phi_{d-1}} \in \mathcal{N}_{T,\mathrm{span}\{\ket{\Phi_1},\cdots,\ket{\Phi_{d-2}}\}^{\perp}}, \ket{\Phi_d}\in \mathrm{span}\{\ket{\Phi_1},\cdots,\ket{\Phi_{d-1}}\}^{\perp}
\Biggr\}.
\end{align*}
Here, for all $m \in I_{\mathrm{coff}} \cup \{0\}$,
\begin{align*}
B_m&=\sum_{(i_1,\cdots, i_n)}
a_{(i_1,\cdots, i_n),m}
\left(
\bigotimes_{j \in I_{\mathrm{basis}}}\{\ket{\Phi_k}_j\bra{\Phi_k}_j\ \mathrm{if} \ i_j=k\}
\right) \otimes \left(
\bigotimes_{j \in I_{\mathrm{basis}}^c}\{\ket{k}_j\bra{k}_j\ \mathrm{if} \ i_j=k\}
\right),
\end{align*}
where $(i_1,\cdots, i_n) \in  \{1,2,\cdots,d\} \times \cdots \times \{1,2,\cdots,d\}$.
Note that an underscript $j$ ($j=1,\cdots,n$) means 
being a quantum state in the $j$-th quantum system
$\mathbb{C}^d$
(the whole quantum system is $(\mathbb{C}^d)^{\otimes n}$).

Then, a set $\mathcal{N}_S$ defined above has the following property.

\begin{proposition}\label{onbchikai}
(The property of a set $\mathcal{N}_S$)
Let 
$\{\ket{\phi_1}, \cdots, \ket{\phi_d}\}$
be an arbitrary orthonormal basis of $\mathbb{C}^d$.
Then, there exists
\begin{align*}
\ket{\Phi_1} &\in \mathcal{N}_{T,\mathbb{C}^d},\\
\ket{\Phi_2} &\in \mathcal{N}_{T,\mathrm{span}\{\ket{\Phi_1}\}^{\perp}},\\
&\cdots\\
\ket{\Phi_{d-1}} &\in \mathcal{N}_{T,\mathrm{span}\{\ket{\Phi_1},\cdots,\ket{\Phi_{d-2}}\}^{\perp}},\\
\ket{\Phi_d}&\in \mathrm{span}\{\ket{\Phi_1},\cdots,\ket{\Phi_{d-1}}\}^{\perp}
\end{align*}
such that for all $k=1,\cdots,d$,
\begin{equation}\label{1-norm}
\|\ket{\phi_j}\bra{\phi_j}-\ket{\Phi_j}\bra{\Phi_j}\|_1 \leq C\epsilon_p
\end{equation}
and
\begin{equation}\label{2-norm}
\|\ket{\phi_j}-\ket{\Phi_j}\|_2 \leq \frac{C\epsilon_p}{2},
\end{equation}
where $C$ is a constant, which is independent of $n$.
\hfill $\blacksquare$
\end{proposition}

Denote an element of $\mathcal{N}_S$ by $H_\mathrm{rep}$
(a representative of $S_{LD}^{\mathrm{Result\ 1}},S_{LD}^{\mathrm{Result\ 3}}$).
Then,
an upper bound on the number of elements in $\mathcal{N}_S$
(a representative of $S_{LD}^{Result\ 3}$)
is given as follows:
\begin{equation}\label{kosuu2}
|\mathcal{N}_S| \leq 
\left(\frac{B-A}{\epsilon_c}+4\right)^{s_{\mathrm{coff}}}
\left(\frac{5}{\epsilon_p}\right)^{d(d+1) s_{\mathrm{basis}}}{\red.}
\end{equation}
Also,
an upper bound on the number of elements in $\mathcal{N}_S$
(a representative of $S_{LD}^{Result\ 1}$)
is given as follows:
\begin{equation}\label{kosuu}
|\mathcal{N}_S| \leq 
\left(\frac{B-A}{\epsilon_c}+4\right)^{dn}
\left(\frac{5}{\epsilon_p}\right)^{d(d+1)}{\red.}
\end{equation}
Thus, a set $\mathcal{N}_S$ is finite and (I) follows.

Moreover,
(II) holds and 
it follows that 
a set $\mathcal{N}_S$ is an $\epsilon$-net of $S_{LD}^{\mathrm{Result\ 1}},S_{LD}^{\mathrm{Result\ 3}}$:

\begin{proposition}\label{epsilonnet}
($\mathcal{N}_S$ satisfies (II) and 
a set $\mathcal{N}_S$ is an $\epsilon$-net of $S_{LD}^{\mathrm{Result\ 1}},S_{LD}^{\mathrm{Result\ 3}}$)
Fix $\epsilon>0$.
Set $\epsilon_p$ and $\epsilon_c$ as
\[
\epsilon_p=\frac{\epsilon}{2{\red\sqrt{2}}d C s_{\mathrm{basis}}
(s_{\mathrm{coff}}B a
+\|A_0\|_\infty)},\ 
\epsilon_c=\frac{\epsilon}{2 s_{\mathrm{coff}} a},
\]
where $a=\max_{m \in I_{\mathrm{coff}}} \|A_m\|_\infty$.
Then, for any $H \in S_{LD}$,
there exists
$H_{\mathrm{rep}} \in \mathcal{N}_S$ such that
\[
\|H- H_{\mathrm{rep}}\|_\infty \leq \epsilon.
\]
\hfill $\blacksquare$
\end{proposition}

\subsection{Property of an $\epsilon$-net $\mathcal{N}_S$}

We give the property of an $\epsilon$-net $\mathcal{N}_S$.
We prove
the following proposition 
related to Result~1,
by the similar discussion in 
a proof of Proposition~{\ref{epsilonnet}}:

\begin{proposition}\label{property1}
(Property of an $\epsilon$-net $\mathcal{N}_S$ in Result~1)
Fix $\epsilon>0$.
Set 
$\epsilon_p$ and $\epsilon_c$ as 
\begin{align*}
    \epsilon_p&=\frac{\epsilon}{8 \ {\red(1+2\sqrt{2})} d C s_{\mathrm{basis}}
(s_{\mathrm{coff}}Ba
+\|A_0\|_\infty)^2},\\
\epsilon_c&=\frac{\epsilon}{8
(2 s_{\mathrm{coff}} Ba
+ s_{\mathrm{coff}}^2(2B+{\red 2})a^2
+
2s_{\mathrm{coff}}
\|A_0\|_\infty
{\red a}
+{\red 4Bn(n+d)/d})},
\end{align*}
where $a=\max_{m \in I_{\mathrm{coff}}} \|A_m\|_\infty$.
Then,
for any $H \in S_{LD}^{\mathrm{Result\ 1}}$,
there exists $H_{\mathrm{rep}} \in \mathcal{N}_S$ such that
\begin{align*}
\forall \psi, 
\left|\left(F_Q(\ket\psi,H)-\underset{\ket{\psi} \leftarrow
Sym^n(\mathbb{C}^d)}{E}[F_Q(\ket{\psi},H_S)]\right)
-
\left(F_Q(\ket\psi,H_{\mathrm{rep}})-\underset{\ket{\psi} \leftarrow
Sym^n(\mathbb{C}^d)}{E}[F_Q(\ket{\psi},H_{\mathrm{rep},S})]\right)
\right|\leq
\epsilon,
\end{align*}
where 
\begin{equation}\label{H_S1}
H_S=\frac{1}{|S_n|} \sum_{\pi \in S_n} V_d(\pi) \ H \ V_d(\pi),
\end{equation}
and
\begin{equation}\label{H_S2}
H_{\mathrm{rep},S}=\frac{1}{|S_n|} \sum_{\pi \in S_n} V_d(\pi) \ H_{\mathrm{rep}} \ V_d(\pi).
\end{equation}
\hfill $\blacksquare$
\end{proposition}

Note that $H$ and $H_S$
 (\ref{H_S1}) are represented respectively as follows:
\begin{align*}
H&=h_{\red 1} \otimes I \otimes \cdots \otimes I
+ I \otimes h_{\red 2} \otimes I \otimes \cdots \otimes I
+\cdots 
+ I \otimes \cdots \otimes I \otimes h_{\red n}\\
H_S&=h_S \otimes I \otimes \cdots \otimes I
+ I \otimes h_S \otimes I \otimes \cdots \otimes I
+\cdots 
+ I \otimes \cdots \otimes I \otimes h_S
\end{align*}
for the following single-qudit Hermitian operators
$
h_{\red i}=\sum_{j=1}^{d} \mu_{i,j} 
\ket{\phi_j}\bra{\phi_j}$ and
$
h_S=\sum_{j=1}^{d} 
\frac{\sum_{i=1}^n \mu_{i,j}}{n} 
\ket{\phi_j}\bra{\phi_j}.
$
Similarly, 
$H_{\mathrm{rep}}$ and $H_{\mathrm{rep},S}$
 (\ref{H_S2}) are represented respectively as follows:
\begin{align*}
H_{\mathrm{rep}}&=h_{\mathrm{rep}{\red ,1}} \otimes I \otimes \cdots \otimes I
+ I \otimes h_{\mathrm{rep}{\red ,2}} \otimes I \otimes \cdots \otimes I
+\cdots 
+ I \otimes \cdots \otimes I \otimes h_{\mathrm{rep}{\red ,n}}\\
H_{\mathrm{rep},S}&=h_{\mathrm{rep},S} \otimes I \otimes \cdots \otimes I
+ I \otimes h_{\mathrm{rep},S} \otimes I \otimes \cdots \otimes I
+\cdots 
+ I \otimes \cdots \otimes I \otimes h_{\mathrm{rep},S}
\end{align*}
for the following single-qudit Hermitian operators
$
h_{\mathrm{rep}{\red, i}}=\sum_{j=1}^{d} ({\blue {\blue \pm B}} {\blue \mp 2\epsilon_c} k_{i,j})
\ket{\Phi_j}\bra{\Phi_j}$ and
$
h_{\mathrm{rep},S}=\sum_{j=1}^{d} 
\frac{\sum_{i=1}^n ({\blue {\blue \pm B}} {\blue \mp 2\epsilon_c} k_{i,j})}{n} 
\ket{\Phi_j}\bra{\Phi_j}.
$

We prove
the following proposition 
related to Result~3,
by the similar discussion in 
a proof of Proposition~{\ref{epsilonnet}}:

\begin{proposition}\label{property3}
(Property of an $\epsilon$-net $\mathcal{N}_S$ in Result~3)
Fix $\epsilon>0$.
Set 
$\epsilon_p$ and $\epsilon_c$ as 
\begin{align*}
    \epsilon_p&=\frac{\epsilon}{8 \ {\red(1+2\sqrt{2})} d C s_{\mathrm{basis}}
(s_{\mathrm{coff}}Ba
+\|A_0\|_\infty)^2},\\
\epsilon_c&=\frac{\epsilon}{8
(2 s_{\mathrm{coff}} Ba
+ s_{\mathrm{coff}}^2(2B+{\red 2})a^2
+
2s_{\mathrm{coff}}
\|A_0\|_\infty
{\red a}
+4(s_{\mathrm{coff}}Ba+
\|A_0\|_\infty)
s_{\mathrm{coff}} a)},
\end{align*}
where $a=\max_{m \in I_{\mathrm{coff}}} \|A_m\|_\infty$.
Then,
for any $H \in S_{LD}^{\mathrm{Result\ 3}}$,
there exists $H_{\mathrm{rep}} \in \mathcal{N}_S$ such that
\begin{align*}
\forall \psi, 
\left|\Big(F_Q(\ket\psi,H)-\max_{\ket\phi: \mathrm{separable}}F_Q(\ket\phi,H)\Big)
-
\Big(F_Q(\ket\psi,H_{\mathrm{rep}})-
\max_{\ket\phi: \mathrm{separable}}F_Q(\ket\phi,H_{\mathrm{rep}})\Big)
\right|&\leq\epsilon.
\end{align*}
\hfill $\blacksquare$
\end{proposition}

\subsection{Proof of Result~1}
Let $D_{\mathrm{mean-lower}}$ be 
{\red the} difference between
the expected QFI of random symmetric states 
and
its lower bound.
That is,
\[
D_{\mathrm{mean-lower}}^{H_{\mathrm{rep}}}
=\underset{\ket{\psi} \leftarrow Sym^n(\mathbb{C}^d)}{E}[ F_Q(\ket\psi,H_{\mathrm{rep}})]
-
\underset{\ket{\psi} \leftarrow Sym^n(\mathbb{C}^d)}{E}[ F_Q(\ket\psi,H_{\mathrm{rep},S})]
\geq 0,
\]
which is non-negative from
Proposition~\ref{largerthanHS}.

Let $c$ be a positive number.
From Lemma~\ref{expectation},
Lemma~\ref{concentration},
(\ref{kosuu}) and 
Proposition~\ref{property1},
we can evaluate an
upper bound on
the probability that
for any element of $S_L$
the QFI of random symmetric states
is less than
the expectation of random symmetric states
for linear Hamiltonian which has 
the form of (\ref{H_S})
 as follows:

\begin{align*}
&\mathrm{Pr}\left(\inf_{H \in S_{L}} \Big(F_Q(\ket{\psi},H)
-
\underset{\ket{\psi} \leftarrow
Sym^n(\mathbb{C}^d)}{E}[F_Q(\ket{\psi},H_S)]\Big)< -c\right)\\
\leq &\mathrm{Pr}\left(\inf_{H \in S_{LD}^{\mathrm{Result\ 1}}} \Big(F_Q(\ket{\psi},H)
-
\underset{\ket{\psi} \leftarrow
Sym^n(\mathbb{C}^d)}{E}[F_Q(\ket{\psi},H_S)]\Big)< -c\right)\\
\leq &\mathrm{Pr}\left(\min_{H_{\mathrm{rep}} \in \mathcal{N}_S} \Big(F_Q(\ket{\psi},H_{\mathrm{rep}})-\underset{\ket{\psi} \leftarrow
Sym^n(\mathbb{C}^d)}{E}[F_Q(\ket{\psi},H_{\mathrm{rep},S})]\Big)
<-c{\red+}\epsilon \right)\\
\leq & \sum_{H_{\mathrm{rep}} \in \mathcal{N}_S} \mathrm{Pr}\left(F_Q(\ket{\psi},H_{\mathrm{rep}})-\underset{\ket{\psi} \leftarrow
Sym^n(\mathbb{C}^d)}{E}[F_Q(\ket{\psi},H_{\mathrm{rep},S}){\red]}
<-c{\red+}\epsilon \right)\\
\leq &|\mathcal{N}_S| \max_{H_{\mathrm{rep}} \in \mathcal{N}_S} \mathrm{Pr}\left(F_Q(\ket{\psi},H_{\mathrm{rep}})-\underset{\ket{\psi} \leftarrow
Sym^n(\mathbb{C}^d)}{E}[F_Q(\ket{\psi},H_{\mathrm{rep},S})]
<-c{\red+}\epsilon\right)\\
\leq &
\left(\frac{B-A}{\epsilon_c}+4\right)^{dn}
\left(\frac{5}{\epsilon_p}\right)^{{\red d}(d+1)}
\max_{H_{\mathrm{rep}} \in \mathcal{N}_S} \mathrm{Pr}
\left(
F_Q(\ket{\psi},H_{\mathrm{rep}})
-\underset{\ket{\psi} \leftarrow 
Sym^n(\mathbb{C}^d)
}{E}[ F_Q(\ket\psi,H_{\mathrm{rep}})]
<-c{\red+}\epsilon-D_{\mathrm{mean-lower}}^{H_{\mathrm{rep}}}
\right)\\
\leq &
\left(\frac{B-A}{\epsilon_c}+4\right)^{dn}
\left(\frac{5}{\epsilon_p}\right)^{{\red d}(d+1)}
2 \mathrm{exp}
\left(-
\frac{2 _{n+d-1}C_n
\left(c{\red-}\epsilon+\underset{H_{\mathrm{rep}} \in \mathcal{N}_S}{\min} D_{\mathrm{mean-lower}}^{H_{\mathrm{rep}}}
\right)^2}
{{\red 144} \pi^3 \log_e 2(2+ 2\sqrt{2})^2
\Theta(n)^4
}
\right){\red.}
\end{align*}

We summarize as the following theorem:
\begin{theorem}\label{upperbound_symmetric}
(Upper bound on
the probability that
there exists
an element of $S_{LD}^{\mathrm{Result\ 1}}$ such that
the QFI of random {\red symmetric} states
is less than
the expectation of random  symmetric {\red states})
An upper bound on
\[
\mathrm{Pr}\left(\inf_{H \in S_{L}} \Big(F_Q(\ket{\psi},H)
-
\underset{\ket{\psi} \leftarrow
Sym^n(\mathbb{C}^d)}{E}[F_Q(\ket{\psi},H_S)]\Big)< -c\right)
\]
is given as
\begin{equation}\label{bound_symmetric}
\left(\frac{B-A}{\epsilon_c}+4\right)^{dn}
\left(\frac{5}{\epsilon_p}\right)^{{\red d}(d+1)}
2 \mathrm{exp}
\left(-
\frac{2 _{n+d-1}C_n
\left(c{\red-}\epsilon+\underset{H_{\mathrm{rep}} \in \mathcal{N}_S}{\min} D_{\mathrm{mean-lower}}^{H_{\mathrm{rep}}}
\right)^2}
{{\red 144} \pi^3 \log_e 2(2+ 2\sqrt{2})^2
\Theta(n)^4
}
\right).
\end{equation}
If (\ref{bound_symmetric}) is less than 1,
upper bound above exists.
If $d>13$,
(\ref{bound_symmetric}) {\red converges} to 0
in the limit of $n \rightarrow \infty$.
\hfill $\blacksquare$
\end{theorem}

From
Theorem~\ref{upperbound_symmetric}, 
the probability that
for all {\red elements} of $S_L \subset S_{LD}$,
the QFI of random symmetric {\red states}
is less than
the expectation of random symmetric states
is extremely small
when $n$ is enough large.
For linear Hamiltonians
which has the form:
\begin{equation}\tag{\ref{LH''}}
H_{L}=h_1 \otimes I \otimes \cdots \otimes I
+ I \otimes h_2 \otimes I \otimes \cdots \otimes I
+\cdots 
+ I \otimes \cdots \otimes I \otimes h_n,
\end{equation}
for some single-qudit Hermitian operator
$h_i=\sum_{j=1}^{d} \lambda_{i,j} \ket{\phi_j}\bra{\phi_j}$
and there exists $j \neq j'$ such that
$\sum_{i=1}^n \lambda_{i,j} - \sum_{i=1}^n \lambda_{i,j'}=\Theta(n)$,
the expectation of random symmetric states
is $\Theta(n^2)$.

Here, we summarize this as the following result:
\setcounter{Result}{0}
\begin{Result}
Let $S_{L}$ be 
the following set of linear Hamiltonians:
\begin{align*}
S_{L}
=\Biggl\{
H
=h_1 \otimes I \otimes \cdots \otimes I+ \cdots+ I \otimes \cdots \otimes I \otimes h_n,
&:\mu_{j,k} \in [-B,-A] \cup [A,B],\\
&\ \ \{\ket{\phi_1}, \cdots, \ket{\phi_d}\}\ 
\mathrm{is} \ \ \mathrm{ONB}\ \ \mathrm{of} \ \ 
\mathbb{C}^d,\\
&\ \ \exists j \neq j' \ \mathrm{s.t.}
\sum_{i=1}^n \lambda_{i,j} - \sum_{i=1}^n \lambda_{i,j'}=\Theta(n)
\Biggr\}{\red,}
\end{align*}
where $B>A>0$.
Denote by $\ket{\psi} \leftarrow
Sym^n(\mathbb{C}^d)$,
a quantum state
sampled uniformly at random
from all $n$-qudit symmetric states.
For any positive constant $c$,
an upper bound on
\[
\mathrm{Pr}\left(\sup_{H_L \in S_{L}} 
\Big(
\Theta(n^2)
-
{\red F_Q}(\ket{\psi},H_L)
\Big)> c\right)
\]
which is a probability 
that there exists an element of 
$S_{L}$
 such that
the quantum Fisher information of
 $\ket{\psi}$ is less than
the expectation of random symmetric states,
converges to $0$ in the limit of $n \rightarrow \infty$.
\hfill $\blacksquare$
\end{Result}

This implies that
the accuracy attained by random symmetric states {\red is} almost the same as
that of a truly optimal state with a high probability
when $n$ is enough large.

\subsection{Proof of Result~3}

Let $D_{\mathrm{optimized}\ \mathrm{sep.-mean}}^{H_{\mathrm{rep}}}$ be 
{\red the} difference between
the QFI of an {\blue optimal separable state}
and {\red the expected} QFI of random pure states,
that is,
\[
D_{\mathrm{optimized}\ \mathrm{sep.-mean}}^{H_{\mathrm{rep}}}
=\max_{\ket{\Phi}: \mathrm{separable}}F_Q(\ket{\Phi},H_{\mathrm{rep}}) 
-
\underset{\ket{\psi} \leftarrow (\mathbb{C}^d)^{\otimes n}}{E}[ F_Q(\ket\psi,H_{\mathrm{rep}})]
\geq 0.
\]

Let $c$ be a positive number.
From Lemma~\ref{expectation},
Lemma~\ref{concentration},
{\red(\ref{kosuu2})} and 
Proposition~\ref{property3},
we can evaluate an
upper bound on
the probability that
there exists
an element of $S_{LD}^{\mathrm{Result\ 3}}$ such that
the QFI of random pure states
is higher than
the nearby value of 
that of an {\blue optimal separable state} as follows:

\begin{align*}
&\mathrm{Pr}\left(\sup_{H \in S_{LD}^{\mathrm{Result\ 3}}} \Big(F_Q(\ket{\psi},H)
-
\max_{\ket{\Phi}: \mathrm{separable}}F_Q(\ket{\Phi},H)\Big)> c\right)\\
\leq &\mathrm{Pr}\left(\max_{H_{\mathrm{rep}} \in \mathcal{N}_S} \Big(F_Q(\ket{\psi},H_{\mathrm{rep}})-\max_{\ket{\Phi}: \mathrm{separable}}F_Q(\ket{\Phi},H_{\mathrm{rep}}) )\Big)
{\red+}\epsilon > c\right)\\
\leq & \sum_{H_{\mathrm{rep}} \in \mathcal{N}_S} \mathrm{Pr}\left(F_Q(\ket{\psi},H_{\mathrm{rep}})-\max_{\ket{\Phi}: \mathrm{separable}}F_Q(\ket{\Phi},H_{\mathrm{rep}}) )
{\red+}\epsilon > c\right)\\
\leq &|\mathcal{N}_S| \max_{H_{\mathrm{rep}} \in \mathcal{N}_S} \mathrm{Pr}\left(F_Q(\ket{\psi},H_{\mathrm{rep}})-\max_{\ket{\Phi}: \mathrm{separable}}F_Q(\ket{\Phi},H_{\mathrm{rep}}) )
{\red+}\epsilon > c\right)\\
\leq &
\left(\frac{B-A}{\epsilon_c}+4\right)^{s_{\mathrm{coff}}}
\left(\frac{5}{\epsilon_p}\right)^{{\red d}(d+1)s_{\mathrm{basis}}}
\max_{H_{\mathrm{rep}} \in \mathcal{N}_S} \mathrm{Pr}
\left(
F_Q(\ket{\psi},H_{\mathrm{rep}})
-\underset{\ket{\psi} \leftarrow (\mathbb{C}^d)^{\otimes n}}{E}[ F_Q(\ket\psi,H_{\mathrm{rep}})]
> c{\red-}\epsilon+D_{\mathrm{optimized}\ \mathrm{sep.-mean}}^{H_{\mathrm{rep}}}
\right)\\
\leq &
\left(\frac{B-A}{\epsilon_c}+4\right)^{s_{\mathrm{coff}}}
\left(\frac{5}{\epsilon_p}\right)^{{\red d}(d+1)s_{\mathrm{basis}}}
2 \mathrm{exp}
\left(-
\frac{2d^{n} \left(c{\red -}\epsilon+\underset{H_{\mathrm{rep}} \in \mathcal{N}_S}{\min} D_{\mathrm{optimized}\ \mathrm{sep.-mean}}^{H_{\mathrm{rep}}}
\right)^2}
{{\red 144} \pi^3 \log_e 2(2+ 2\sqrt{2})^2
(s_{\mathrm{coff}}Ba+\|A_0\|_\infty)^4
}
\right)
\end{align*}

We summarize as the following theorem:
\begin{theorem}\label{upperbound_random}
(Upper bound on
the probability that
there exists
an element of $S_{LD}^{\mathrm{Result\ 3}}$ such that
the QFI of random pure states
is higher than
the nearby value of 
that of an {\blue optimal separable state})
An upper bound on
\[
\mathrm{Pr}\left(\sup_{H \in S_{LD}}\Big( F_Q(\ket{\psi},H)
-
\max_{\ket{\Phi}: \mathrm{separable}}F_Q(\ket{\Phi},H)
\Big)> c\right)
\]
is given as
\begin{equation}\label{bound_random}
\left(\frac{B-A}{\epsilon_c}+4\right)^{s_{\mathrm{coff}}}
\left(\frac{5}{\epsilon_p}\right)^{{\red d}(d+1)s_{\mathrm{basis}}}
2 \mathrm{exp}
\left(-
\frac{2d^{n} \left(c{\red -}\epsilon+\underset{H_{\mathrm{rep}} \in \mathcal{N}_S}{\min} D_{\mathrm{optimized}\ \mathrm{sep.-mean}}^{H_{\mathrm{rep}}}
\right)^2}
{{\red 144} \pi^3 \log_e 2(2+ 2\sqrt{2})^2
(s_{\mathrm{coff}}Ba+\|A_0\|_\infty)^4
}
\right).
\end{equation}
If (\ref{bound_random}) is less than 1,
upper bound above exists.
For 
$s_{\mathrm{basis}}=n$,
$s_\mathrm{coff}=d^{o(n)}$,
$\|A_0\|_\infty=d^{o(n)}$ and
$a=\max_{m \in I_{\mathrm{coff}}} \|A_m\|_\infty=\Theta(n)$,
(\ref{bound_random}) {\red converges} to 0
in the limit of $n \rightarrow \infty$.
\hfill $\blacksquare$
\end{theorem}

From
Theorem~\ref{upperbound_random}, 
the probability that
there exists
an element of $S_{LD}^{\mathrm{Result\ 3}}$ such that
the QFI of random pure states
is higher than
the nearby value of 
that of an {\blue optimal separable state}
is extremely small
when $n$ is enough large.
That is,
the accuracy attained by random pure states is
at most almost the same as
that of an optimal separable state with a high probability
when $n$ is enough large.

We summarize this as the following result:
\setcounter{Result}{2}
\begin{Result}
Let $S_{LD}^{\mathrm{Result\ 3}}$ be 
the following set of 
locally diagonalizable Hamiltonians:
\begin{align*}
S_{LD}
=\Biggl\{
H_{LD}
=A_0+\sum_{m \in I_{\mathrm{coff}}} \mu_m A_m
&:\mu_m \in [-B,-A]\cup [A,B],
\ket{\phi_k}_j \in \mathbb{C}^d
\Biggr\}
\end{align*}
where $A,B>0$,
$A_m$ is diagonalizable by $\otimes_{j=1}^{n} \ket{\phi_j}$,
$a=\max_{m \in I_{\mathrm{coff}}} \|A_m\|_\infty=\Theta(n)$,
and $A_0$ is a fixed Hamiltonian
with $\|A_0\|_\infty=d^{o(n)}$
and $|I_{\mathrm{coff}}|=d^{o(n)}$.
Denote by $\ket{\psi} \leftarrow
(\mathbb{C}^{\red d})^{\otimes n}$,
a quantum state
sampled uniformly at random
from all $n$-{\red qudit} pure states.
For any positive constant $c$,
an upper bound on
\[
\underset{\ket{\psi} \leftarrow
(\mathbb{C}^{\red d})^{\otimes n}}{\mathrm{Pr}}\left(\sup_{H_{\red LD} \in S_{LD}} \Big(F_Q(\ket{\psi},H_{LD})
-
\max_{\ket{\Phi}: \mathrm{separable}}F_Q(\ket{\Phi},H_{LD})\Big) > c\right),
\]
which is a probability 
that there exists an element of 
$S_{LD}^{\mathrm{Result\ 3}}$
 such that
the quantum Fisher information of
 $\ket{\psi}$ is higher than
the nearby value of that of 
an {\blue optimal separable state},
converges to $0$ in the limit of $n \rightarrow \infty$.
\hfill $\blacksquare$
\end{Result}

Next,
we prove that
a set $S_{LD}^{\mathrm{Result\ 3}}$
can be the set of
all linear Hamiltonians
and all $k$-body Hamiltonians 
with {\red$k=o(n/\log{n})$}.
Define
a Hamiltonian
$h_j$($j=1,\cdots,n$)
which operates on the $j$th qubit system:
\[
h_j=
\sum_{k=1}^{d} \lambda_{j,k} \ket{\phi_{\red k}}\bra{\phi_{\red k}},
\]
where
$\{\ket{\phi_1}, \cdots, \ket{\phi_d}\}$
is an orthonormal basis of $\mathbb{C}^d$.
Define a Hamiltonian $A_m(m \in I_{\mathrm{coff}})$ as follows:
\begin{align*}
A_{1,1} =&h_1 \otimes I \otimes I \otimes \cdots \otimes I \otimes I,\\
A_{1,2} =&I \otimes h_2 \otimes I \otimes \cdots \otimes I \otimes I,\\
&\vdots \\
A_{1,n}=&I \otimes I \otimes I \otimes \cdots \otimes I \otimes h_n,\\
A_{2,1} =&h_1 \otimes h_2 \otimes I \otimes \cdots \otimes I \otimes I,\\
&\vdots \\
A_{2,n(n-1)/2}=&I \otimes I \otimes I \otimes \cdots \otimes h_{n-1} \otimes h_n,\\
A_{k,1} =&h_1 \otimes \cdots \otimes h_k \otimes \cdots \otimes I,\\
&\vdots \\
A_{k,n(n-1)\cdots(n-k+1)/k!}=& I \otimes \cdots \otimes I \otimes h_{n-k+1} \otimes \cdots \otimes h_n,
&\vdots \\
A_{n,1}=& h_1 \otimes h_2 \otimes \cdots \otimes h_n.
\end{align*}
Then, 
$I_{\mathrm{coff}}=O({\red 2}^n)$.
Let $A_0=0$.
Define
\begin{align*}
S_{LD}
=\Biggl\{
H
=\sum_{m \in I_{\mathrm{coff}}} \mu_m A_m
&:\mu_m \in \{0,1\},
\ket{\phi_j} \in \mathbb{C}^d
\Biggr\}.
\end{align*}
This set $S_{LD}^{\mathrm{Result\ 3}}$ includes
the following Hamiltonians:
\begin{align*}
&H_1=\sum_{i=1}^{n} h_i =h_1 \otimes I \otimes \cdots \otimes I+ \cdots+ I \otimes \cdots \otimes I \otimes h_n,\\
&H_{2,m}=\sum_{\red (i,j) \in S_{2,m}} h_i \otimes h_j =h_1 \otimes h_2 \otimes \cdots \otimes I+ \cdots+ I \otimes \cdots \otimes I \otimes h_{n-1} \otimes h_n{\red,} \\
&H_{k,m}=\sum_{\red (i_1,\cdots,i_k) \in S_{k,m}} h_{i_1} \otimes \cdots \otimes h_{i_k} =h_1 \otimes \cdots \otimes h_k \otimes \cdots \otimes I+ \cdots+ I \otimes \cdots \otimes I \otimes h_{n-k+1} \otimes \cdots \otimes h_n
\end{align*}
where $S_{2,m}$ has an element
$(i,j)$ ($i,j \in \{1,\cdots,n\}$ and
$i \neq j$) and 
$S_{k,m}$ has an element
$(i_1,\cdots,i_k)$ ($i_j \in \{1,\cdots,n\}$ and
all $i_j$ are disjoint).

Thus, we have the following corollary:
\begin{corollary}\label{upperbound_cor}
Let $S_{LD}^{\mathrm{Result\ 3}}$ be
the set of
all linear $n$-qudit Hamiltonians
and all $k$-body 
locally diagonalizable $n$-{\red qudit} Hamiltonians with {\red$k=o(n/\log{n})$}.
For any positive constant $c$,
an upper bound on
\[
\underset{\ket{\psi} \leftarrow
(\mathbb{C}^{\red d})^{\otimes n}}{\mathrm{Pr}}\left(\sup_{H \in S_{LD}} \Big(F_Q(\ket{\psi},H)
-
\max_{\ket{\Phi}: \mathrm{separable}}F_Q(\ket{\Phi},H)\Big)  > c\right)
\]
converges to $0$ in the limit of $n \rightarrow \infty$.
\hfill $\blacksquare$
\end{corollary}
Corollary~\ref{upperbound_cor} implies the following:
Even if the most suitable Hamiltonian 
is chosen (from the set $S_{LD}^{\mathrm{Result\ 3}}$ of linear Hamiltonians
and $k$-body Hamiltonians) 
for each sampled quantum state,
%(the most suitable Hamiltonian means Hamiltonian which can bring out ability each quantum state),
random pure states can only achieve at most almost the same accuracy as that of an {\blue optimal separable state} with high probability.

\subsection{Tightness of Result~3}

For any locally diagonalizable Hamiltonians $H$
\begin{align*}
H
=\sum_{(i_1,\cdots, i_n)}
a_{(i_1,\cdots, i_n),m}
\left(
\bigotimes_{j ={\red 1}}^{n}\{\ket{\phi_k}_j\bra{\phi_k}_j\ \mathrm{if} \ i_j=k\}
\right),
\end{align*}
there exists 
a local unitary $\otimes_{j \in \{1,\cdots,n\}} U_j$
such that
\begin{align*}
&\left(
\bigotimes_{j=1}^n U_j
\right)
H
\left(
\bigotimes_{j=1}^n U_j^\dag
\right)
=\sum_{(i_1,\cdots, i_n)}
b_{(i_1,\cdots, i_n),m}
\left(
\bigotimes_{j ={\red 1}}^{n}\{\ket{k}_j\bra{k}_j\ \mathrm{if} \ i_j=k\}
\right){\red.}
\end{align*}
In other words,
there exists a
local unitary $\otimes_{j \in \{1,\cdots,n\}} U_j$
which can transform $H$
into a Hamiltonian
which has the following vector
as eigenbasis:
\[
\{\ket{i_1}\otimes \ket{i_2}\otimes \cdots
\ket{i_n}
: i_j \in \{1,2,\cdots,d\}\}{\red.}
\]

In the discussion above,
we consider operating
local unitary operators 
$\otimes_{j \in \{1,\cdots,n\}} U_j$
on locally diagonalizable Hamiltonian $H$ such as (\ref{localhamiltonian}).
Here,
we consider operating
global unitary operators
which can not necessarily be written 
in the form $\otimes_{j \in \{1,\cdots,n\}} U_j$,
on locally diagonalizable Hamiltonian $H$ such as (\ref{localhamiltonian}).
By allowing operating global unitary operator on
locally diagonalizable Hamiltonians $H$, 
we prove that
the QFI of pure states 
can be the same
as that of optimal states in all quantum states.
This is the contrary
to the result
that the QFI of random pure states
is at most almost the same as
an {\blue optimal separable state}.

\begin{theorem}\label{globalunitary}
(By allowing operating global unitary operator on
locally diagonalizable Hamiltonians $H$, 
the QFI of pure states 
can be the same
as that of optimal states in all quantum states.)
Let $\psi$ be an arbitrary pure state.
For any locally diagonalizable Hamiltonians such as
(\ref{localhamiltonian}),
there exists a unitary operator
$U$ such that
\[
F_Q(\ket{\psi}, UHU^\dag)
=
\max_{\phi: \mathrm{arbitrary}}F_Q(\ket{\phi},UHU^\dag){\red.}
\]
That is,
for a Hamiltonian $UHU^\dag$,
the quantum Fisher information
of random pure states $\psi$
is the same as
that of a truly optimal state.
\hfill $\blacksquare$
\end{theorem}

From Theorem~\ref{globalunitary},
{Result~3 is tight in the sense that}
if {we} extend the class of Hamiltonians from {locally} diagonalizable Hamiltonians to a slightly wider class of Hamiltonians{, the statement in} Result {\red 3} does not hold.

\section{Details of Result~2}

In this section,
we consider an $n$-qubit quantum system $(\mathbb{C}^2)^{\otimes n}$ 
and focus on linear Hamiltonians
which have the following form:
\begin{equation}\tag{\ref{H_S}}
H_{\mathrm{S}}=h_S \otimes I \otimes \cdots \otimes I
+ I \otimes h_S \otimes I \cdots \otimes I
+\cdots 
+ I \otimes \cdots \otimes I \otimes h_S,
\end{equation}
for some single-qubit non-degenerate Hermitian operator $h_S=\lambda_0 \ket{\phi_0}\bra{\phi_0}
+\lambda_1 \ket{\phi_1}\bra{\phi_1}$.
Let $E_g$ be geometric measure of entanglement (GME):
\[
E_g(\ket{\Psi})=-\log_2 \sup_{\ket{\alpha}:\mathrm{product}}|\langle \alpha|\Psi\rangle|^2.
\]
Denote $F_Q(\ket{\Phi},H_S)$ by $F(\Phi)$.
Denote 
an $n$-qubit quantum state
$\ket{\Psi}$ by
\[
\ket{\Psi}=\sum_{i_1,\cdots,i_n \in \{0,1\}} 
c_{i_1,\cdots,i_n} \ket{i_1,i_2,\cdots,i_n},
\]
where
\[
c_{i_1,\cdots,i_n} \in \mathbb{C},
\sum_{i_1,\cdots,i_n \in \{0,1\}} 
|c_{i_1,\cdots,i_n}|^2=1.
\]

In this section, we show that
very high GME
leads to low values in QFI (not useful)
for linear Hamiltonians.
For proofs, see Appendix.

First, we show that 
for an arbitrary $n$-qubit quantum state $\ket{\Psi}$,
there exists 
an $n$-qubit symmetric state $\ket{\Psi_{\mathrm{symmetric}}}$
such that 
the QFI of $\ket{\Psi}$ for a linear Hamiltonian such as (\ref{H_S})
is {\red less than or equal to}
that of $\ket{\Psi_{\mathrm{symmetric}}}$.

We have the following proposision:
\begin{proposition}\label{symmetric_takai}
Let $\ket{\Psi}$ 
an arbitrary $n$-qubit quantum state,
denoted by
\[
\ket{\Psi}=\sum_{i_1,\cdots,i_n \in \{0,1\}} 
c_{i_1,\cdots,i_n} \ket{i_1,i_2,\cdots,i_n}.
\]
For all $k=0,1,\cdots,n$,
let
\[
a_k=\sqrt{\frac{\sum_{i_1+\cdots+i_n=k}|c_{i_1,\cdots,i_n}|^2}{\# \{(i_1,i_2,\cdots,i_n)|i_1+\cdots+i_n=k\}}}
=\sqrt{\frac{\sum_{i_1+\cdots+i_n=k}|c_{i_1,\cdots,i_n}|^2}{{}_n C_k}}.
\]
Define
an $n$-qubit quantum state
$\ket{\Psi_{\mathrm{symmetric}}}$
as follows:
\[
\ket{\Psi_{\mathrm{symmetric}}}
=\sum_{k=0}^{n}
\sum_{i_1+\cdots+i_n=k}
\sqrt{\frac{a_k^2+a_{n-k}^2}{2}} \ket{i_1,i_2,\cdots,i_n}.
\]
Then, the following inequality holds:
\[
F(\Psi) \leq
F(\Psi_{\mathrm{symmetric}}).
\]
\hfill $\blacksquare$
\end{proposition}

Let $c<2$.
Assume that
a geometric measure of entanglement $E_g(\ket{\Psi})$ is 
larger than $n-\frac{2n^{c-1}}{\log_e 2}+\frac{(2-c)\log_e n}{\log_e 2}$.
Then, the following inequality holds:
\[
|\langle \Psi |i_1,i_2,\cdots,i_n \rangle|^2
\leq 2^{-E_g(\ket{\Psi})}
\leq 2^{-n+\frac{2n^{c-1}}{\log_e 2}-\frac{(2-c)\log_e n}{\log_e 2}}.
\]
From Proposition~\ref{symmetric_takai},
an $n$-qubit quantum state $\ket{\Psi_{\mathrm{symmetric}}}$ 
which has the following form:
\[
\ket{\Psi_{\mathrm{symmetric}}}
=\sum_{k=0}^{n}
\sum_{i_1+\cdots+i_n=k} b_k
\ket{i_1,i_2,\cdots,i_n},
\]
where
\[
b_k=\sqrt{\frac{a_k^2+a_{n-k}^2}{2}},
\]
satisfies that for all $k=0,1,\cdots,n$,
\[
b_k^2
\leq 2^{-E_g(\ket{\Psi})}
\leq 2^{-n+\frac{2n^{c-1}}{\log_e 2}-\frac{(2-c)\log_e n}{\log_e 2}}
\]
and
\[
F(\Psi) \leq
F(\Psi_{\mathrm{symmetric}}).
\]

Furthermore, we give
an upper bound on the QFI
of a symmetrized state $\ket{\Psi_{\mathrm{symmetric}}}$
for a linear Hamiltonian such as (\ref{H_S}).

We have the following Proposition:
\begin{proposition}\label{symmetric_bound}
Let $c<2$.
Let $\ket{\Psi_{\mathrm{symmetric}}}$ 
be an arbitrary
$n$-qubit quantum state
such that 
\[
\ket{\Psi_{\mathrm{symmetric}}}
=\sum_{k=0}^{n}
\sum_{i_1+\cdots+i_n=k} b_k
\ket{i_1,i_2,\cdots,i_n}
\]
and for all $k=0,1,\cdots,n$,
\[
b_k^2
\leq 2^{-E_g(\ket{\Psi})}
\leq 2^{-n+\frac{2n^{c-1}}{\log_e 2}-\frac{(2-c)\log_e n}{\log_e 2}}.
\]
Then, the following inequality holds:
\[
F(\Psi_{\mathrm{symmetric}})
\leq 
{\red 6} (\lambda_1-\lambda_0)^2 n^c{\red.}
\]
\hfill $\blacksquare$
\end{proposition}

By Proposition~\ref{symmetric_takai} and 
Proposition~\ref{symmetric_bound},
{\red if} a geometric measure of entanglement $E_g(\ket{\psi})$ is 
larger than $n-\{2(n^{c-1}-\log_e n)+c\log_e n\}/\log_e 2$,
then the quantum Fisher information 
$F_Q(\ket{\psi},H_S)$ 
for linear $n$-qubit Hamiltonians $H_S$ is
\[
F(\Psi)\leq
F(\Psi_{\mathrm{symmetric}})
\leq {\red 6} (\lambda_1-\lambda_0)^2 n^c
\]
and thus less than $n^2$ (HL).

We summarize this as the following result:

\setcounter{Result}{1}
\begin{Result}
Let ${\red 1}<c<2$ and $n^{c-1}>\log_e n$.
If a geometric measure of entanglement $E_g(\ket{\Psi})$ is 
larger than $n-\{2(n^{c-1}-\log_e n)+c\log_e n\}/\log_e 2$,
then the QFI 
$F_Q(\ket{\Psi},H_S)$ is less than $n^c$
for linear $n$-qubit Hamiltonians $H_S$.
Here, a geometric measure of entanglement
is defined as $E_g(\ket{\Psi})=-\log_2 \sup_{\ket{\alpha}:\mathrm{product}}|\langle \alpha|\Psi\rangle|^2$,
the QFI 
is defined as (\ref{qfi}), and 
$H_S=h_S \otimes I \otimes \cdots \otimes I
+ I \otimes h_S \otimes I \otimes \cdots \otimes I
+\cdots 
+ I \otimes \cdots \otimes I \otimes h_S$,
where $h_S=\lambda_0 \ket{\phi_{0}}\bra{\phi_{0}}+\lambda_1 \ket{\phi_{1}}\bra{\phi_{1}}$ ($\lambda_0 \neq \lambda_1$)
is some non-degenerate single-qubit Hermitian operator.
\end{Result}

Thus, high GME is not useful
in quantum metrology
of linear Hamiltonians.

\section{Details of Result~4}

In the previous section,
we prove that
for an arbitrary Hamiltonian
in a set of locally diagonalizable Hamiltonians,
the accuracy achieved by
random pure states
is at most almost the same as
that of the {\blue optimal separable state} with high probability.
However,
even if
the accuracy achieved by
random pure states is not significantly higher than
that of an {\blue optimal separable state},
the accuracy attained by
an {\blue optimal separable state}
may be almost the same as
that of a truly optimal state.
In this case,
one cannot say that
random pure states are not useful.
In this section,
we tackle this problem and
identify
the set of Hamiltonians
in which 
the accuracy attained by an {\blue optimal symmetric product state}
 is much lower than
that of an optimal state in all quantum states.

In this section,
we consider an $n$-qubit quantum system $(\mathbb{C}^2)^{\otimes n}$.

\subsection{The class of Hamiltonians for our analysis}
Define
a Hamiltonian
$h_i$($i=1,\cdots,n$)
which operates on 
the $i$th qubit quantum system:
\[
h_i=\lambda_0 \ket{\phi_i}\bra{\phi_i}+\lambda_1 \ket{\phi_i^{\perp}}\bra{\phi_i^{\perp}}{\red,}
\]
where $0<\lambda_0<\lambda_1$,
$\ket{\phi_i} \in \mathbb{C}^2$ and 
$\ket{\phi_i^{\perp}} \in \mathbb{C}^2$ is 
orthogonal to $\ket{\phi_i}$.

In this section,
we analyze the following locally diagonalizable $k$-body Hamiltonian $H_{k,m}$ ($k=O(1)$),
especially locally diagonalizable 2-body Hamiltonian $H_{2,m}$:
\begin{align*}
&H_{2,m}=\sum_{\red (i,j) \in S_{2,m}} h_i \otimes h_j =h_1 \otimes h_2 \otimes \cdots \otimes I+ \cdots+ I \otimes \cdots \otimes I \otimes h_{n-1} \otimes h_n{\red,} \\
&H_{k,m}=\sum_{\red (i_1,\cdots,i_k) \in S_{k,m}} h_{i_1} \otimes \cdots \otimes h_{i_k} =h_1 \otimes \cdots \otimes h_k \otimes \cdots \otimes I+ \cdots+ I \otimes \cdots \otimes I \otimes h_{n-k+1} \otimes \cdots \otimes h_n
\end{align*}
where $S_{2,m}$ can be 
defined as follows:
\begin{align*}
S_{2,1}&=\{(1,2),(1,3),(1,4),\cdots,(1,n-1),(1,n)\}(\mathrm{star})\\
S_{2,2}&=\{(1,2),(2,3),(3,4),\cdots,(n-1,n),(n,1)\}(\mathrm{ring})\\
S_{2,3}&=\{(1,2),(2,3),(3,4),\cdots,(n-1,n)\}(\mathrm{chain})\\
S_{2,4}&=\{(i,j)|i,j=1,\cdots,n\mathrm{\ and\ }i\neq j\}(\mathrm{fully \ connected}){\red.}
\end{align*}

\subsection{The QFI of truly optimal state and
that of an {\blue optimal symmetric product state}
}

To identify
the set of Hamiltonians
in which 
the accuracy attained by an {\blue optimal separable state}
 is much lower than
that of an optimal state in all quantum states,
{\red we} compute
the maximal QFI
of all quantum states
and
the maximal QFI
of all symmetric product {\red states}.

First, we consider 2-body Hamiltnians:
\[
H_{2,m}=\sum_{\red (i,j) \in S_{2,m}} h_i \otimes h_j =h_1 \otimes h_2 \otimes \cdots \otimes I+ \cdots+ I \otimes \cdots \otimes I \otimes h_{n-1} \otimes h_n{\red.}
\]
For simplicity, let 
$h_i$ be 
the following non-degenerate single-qubit Hermitian operator
for any $i=1,\cdots,n$:
\[
h_i=\lambda_0 \ket{0}\bra{0}+\lambda_1 \ket{1}\bra{1}.
\]
Since the maximal eigenvalue and
minimal eigenvalue {\red are} $\Theta(|S_{2,m}|)$,
\begin{equation}\label{max_arbitrary}
\max_{\phi:\mathrm{arbitrary}} F_Q(H_{2,m},\ket{\psi})=\Theta(|S_{2,m}|^2){\red.}
\end{equation}

Here, we consider the following 
symmetric product state:
\[
\ket{\psi}=\ket\phi^{\otimes n}=(\sqrt{p}\ket{0}+e^{i \psi}\sqrt{1-p}\ket{1})^{\otimes n}{\red.}
\]
Since
\[
H_{2,m}=\sum_{(i,j) \in S_{2,m}} h_i \otimes h_j,
\]
it holds that
\[
\langle \psi| H_{2,m} |\psi \rangle =\sum_{(i,j) \in S_{2,m}}\langle\psi| h_i |\psi\rangle \langle \psi|h_j| \psi \rangle.
\]
Here, we define
\begin{align*}
T_{\mathrm{all}}&=\{((i,j),(k,l)):(i,j),(k,l) \in S_{2,m} \}\\
T_{\mathrm{same}}&=\{((i,j),(i,j)):(i,j) \in S_{2,m} \}\\
T_{\mathrm{disjoint}}&=\{
((i,j),(k,l)):
(i,j),(k,l) \in S_{2,m}
\mathrm{\ and\ }i,j,k,l\ \mathrm{are} \ \mathrm{disjoint}\}{\red.}  
\end{align*}
Then,
\begin{align*}
\langle \psi| H_{2,m} |\psi \rangle^2 &=\sum_{(i,j) \in S_{2,m}}\langle h_i \rangle^2 \langle h_j \rangle^2
+\sum_{((i,j),(k,l)) \in T_{\mathrm{disjoint}}} \langle h_i \rangle \langle h_j \rangle \langle h_k \rangle \langle h_l \rangle
+\sum_{((i,j),(k=j,l)) \in T_{\mathrm{all}}-T_{\mathrm{same}}-T_{\mathrm{disjoint}}} \langle h_i \rangle \langle h_j \rangle^2 \langle h_l \rangle\\
\langle \psi| H_{2,m}^2 |\psi \rangle &=\sum_{(i,j) \in S_{2,m}}\langle h_i^2 \rangle \langle h_j^2 \rangle
+\sum_{((i,j),(k,l)) \in T_{\mathrm{disjoint}}} \langle h_i \rangle \langle h_{j} \rangle \langle h_k \rangle \langle h_{l} \rangle
+\sum_{((i,j),(k=j,l)) \in T_{\mathrm{all}}-T_{\mathrm{same}}-T_{\mathrm{disjoint}}}\langle h_i \rangle \langle h_j^2 \rangle \langle h_l \rangle{\red.}
\end{align*}
Thus,
\begin{align*}
\langle \psi| H_{2,m}^2 |\psi \rangle -\langle \psi| H_{2,m} |\psi \rangle^2 
&=\sum_{(i,j) \in S_{2,m}}{\red(}\langle h_i^2 \rangle \langle h_j^2 \rangle
-\langle h_i \rangle^2 \langle h_j \rangle^2)
+\sum_{((i,j),(k=j,l)) \in T_{\mathrm{all}}-T_{\mathrm{same}}-T_{\mathrm{disjoint}}}\langle h_i \rangle (\langle h_j^2 \rangle-\langle h_j \rangle^2) \langle h_l \rangle\\
&=\sum_{(i,j) \in S_{2,m}}(\langle (h_i \otimes h_j)^2 \rangle
-\langle h_i \otimes h_j \rangle^2)
+\sum_{((i,j),(k=j,l)) \in T_{\mathrm{all}}-T_{\mathrm{same}}-T_{\mathrm{disjoint}}}\langle h_i \rangle (\langle h_j^2 \rangle-\langle h_j \rangle^2) \langle h_l \rangle{\red.}
\end{align*}
Since
\begin{align*}
h_i \otimes h_j&=(\lambda_0 \ket{0}\bra{0}+\lambda_1 \ket{1}\bra{1})\otimes (\lambda_0 \ket{0}\bra{0}+\lambda_1 \ket{1}\bra{1})\\
&=\lambda_0^2\ket{00}\bra{00}+\lambda_0\lambda_1(\ket{01}\bra{01}+\ket{10}\bra{10})
+\lambda_1^2\ket{11}\bra{11},
\end{align*}
it holds that
\begin{align*}
\langle h_i \otimes h_j \rangle
&=\lambda_0^2 p^2 + \lambda_0 \lambda_1 2p(1-p)+\lambda_1^2 (1-p)^2=((\lambda_0-\lambda_1)p+\lambda_1)^2\\
\langle (h_i \otimes h_j)^2 \rangle
&=\lambda_0^4 p^2 + \lambda_0^2 \lambda_1^2 2p(1-p)+\lambda_1^4 (1-p)^2=((\lambda_0^2-\lambda_1^2)p+\lambda_1^2)^2{\red.}
\end{align*}
Also,
\[
\langle h_i \rangle (\langle h_j^2 \rangle-\langle h_j \rangle^2) \langle h_l \rangle
=(p\lambda_0+(1-p)\lambda_1)^2
\{(p\lambda_0^2+(1-p)\lambda_1^2)-(p\lambda_0+(1-p)\lambda_1)^2\}{\red.}
\]
Therefore,
\begin{align*}
\frac{1}{4}F_Q(\ket\psi,H_{2,m})
&=\langle \psi| H_{2,m}^2 |\psi \rangle -\langle \psi| H_{2,m} |\psi \rangle^2\\
&=\sum_{(i,j) \in S_{2,m}}(\langle (h_i \otimes h_j)^2 \rangle
-\langle h_i \otimes h_j \rangle^2)
+\sum_{((i,j),(k=j,l)) \in T_{\mathrm{all}}-T_{\mathrm{same}}-T_{\mathrm{disjoint}}}\langle h_i \rangle (\langle h_j^2 \rangle-\langle h_j \rangle^2) \langle h_l \rangle\\
&=|S_{2,m}|\{((\lambda_0^2-\lambda_1^2)p+\lambda_1^2)^2-((\lambda_0-\lambda_1)p+\lambda_1)^4\}\\
&\ \ \ \ +|T_{\mathrm{all}}-T_{\mathrm{same}}-T_{\mathrm{disjoint}}|(p\lambda_0+(1-p)\lambda_1)^2
\{(p\lambda_0^2+(1-p)\lambda_1^2)-(p\lambda_0+(1-p)\lambda_1)^2\}{\red.}
\end{align*}
By using
\[
c=(p\lambda_0^2+(1-p)\lambda_1^2)-(p\lambda_0+(1-p)\lambda_1)^2,
\]
we can evaluate as follows:
\[
4c (p\lambda_0+(1-p)\lambda_1)^2 \times |T_{\mathrm{all}}-T_{\mathrm{disjoint}}| \leq
F_Q(\ket\psi,H_{2,m})
\leq 
4c \{(p\lambda_0^2+(1-p)\lambda_1^2)+(p\lambda_0+(1-p)\lambda_1)^2\}\times |T_{\mathrm{all}}-T_{\mathrm{disjoint}}|{\red.} 
\]
Therefore,
\begin{equation}\label{max_sep}
F_Q(\ket\psi,H_{2,m})=\Theta(|T_{\mathrm{all}}-T_{\mathrm{disjoint}}|).    
\end{equation}

\subsubsection{2-body 
star-shaped,
chain-shaped,
ring-shaped and
fully connected
Hamiltonians}

We consider 2-body 
star-shaped,
chain-shaped,
ring-shaped and
fully connected
Hamiltonians.
From (\ref{max_arbitrary}) and (\ref{max_sep}),
we have
$|S_{2,m}|=|T_{\mathrm{same}}|, 
|T_{\mathrm{disjoint}}|,
|T_{\mathrm{all}}-T_{\mathrm{same}}-T_{\mathrm{disjoint}}|$,
the maximal QFI
of all quantum {\red states},
and 
the maximal QFI
of all separable {\red states}:

\begin{table}[H]
    \centering
    \caption{
    The {\red values} of $|S_{2,m}|=|T_{\mathrm{same}}|, 
|T_{\mathrm{disjoint}}|,
|T_{\mathrm{all}}-T_{\mathrm{same}}-T_{\mathrm{disjoint}}|${\red.}}
    \label{tab:hogehoge}
    \begin{tabular}{|c|c|c|c|c|c||}
        \hline
      shape & $|S_{2,m}|=|T_{\mathrm{same}}|$ & $|T_{\mathrm{disjoint}}|$ &
$|T_{\mathrm{all}}-T_{\mathrm{same}}-T_{\mathrm{disjoint}}|$  \\ \hline\hline
 star & $n-1$ & $0$ & $(n-1)(n-2)$\\ \hline
       chain & $n-1$ & $n^2-5n+6$ & $2n-4$\\ \hline
	ring & $n$ & $n^2-3n$ & $2n$\\ \hline
fully connected & $\displaystyle \frac{n(n-1)}{2}$ & $\displaystyle \frac{n(n-1)}{2}\frac{(n-2)(n-3)}{2}$ & $\displaystyle \frac{n(n-1)}{2}2(n-2)$\\ \hline
    \end{tabular}
\end{table}
\begin{table}[H]
    \centering
\caption{The {\red values} $f(n)$ for
$\max_{\mathrm{arbitrary}}F_Q({\red \ket{\psi},H_{2,m}})$ and
$\max_{\mathrm{product}}F_Q({\red \ket{\psi},H_{2.m}})$ ({\blue $k=O(1)$}){\red.}}
    \label{tab:hogehoge}
    \begin{tabular}{|c|c|c|c|c|c||}
        \hline
      shape & 
$\max_{\mathrm{arbitrary}}F_Q({\red \ket{\psi},H_{2,m}})=\Theta(f(n))$ &
$\max_{\mathrm{product}}F_Q({\red \ket{\psi},H_{2,m}})=\Theta(f(n))$ \\ \hline\hline
star & $(n-1)^2$ & $(n-1)^2$ \\ \hline
       chain & $(n-1)^2$ & $3n-5$ \\ \hline
	ring & $n^2$ & $3n$ \\ \hline
fully connected & $\displaystyle \left(\frac{n(n-1)}{2}\right)^2$ & $\displaystyle n(n-1)\left(n-\frac{3}{2}\right)$ \\ \hline
    \end{tabular}
\end{table}

Table~2 follows from
$\max_{\mathrm{arbitrary}}F_Q({\red \ket{\psi},H_{2,m}})=\Theta(|S_{2,m}|^2)$ and
$\max_{\mathrm{product}}F_Q({\red \ket{\psi},H_{2,m}})=\Theta(|T_{\mathrm{all}}-T_{\mathrm{disjoint}}|)$.

\subsubsection{$k$-body 
star-shaped,
chain-shaped,
ring-shaped and
fully connected
Hamiltonians}

Here, we define
\begin{align*}
T_{\mathrm{all}}&=\{((i_1,i_2,\cdots,i_k),(j_1,j_2,\cdots,j_k)):(i_1,i_2,\cdots,i_k),(j_1,j_2,\cdots,j_k) \in S_{k,m} \}\\
T_{\mathrm{same}}&=\{((i_1,i_2,\cdots,i_k),(i_1,i_2,\cdots,i_k)):(i_1,i_2,\cdots,i_k) \in S_{{\red k},m} \}\\
T_{\mathrm{disjoint}}&=\{
((i_1,i_2,\cdots,i_k),(j_1,j_2,\cdots,j_k)):
(i_1,i_2,\cdots,i_k),(j_1,j_2,\cdots,j_k) \in S_{{\red k},m}
\mathrm{\ and\ }i_1,i_2,\cdots,i_k,j_1,j_2,\cdots,j_k\ \mathrm{are} \ \mathrm{disjoint}\}{\red.}    
\end{align*}

We consider $k$-body 
star-shaped,
chain-shaped,
ring-shaped and
fully connected
Hamiltonians ($k=O(1)$).
From the similar discussion to proof of
(\ref{max_arbitrary}) and (\ref{max_sep}),
we have
$|S_{{\red k},m}|=|T_{\mathrm{same}}|, 
|T_{\mathrm{disjoint}}|,
|T_{\mathrm{all}}-T_{\mathrm{same}}-T_{\mathrm{disjoint}}|$,
the maximal QFI
of all quantum {\red states}
and 
the maximal QFI
of all separable {\red states}:

\begin{table}[H]
    \centering
    \caption{The {\red values} of $|S_{{\red k},m}|=|T_{\mathrm{same}}|, 
|T_{\mathrm{disjoint}}|${\red.}}
    \label{tab:hogehoge}
    \begin{tabular}{|c|c|c|c|c|c||}
        \hline
      shape & $|S_{{\red k},m}|=|T_{\mathrm{same}}|$ & $|T_{\mathrm{disjoint}}|$   \\ \hline\hline
       chain & $n-(k-1)$ & $(n-(k-1))O(n)$ \\ \hline
	ring & $n$ & $n(n-(2k-1))$  \\ \hline
fully connected & $\displaystyle \frac{n(n-1)\cdots(n-(k-1))}{k!}$ &
$\displaystyle \frac{n(n-1)\cdots(n-(k-1))}{k!} \frac{(n-k)\cdots(n-(2k-1))}{k!}$ 
\\ \hline
    \end{tabular}
\end{table}
\begin{table}[H]
    \centering
\caption{The {\red values} $f(n)$ for
$\max_{\mathrm{arbitrary}}F_Q({\red \ket{\psi},H_{k,m}})$ and
$\max_{\mathrm{product}}F_Q({\red \ket{\psi},H_{k,m}})$ ({\blue $k=O(1)$}){\red.}}
    \label{tab:hogehoge}
    \begin{tabular}{|c|c|c|c|c|c||}
        \hline
      shape & 
$\max_{\mathrm{arbitrary}}F_Q({\red \ket{\psi},H_{k,m}})=\Theta(f(n))$ &
$\max_{\mathrm{product}}F_Q({\red \ket{\psi},H_{k,m}})=\Theta(f(n))$
\\ \hline\hline
        chain & $(n-(k-1))^2$ & $O(n)$ \\ \hline
	ring & $n^2$ & $(2k-1)n$ \\ \hline
fully connected & $\displaystyle \left( \frac{n(n-1)\cdots(n-(k-1))}{k!}\right)^2$ & 
$O(n^{2k-1})$ \\ \hline
    \end{tabular}
\end{table}

%takeuchi
Table~4 follows from
$\max_{\mathrm{arbitrary}}F_Q({\red \ket{\psi},H_{k,m}})=\Theta(|S_{{\red k},m}|^2)$ and
$\max_{\mathrm{product}}F_Q({\red \ket{\psi},H_{k,m}})=\Theta(|T_{\mathrm{all}}-T_{\mathrm{disjoint}}|)$.

{\red Tables}~1 and 2 are the special case
{\red of Tables}~3 and 4.
From above table,
{\red then}, we concretely show the following results:
\begin{itemize}
  \setlength{\itemsep}{-0.1mm} % 項目の隙間
\item For star-shaped Hamiltonians, 
the maximal QFI of all symmetric product states
with respect to $n$
are the same as that of all quantum states.
That is,
the accuracy attained by a symmetric product state
is the same as
that of an optimal state in all quantum states.
\item For ring-shaped, chain-shaped, and {\red fully connected} Hamiltonians, 
the scaling of
the maximal QFI of all symmetric product states
with respect to $n$
is different
from that of all quantum states.
That is,
the accuracy attained by a symmetric product state
 is much lower than
that of an optimal state in all quantum states.
\end{itemize}

\subsubsection{Arbitrary locally diagonalizable 2-body Hamiltonians}

Let $H$ be {\red a} locally diagonalizable 2-body 
Hamiltonian.
For a Hamiltonian $H$ above,
let $d_k$ be 
the number of appearing $h_k$
which operates on $k$th qubit system.
For example,
for 2-body 
chain-shaped,
star-shaped,
ring-shaped and
fully connected
Hamiltonians,
the value $d_k$ is 
given as follows:

\begin{table}[h]
    \centering
    \caption{The value of {\red $d_k$}
    of 2-body
    star-shaped,
chain-shaped,
ring-shaped and
fully connected
Hamiltonians{\red.}}
    \label{tab:hogehoge}
    \begin{tabular}{|c|c|c|c|c|c|c|}
        \hline
      shape & $d_1$ & $d_2$ & $d_3$ & $\cdots$ & $d_{n-1}$ & $d_n$   \\ \hline\hline
      star & $n-1$ & $1$ & $1$ & $\cdots$ & $1$ & $1$   \\ \hline
        chain & $1$ & $2$ & $2$ & $\cdots$ & $2$ & $1$   \\ \hline
	 ring & $2$ & $2$ & $2$ & $\cdots$ & $2$ & $2$   \\ \hline
 fully connected & $n-1$ & $n-1$ & $n-1$ & $\cdots$ & $n-1$ & $n-1$   \\ \hline
    \end{tabular}
\end{table}

For any locally diagonalizable 2-body Hamiltonian $H$,
we compute
$|S_{2,m}|=|T_{\mathrm{same}}|, 
|T_{\mathrm{disjoint}}|,
|T_{\mathrm{all}}-T_{\mathrm{same}}-T_{\mathrm{disjoint}}|$,
the maximal QFI
of all quantum {\red states},
and 
the order of maximal QFI
of all separable {\red states}
as follows:

\begin{table}[H]
    \centering
    \caption{
    The {\red values} of $|S_{2,m}|=|T_{\mathrm{same}}|, 
|T_{\mathrm{disjoint}}|,
|T_{\mathrm{all}}-T_{\mathrm{same}}-T_{\mathrm{disjoint}}|${\red.}}
    \label{tab:hogehoge}
    \begin{tabular}{|c|c|c|}
        \hline
$|S_{2,m}|=|T_{\mathrm{same}}|$ & $|T_{\mathrm{disjoint}}|$ &
$|T_{\mathrm{all}}-T_{\mathrm{same}}-T_{\mathrm{disjoint}}|$  \\ \hline\hline
$\displaystyle \frac{1}{2} \sum_{k=1}^{n} d_k$ & 
$\displaystyle \left(\frac{1}{2} \sum_{k=1}^{n} d_k \right)^2-\frac{1}{2} \sum_{k=1}^{n} d_k-\sum_{k=1}^{n} d_k(d_k-1)$ & 
$\displaystyle \sum_{k=1}^{n} d_k(d_k-1)$\\ \hline
	
    \end{tabular}
\end{table}

\begin{table}[H]
    \centering
    \caption{The {\red values} $f(n)$ of $\max_{\mathrm{arbitrary}}F_Q({\red \ket{\psi},H_{2,m}})$ and
$\max_{\mathrm{product}}F_Q({\red \ket{\psi},H_{2,m}})$ ({\blue $d_k$ is a function of $n$}){\red.}}
    \label{tab:hogehoge}
    \begin{tabular}{|c|c|}
        \hline $\max_{\mathrm{arbitrary}}F_Q({\red \ket{\psi},H_{2,m})}=\Theta(f(n))$ &
$\max_{\mathrm{product}}F_Q({\red \ket{\psi},H_{2,m}})=\Theta(f(n))$ \\ \hline\hline
$\displaystyle \left(\frac{1}{2} \sum_{k=1}^{n} d_k \right)^2$ 
& $\displaystyle \frac{1}{2} \sum_{k=1}^{n} d_k+\sum_{k=1}^{n} d_k(d_k-1)$ \\ \hline
    \end{tabular}
\end{table}

Table~7 follows from
$\max_{\mathrm{arbitrary}}F_Q({\red \ket{\psi},H_{2,m}})=\Theta(|S_{2,m}|^2)$ and
$\max_{\mathrm{product}}F_Q({\red \ket{\psi},H_{2,m}})=\Theta(|T_{\mathrm{all}}-T_{\mathrm{disjoint}}|)$.

Here,
we treat a 2-body Hamiltonian as a graph 
by regarding qubits and couplings as vertices and edges, respectively (see also Fig.~{\red 2} in Section~2).
In a graph,
the degree of a vertex is the number of edges connected to the vertex.
The degree of the $k$th vertex $v_k$
which corresponds to the $k$th qubit
is equal to $d_k$.
Thus, 
the order of maximal QFI
of all quantum {\red states}
(the same as $|S_{2,m}|^2$),
and 
the order of maximal QFI
of all separable {\red states}
(the same as
$|T_{\mathrm{all}}-T_{\mathrm{disjoint}}|$)
can be determined by
the degree of the vertex which corresponds to a qubit.

{\red Furthermore}, the {\red values}
$|S_{2,m}|=|T_{\mathrm{same}}|, 
|T_{\mathrm{disjoint}}|,
|T_{\mathrm{all}}-T_{\mathrm{same}}-T_{\mathrm{disjoint}}|$
can be interpreted
{\red as} follows:

\begin{table}[H]
    \centering
    \caption{The {\red values} of $|S_{2,m}|=|T_{\mathrm{same}}|, 
|T_{\mathrm{disjoint}}|,
|T_{\mathrm{all}}-T_{\mathrm{same}}-T_{\mathrm{disjoint}}|${\red.}}
    \label{tab:hogehoge}
    \begin{tabular}{|c|c|c|}
        \hline
$|S_{2,m}|=|T_{\mathrm{same}}|$ & $|T_{\mathrm{disjoint}}|$ &
$|T_{\mathrm{all}}-T_{\mathrm{same}}-T_{\mathrm{disjoint}}|$  \\ \hline\hline
total of the number of edges & 
the number of pairs of edges disconnected $\times 2$ & 
the number of pairs of edges connected
$\times 2$ \\ \hline
	
    \end{tabular}
\end{table}

We summarize as follows:
\setcounter{Result}{3}
\begin{Result}
Consider a 2-body locally diagonalizable
$n$-qubit Hamiltonian $H_{LD}$, as a graph 
by regarding qubits and couplings as vertices and edges, respectively (see also Fig.~2).
Let $d_k$ be the degree of a vertex $v_k$
which corresponds to 
{the $k$th qubit}
(i.e., the number of edges connected 
to the vertex $v_k$) and 
$d=(d_1,d_2,\cdots,d_n)$.
The maximal QFI of all quantum states 
{and} the maximal QFI of 
all {symmetric product states}
are {respectively} given
as follows:
\begin{align}
\tag{\ref{max_arbitrary_intro}}
\max_{{\ket{\Psi}}:\mathrm{arbitrary}}
F_Q \left(\ket{\Psi},H_{LD}\right)
&=\Theta \left((\|d\|_1 )^2\right)
=\Theta \left((|d_1|+\cdots+|d_n|)^2\right),\\
\tag{\ref{max_sep_intro}}
\max_{{\ket{\phi}^{\otimes n}}: \mathrm{product}}F_Q\left({\ket{\phi}^{\otimes n}},H_{LD}\right) 
&=\Theta \left((\|d\|_2 )^2\right)
=\Theta \left(|d_1|^2+\cdots+|d_n|^2\right).    
\end{align}
This means that
for arbitrary 
2-body locally diagonalizable
$n$-qubit Hamiltonians $H_{LD}$ 
such that
{the scaling of a 2-norm $\|d\|_2=\sqrt{|d_1|^2+\cdots+|d_n|^2}$ with respect to $n$
is  different
from that of a 1-norm $\|d\|_1=|d_1|+\cdots+|d_n|$,
the scaling of
the maximal QFI of all {symmetric product states}
with respect to $n$
is different
from that of all quantum states.}
That is,
the accuracy attained by a {symmetric product state}
 is much lower than
that of an optimal state in all quantum states.
\hfill $\blacksquare$
\end{Result}

\section{{\red Conclusion}}

In Section~4, we present the details of Result~1 and Result~3.
We adopt the notion of $\epsilon$-net \cite{epsilon_net,aspects} and
show that for an arbitrary linear Hamiltonian
such as (\ref{LH''}),
the QFI of random symmetric states
is $\Theta(n^2)$ (HL)
with high probability.
It implies the existence of { universal resource} states for quantum metrology 
for a certain class of linear Hamiltonians.
{\blue The existence of a universal resource state for quantum metrology was independently discovered by \cite{aragon2025universality}}.
By the same discussion as the proof of Result~1,
we show that for an arbitrary Hamiltonian
in a set of locally diagonalizable Hamiltonians
parameterized by at most ${\red d}^{o(n)}$
parameters,
the QFI of random pure states
is at most almost the same as
that of the {\blue optimal separable state} with high probability.

In Section~5, we present the details of Result~2.
We show that
very high GME
leads to low values in QFI (not useful)
for linear Hamiltonians.
In Section~6, we present the details of Result~4.
%We prove that in a locally diagonalizable 2-body Hamiltonian on a regular graph, the difference between the maximal QFI of all quantum states and the maximal QFI of all symmetric product states is the largest.
{\red We} clarify
the class of locally diagonalizable 
2-body $n$-qubit Hamiltonians
in which
{\red the maximal QFI of all quantum states and that of all symmetric product states} have different {\red scalings}
with respect to $n$.

As a potential application of Result~1,
we give the delegation 
of quantum metrology~\cite{shettell2022quantum}.
We consider the following situation:
{\red there are a server and a client.
The client has a quantum register, and his/her magnetic field includes the confidential information which the client wishes to conceal from the server.
The client can perform phase estimation of $U=e^{-iH\theta}$ securely as follows:
(1)~The server prepares a ``{universal resource} state'' and sends it to the client.
(2)~The client interacts the state in (1)
with his/her magnetic field and then measures it.
(3)~The client obtains an estimated value by repeating (1) and (2).
This protocol is a quantum-metrology analogue of \cite{morimae2013blind} and improves \cite{takeuchi2019quantum}.
A thorough analysis is beyond the scope of this paper.}
%there are the server and 3 clients, $A$, $B$, and $C$.
%Each client has the quantum registers, and their magnetic fields $h_A$, $h_B$, and $h_c$ have the confidential information which the clients wish to conceal from the server.
%The clients can perform phase estimation of $U=e^{-iH\theta}$ where $H=h_A \otimes I \otimes  I+I \otimes h_B \otimes I+I \otimes I \otimes h_c$ as follows:
%(1)~The server prepares a ``{universal resource} state''.
%(2)~Each client intaracts the state in (1) with his/her magnetic field.
%(3)~The clients calculate the estimated value by LOCC between every client. 
%From \cite{hayashi2004asymptotic}, the highest accuracy of the estimated value of a phase $\theta$, computed in (3) can be attained by LOCC.
%But explicit construction is still open.

In conclusion, 
we demonstrate the existence of { universal resource} states for quantum metrology 
for a certain class of linear Hamiltonians.
In addition, we show that too entangled states are not useful in quantum metrology for a wider class 
of Hamiltonians including
linear Hamiltonians.
Since we analyze a wider class of 
Hamiltonians than \cite{random_qfi},
experimenters
will be one step closer to the implementation of quantum metrology.

\section{Appendix~A : Proofs of Section~4}

\subsection{Proof of Lemma~\ref{expectation}}

Let $\mathcal{H}$ be a $\mathbb{C}$-linear vector space of a dimension $|D|$.
The expectation of QFI
of Haar random states follows
if we set $\mathcal{H}=(\mathbb{C}^d)^{\otimes n}$.
The expectation of QFI
of random symmetric states follows
if we set $\mathcal{H}=Sym^n(\mathbb{C}^d)$.

Let $U(\mathcal{H})$
be a unitary group
\[
U(\mathcal{H})=\{U\in GL(|D|,\mathbb{C})\ | \ U:\mathcal{H} \rightarrow \mathcal{H},\ U U^{\dag}=I\}.
\]
Let $\mu$ be 
a Haar measure on a unitary group
$U(\mathcal{H})$.
Here, let
\begin{align*}
M(V) \coloneqq
\int_{U(\mathcal{H})} d\mu\ U^{\otimes 2}
V (U^{\dag})^{\otimes 2}.
\end{align*}
Since
$M(V) U^{\otimes 2} =U^{\otimes 2}M(V)$,
by Schur-Wheyl duality \cite{haar_random},
there exist complex numbers
$\alpha,\beta \in \mathbb{C}$
such that
\[
M(V)=\alpha \Pi_{Sym^2(\mathcal{H})}+\beta \Pi_{Asym^2(\mathcal{H})}.
\]
Let $\Pi_{Sym^2(\mathcal{H})}:\mathcal{H} \otimes \mathcal{H}\rightarrow
Sym^2(\mathcal{H})$ and
$\Pi_{Asym^2(\mathcal{H})}:\mathcal{H} \otimes \mathcal{H} \rightarrow
Asym^2(\mathcal{H})$
be {\red projections}:
\begin{align*}
\Pi_{Sym^2(\mathcal{H})}(x \otimes y)&=\frac{1}{2}(x \otimes y +y \otimes x)\\
\Pi_{Asym^2(\mathcal{H})}(x \otimes y)&=\frac{1}{2}(x \otimes y -y \otimes x){\red.}
\end{align*}
Complex numbers
$\alpha, \beta \in \mathbb{C}$
can be computed as follows:
\begin{align*}
\alpha(V)&=\frac{\mathrm{Tr}(M(V)\Pi_{Sym^2(\mathcal{H})})}{\mathrm{Tr}(\Pi_{Sym^2(\mathcal{H})})}
=\frac{\mathrm{Tr}(VM(\Pi_{Sym^2(\mathcal{H})}))}{\binom{|D|+1}{2}}
=\frac{\mathrm{Tr}(V\Pi_{Sym^2(\mathcal{H})})}{\binom{|D|+1}{2}}\\
\beta(V)&=\frac{\mathrm{Tr}(M(V)\Pi_{Asym^2(\mathcal{H})})}{\mathrm{Tr}(\Pi_{Asym^2(\mathcal{H})})}
=\frac{\mathrm{Tr}(VM(\Pi_{Asym^2(\mathcal{H})}))}{\binom{|D|}{2}}
=\frac{\mathrm{Tr}(V\Pi_{Asym^2(\mathcal{H})})}{\binom{|D|}{2}}{\red.}
\end{align*}
The third equality follows from
\begin{align*}
M(\Pi_\mathcal{H} \otimes \Pi_\mathcal{H})
&=\Pi_\mathcal{H} \otimes \Pi_\mathcal{H},\\
M(F_{\mathcal{H} \otimes \mathcal{H}})
&=F_{\mathcal{H} \otimes \mathcal{H}},\\
\Pi_{Sym^2(\mathcal{H})}&=\frac{1}{2}(\Pi_\mathcal{H} \otimes \Pi_\mathcal{H}+F_{\mathcal{H} \otimes \mathcal{H}}),\\
\Pi_{Asym^2(\mathcal{H})}&=\frac{1}{2}(\Pi_\mathcal{H} \otimes \Pi_\mathcal{H}-F_{\mathcal{H} \otimes \mathcal{H}}),
\end{align*}
where for all $\ket{\psi},\ket{\phi} \in \mathcal{H}$,
\begin{align*}
\Pi_\mathcal{H} \otimes \Pi_\mathcal{H} (\ket{\psi}\otimes \ket{\phi})=\ket{\psi}\otimes \ket{\phi},\\
F_{\mathcal{H}\otimes \mathcal{H}} (\ket{\psi}\otimes \ket{\phi})=\ket{\phi}\otimes \ket{\psi}.
\end{align*}

Set $V=\ket{\psi}\bra{\psi}^{\otimes 2}$.
Namely,
\begin{align*}
M(\ket{\psi}\bra{\psi}^{\otimes 2}) \coloneqq
\int_{U(\mathcal{H})} d\mu\ U^{\otimes 2}
\ket{\psi}\bra{\psi}^{\otimes 2} (U^{\dag})^{\otimes 2}
=\underset{\ket{\psi} \leftarrow \mathcal{H}=(\mathbb{C}^d)^{\otimes n}}{E}[
\ket{\psi}\bra{\psi}^{\otimes 2}].
\end{align*}
Then,
\begin{align*}
\alpha(\ket{\psi}\bra{\psi}^{\otimes 2})&=\frac{\mathrm{Tr}(\ket{\psi}\bra{\psi}^{\otimes 2}\Pi_{Sym^2(\mathcal{H})})}{\binom{|D|+1}{2}}=\frac{1}{\binom{|D|+1}{2}}\\
\beta(\ket{\psi}\bra{\psi}^{\otimes 2})&=\frac{\mathrm{Tr}(\ket{\psi}\bra{\psi}^{\otimes 2}\Pi_{Asym^2(\mathcal{H})})}{\binom{|D|}{2}}
=0.
\end{align*}
Thus,
\begin{align*}
\underset{\ket{\psi} \leftarrow \mathcal{H}}
{E}[
\ket{\psi}\bra{\psi}^{\otimes 2}]
=\frac{\Pi_{Sym^2(\mathcal{H})}}{\binom{|D|+1}{2}}.
\end{align*}
Therefore,
the expectation of the QFI
can be computed as follows:
\begin{align*}
\underset{\ket{\psi} \leftarrow \mathcal{H}}{E}[f(\psi)]
&=\underset{\ket{\psi} \leftarrow \mathcal{H}}{E}[
\langle \psi |H^2| \psi \rangle-
\langle \psi |H| \psi \rangle^2
]\\
&=\mathrm{Tr}[(H^2 \otimes I-H\otimes H)
\underset{\ket{\psi} \leftarrow (\mathbb{C}^d)^{\otimes n}}{E}[
\ket{\psi}\bra{\psi}^{\otimes 2}]{\red]}\\
&=\frac{1}{\binom{|D|+1}{2}}\mathrm{Tr}[(H^2 \otimes I-H\otimes H)
\Pi_{Sym^2(\mathcal{H})}]\\
&=\frac{1}{\binom{|D|+1}{2}}\mathrm{Tr}[(H^2 \otimes I)
\Pi_{Sym^2(\mathcal{H})}]
-\frac{1}{\binom{|D|+1}{2}}\mathrm{Tr}[(H\otimes H)
\Pi_{Sym^2(\mathcal{H})}]\\
&=\frac{\mathrm{Tr}[\Pi_\mathcal{H}H^2\Pi_\mathcal{H}]}{|D|}-\frac{1}{|D|(|D|+1)}(\mathrm{Tr}[\Pi_\mathcal{H}H^2\Pi_\mathcal{H}]+\mathrm{Tr}[\Pi_\mathcal{H}H\Pi_\mathcal{H}]^2)\\
&=\frac{\mathrm{Tr}[\Pi_\mathcal{H}H^2\Pi_\mathcal{H}]}{|D|+1}-\frac{\mathrm{Tr}[\Pi_\mathcal{H}H\Pi_\mathcal{H}]^2}{|D|(|D|+1)}{\red.}
\end{align*}
The {\red fifth} equality follows from
\begin{align*}
\mathrm{Tr}[(H^2 \otimes I)
\Pi_{Sym^2(\mathcal{H})}]&=
\frac{1}{2}\mathrm{Tr}[(H^2 \otimes I)(\Pi_\mathcal{H} \otimes \Pi_\mathcal{H}+F)]\\
&=\frac{1}{2}(\mathrm{Tr}[\Pi_\mathcal{H} H^2 \Pi_\mathcal{H}\otimes \Pi_\mathcal{H}]
+\mathrm{Tr}[(\Pi_\mathcal{H}H^2\Pi_\mathcal{H} \otimes \Pi_\mathcal{H})F])\\
&=\frac{1}{2}(|D|\mathrm{Tr}[\Pi_\mathcal{H}H^2\Pi_\mathcal{H}]
+\mathrm{Tr}[(\Pi_\mathcal{H}H^2\Pi_\mathcal{H}])\\
&=\frac{|D|+1}{2}\mathrm{Tr}[\Pi_\mathcal{H}H^2\Pi_\mathcal{H}]{\red,}
\end{align*}
\begin{align*}
\mathrm{Tr}[(H \otimes H)
\Pi_{Sym^2(\mathcal{H})}]&=
\frac{1}{2}\mathrm{Tr}[(H \otimes H)(\Pi_\mathcal{H} \otimes \Pi_\mathcal{H}+F)]\\
&=\frac{1}{2}(\mathrm{Tr}[\Pi_\mathcal{H}H\Pi_\mathcal{H} \otimes \Pi_\mathcal{H}H\Pi_\mathcal{H}]
+\mathrm{Tr}[(\Pi_\mathcal{H}H \otimes H\Pi_\mathcal{H})F])\\
&=\frac{1}{2}(\mathrm{Tr}[\Pi_\mathcal{H}H\Pi_\mathcal{H}]^2
+\mathrm{Tr}[\Pi_\mathcal{H}H^2\Pi_\mathcal{H}]){\red.}
\end{align*}
Therefore, the expectation of the QFI can be computed as follows:
\begin{align*}
\underset{\ket{\psi} \leftarrow \mathcal{H}}{E}[f(\psi)]
&=\frac{\mathrm{Tr}[\Pi_\mathcal{H}H^2\Pi_\mathcal{H}]}{|D|+1}-\frac{\mathrm{Tr}[\Pi_\mathcal{H}H\Pi_\mathcal{H}]^2}{|D|(|D|+1)}{\red.}
\end{align*}

\subsection{Proof of Lemma~\ref{concentration}}

We compute a Lipschitz constant of
\[
f(\psi)
=\frac{1}{4}{\red F_\mathcal{Q}}(e^{-iH\theta}\rho e^{iH\theta})
=\langle \psi |H^2| \psi \rangle-
\langle \psi |H| \psi \rangle^2.
\]
A Lipschitz constant of
$f(\psi)$ is a constant $L$
such that
\[
|f(v)-f(w)| \leq L\|v-w\|_2.
\]
For an arbitrary
$u,v \in (\mathbb{C}^d)^{\otimes n}$,
\begin{align*}
&\quad |f({\red u})-f({\red v})|\\
&=
|\mathrm{Tr}[(\ket{u}\bra{u}-\ket{v}\bra{v})H^2]-(
\mathrm{Tr}[\ket{u}\bra{u}H]^2-\mathrm{Tr}[\ket{v}\bra{v}H]^2
)|\\
&=
|\mathrm{Tr}[(\ket{u}\bra{u}-\ket{v}\bra{v})H^2]+(
\mathrm{Tr}[\ket{v}\bra{v}^{\otimes 2}H^{\otimes 2}]-\mathrm{Tr}[\ket{u}\bra{u}^{\otimes 2}H^{\otimes 2}]
)|\\
&\leq|\mathrm{Tr}[(\ket{u}\bra{u}-\ket{v}\bra{v})H^2]|+|
\mathrm{Tr}[\ket{v}\bra{v}^{\otimes 2}H^{\otimes 2}]-\mathrm{Tr}[\ket{u}\bra{u}^{\otimes 2}H^{\otimes 2}]|\\
&=|\mathrm{Tr}[(\ket{u}\bra{u}-\ket{v}\bra{v})H^2]|+|
\mathrm{Tr}([\ket{v}\bra{v}^{\otimes 2}-\ket{u}\bra{u}^{\otimes 2})H^{\otimes 2}]|\\
&\leq \|H^2\|_{\infty}\|\ket{u}\bra{u}-\ket{v}\bra{v}\|_1+
\|H^{\otimes 2}\|_{\infty}\|\ket{v}\bra{v}^{\otimes 2}-\ket{u}\bra{u}^{\otimes 2}\|_1\\
&= \|H^2\|_{\infty}\|\ket{u}\bra{u}-\ket{v}\bra{v}\|_1+
\|H\|^2_{\infty}\|\ket{v}\bra{v}^{\otimes 2}-\ket{u}\bra{u}^{\otimes 2}\|_1{\red.}\\
\end{align*}
Here,
the second equality follows
from that for an arbitrary linear operator $A$,
\[
\mathrm{Tr}[A]^2
=\mathrm{Tr}[A^{\otimes 2}].
\]
The fourth equality
follows from that for {\red arbitrary linear operators} $A$ and $B$,
\[
\mathrm{Tr}[A]+\mathrm{Tr}[B]
=\mathrm{Tr}[A+B].
\]
The fifth inequality 
follows from 
H\"{o}lder's inequality.
The sixth equality
follows from
\[
\|H^{\otimes 2}\|_{\infty}=\|H\|^2_{\infty}.
\]
Then,
by the same discussion as
\cite[Example 54]{haar_random},
\[
\|\ket{u}\bra{u}-\ket{v}\bra{v}\|_1
=2\sqrt{1-|\langle u| v\rangle|^2}
\leq 2 \|u-v\|_2{\red.}
\]
Moreover,
\begin{align*}
  \|\ket{u}\bra{u}^{\otimes 2}-\ket{v}\bra{v}^{\otimes 2}\|_1
&=2\sqrt{1-|\langle u|^{\red\otimes 2} {\red|}v\rangle^{\otimes 2}|^2}\\
&=2\sqrt{1-|\langle u| v\rangle|^4}\\
&=\sqrt{1+|\langle u| v\rangle|^2}
2 \sqrt{1-|\langle u| v\rangle|^2}\\
&\leq 2\sqrt{2} \|u-v\|_2{\red.}
\end{align*}
Thus,
\[
|f(v)-f(w)| \leq 
(2\|H^2\|_{\infty}+
2\sqrt{2} \|H^{\otimes 2}\|_{\infty})\|v-w\|_2{\red.}
\]
Therefore,
by Levy's lemma \cite{haar_random,aspects},
\begin{align*}
 \underset{\ket{\psi} \leftarrow (\mathbb{C}^d)^{\otimes n}}{\mathrm{Prob}}\left(
\left|
f(\psi)-
\underset{\ket{\psi} \leftarrow (\mathbb{C}^d)^{\otimes n}}{E}[f(\psi)]
\right|
\geq \epsilon
\right)
\leq &2 \exp
\left(
-\frac{{\red 2d^n} \epsilon^2}{
9\pi^3(2\|H^2\|_{\infty}+
2\sqrt{2} \|H\|^2_{\infty})^2
}
\right),\\
\underset{\ket{\psi} \leftarrow (\mathbb{C}^d)^{\otimes n}}{\mathrm{Prob}}\left(
f(\psi)-
\underset{\ket{\psi} \leftarrow (\mathbb{C}^d)^{\otimes n}}{E}[f(\psi)]
> \epsilon
\right)
\leq &2 \exp
\left(
-\frac{{\red 2d^n} \epsilon^2}{
9\pi^3 \log_e 2 (2\|H^2\|_{\infty}+
2\sqrt{2} \|H\|^2_{\infty})^2
}
\right).
\end{align*}
Furthermore,
\begin{align*}
 \underset{\ket{\psi} \leftarrow
Sym^n(\mathbb{C}^d)}{\mathrm{Prob}}\left(
\left|
f(\psi)-
\underset{\ket{\psi} \leftarrow
Sym^n(\mathbb{C}^d)}{E}[f(\psi)]
\right|
\geq \epsilon
\right)
\leq &2 \exp
\left(
-\frac{2 _{n+d-1}C_n\epsilon^2}{
9\pi^3(2\|H^2\|_{\infty}+
2\sqrt{2} \|H\|^2_{\infty})^2
}
\right),\\
\underset{\ket{\psi} \leftarrow
Sym^n(\mathbb{C}^d)}{\mathrm{Prob}}\left(
f(\psi)-
\underset{\ket{\psi} \leftarrow
Sym^n(\mathbb{C}^d)}{E}[f(\psi)]
<- \epsilon
\right)
\leq &2 \exp
\left(
-\frac{2 _{n+d-1}C_n \epsilon^2}{
9\pi^3 \log_e 2 (2\|H^2\|_{\infty}+
2\sqrt{2} \|H\|^2_{\infty})^2
}
\right).
\end{align*}

\subsection{Proof of Proposition~\ref{lessthansep}}

The expectation$f(\psi)$ is
\[
\underset{\ket{\psi} \leftarrow (\mathbb{C}^{\red d})^{\otimes n}}{E}[f(\psi)]
  =\frac{\mathrm{Tr}[H^2]}{d^n+1}
-\frac{\mathrm{Tr}[H]^2}{d^n(d^n+1)}.
\]
On the other hand, 
\[
\ket{\Phi}^{\otimes n}=\left(\frac{\ket{\phi_1}+\ket{\phi_2}+\cdots+\ket{\phi_d}}{\sqrt{d}}\right)^{\otimes n}
\]
is a separable state and
the QFI of this symmetric product state is
\[
F_Q(\ket{\Phi}^{\otimes n},H)={\red 4\left(\frac{\mathrm{Tr}[H^2]}{d^n}-\frac{\mathrm{Tr}[H]^2}{d^{2n}}\right)}.
\]
Thus,
\begin{align*}
\frac{1}{4}
\max_{\ket{\Phi}: \mathrm{separable}}F_Q(\ket{\Phi},H)
\geq& \frac{1}{4} F_Q(\ket{{\red\Phi}}^{\otimes n},H)\\
 {\red=}& \frac{\mathrm{Tr}[H^2]}{d^n}-\frac{\mathrm{Tr}[H]^2}{d^{2n}}\\
=&\frac{1}{d^{n}}\left(\mathrm{Tr}[H^2]-\frac{\mathrm{Tr}[H]^2}{d^{n}} \right)\\
\geq &\frac{1}{d^{n}+1}\left(\mathrm{Tr}[H^2]-\frac{\mathrm{Tr}[H]^2}{d^{n}} \right)\\
=& \frac{\mathrm{Tr}[H^2]}{d^n+1}-\frac{\mathrm{Tr}[H]^2}{d^n(d^n+1)}
= \underset{\ket{\psi} \leftarrow (\mathbb{C}^{\red d})^{\otimes n}}{E}[f(\psi)]{\red.}
\end{align*}
Therefore,
\[
4\underset{\ket{\psi} \leftarrow (\mathbb{C}^{\red d})^{\otimes n}}{E}[ f(\psi)] \leq \max_{\ket{\Phi}: \mathrm{separable}}F_Q(\ket{\Phi},H).
\]

\subsection{Proof of 
Proposition~\ref{largerthanHS}}

From Lemma~\ref{expectation}, we have
\begin{align*}
{\red\cfrac{1}{4}}\underset{\ket{\psi} \leftarrow Sym^n(\mathbb{C}^d)}{E}[F_Q(\ket{\psi},H_L)]
&=\frac{\mathrm{Tr}[\Pi_{Sym^n(\mathbb{C}^d)}H_L^2\Pi_{Sym^n(\mathbb{C}^d)}]}{|D|+1}-\frac{\mathrm{Tr}[\Pi_{Sym^n(\mathbb{C}^d)}H_L\Pi_{Sym^n(\mathbb{C}^d)}]^2}{|D|(|D|+1)},\\
{\red\cfrac{1}{4}}\underset{\ket{\psi} \leftarrow Sym^n(\mathbb{C}^d)}{E}[F_Q(\ket{\psi},H'_S)]
&=\frac{\mathrm{Tr}[\Pi_{Sym^n(\mathbb{C}^d)}{H'_S}^2\Pi_{Sym^n(\mathbb{C}^d)}]}{|D|+1}-\frac{\mathrm{Tr}[\Pi_{Sym^n(\mathbb{C}^d)}H'_S\Pi_{Sym^n(\mathbb{C}^d)}]^2}{|D|(|D|+1)}.\\
\end{align*}
It is convenient to use the orthonormal basis
of $Sym^n(\mathbb{C}^d)$
consisting of generalized Dicke states.
First, we define the generalized Dicke states.
Let $\vec{k}=(k_1,\cdots,k_d)$ be the vector 
consisting of non-negative integers
which {\red satisfy} the normalization condition
$|\vec{k}|=\sum_{i=1}^{d} k_i={\red n}$.
Let $\ket{\vec{k}}=\ket{{\red\phi_1}}^{\otimes k_1} \otimes \ket{{\red\phi_2}}^{\otimes k_2} \otimes \cdots {\red\otimes}\ket{{\red \phi_d}}^{\otimes k_d}$.
The generalized Dicke states are
given by 
\[
\ket{\vec{k},n}=\sqrt{\frac{n!}{\prod_{i=1}^{d}k_i!}}
\Pi_{Sym^n(\mathbb{C}^d)}\ket{\vec{k}}.
\]

We define $\mu_{{\red\vec{k},}\pi}$($\pi \in S_n$) as follows:
\begin{align*}
\mu_{\vec{k},\pi}
=\lambda_{1,\pi(1)}+\cdots+\lambda_{k_1,\pi(1)} +\lambda_{k_1+1,\pi(2)}+\cdots+\lambda_{k_1+k_2,\pi(2)}+
{\red\cdots}
+\lambda_{k_1+\cdots+k_{d-1}{\red+1},\pi(d)}+\cdots+\lambda_{n,\pi(d)}.
\end{align*}
Then, 
\begin{align*}
\Pi_{Sym^n(\mathbb{C}^d)}H_L\Pi_{Sym^n(\mathbb{C}^d)}\ket{\vec{k},n}
&=\frac{1}{n!}\sum_{\pi \in S_n} \mu_{\vec{k},\pi}
\ket{\vec{k},n}\\
&=k_1\frac{\sum_{i=1}^n \lambda_{i,1}}{n}
+k_2\frac{\sum_{i=1}^n \lambda_{i,2}}{n}
+\cdots
+k_d\frac{\sum_{i=1}^n \lambda_{i,d}}{n}\ket{\vec{k},n}\\
&=\Pi_{Sym^n(\mathbb{C}^d)}H'_S\Pi_{Sym^n(\mathbb{C}^d)}\ket{\vec{k},n}.
\end{align*}
By the fact that the generalized Dicke states form a basis of
$Sym^n(\mathbb{C}^d)$,
we have
\begin{align}\label{HL=HS}
\mathrm{Tr}[\Pi_{Sym^n(\mathbb{C}^d)}H_L\Pi_{Sym^n(\mathbb{C}^d)}]
=\frac{1}{n!}\sum_{\vec{k}:|\vec{k}|={\red n}}
\sum_{\pi \in S_n} \mu_{\vec{k},\pi}
=\mathrm{Tr}[\Pi_{Sym^n(\mathbb{C}^d)}H'_S\Pi_{Sym^n(\mathbb{C}^d)}].        
\end{align}

Furthermore,
\begin{align*}
\Pi_{Sym^n(\mathbb{C}^d)}H_L^2\Pi_{Sym^n(\mathbb{C}^d)}\ket{\vec{k},n}
&=\frac{1}{n!}\sum_{\pi \in S_n} \mu_{\vec{k},\pi}^2
\ket{\vec{k},n}\\
\Pi_{Sym^n(\mathbb{C}^d)}{H'_S}^2\Pi_{Sym^n(\mathbb{C}^d)}\ket{\vec{k},n}
&=\left(\frac{1}{n!}\sum_{\pi \in S_n} \mu_{\vec{k},\pi}\right)^2
\ket{\vec{k},n}{\red.}
\end{align*}
Here,
\begin{align*}
\frac{1}{n!}\sum_{\pi \in S_n} \mu_{\vec{k},\pi}^2
\geq \left(\frac{1}{n!}\sum_{\pi \in S_n} \mu_{\vec{k},\pi}\right)^2.
\end{align*}
By the fact that the generalized Dicke states form a basis of
$Sym^n(\mathbb{C}^d)$,
we have
\begin{equation}\label{HL^2geqHS^2}
\mathrm{Tr}[\Pi_{Sym^n(\mathbb{C}^d)}H_L^2\Pi_{Sym^n(\mathbb{C}^d)}]
=\sum_{\vec{k}:|\vec{k}|={\red n}} \frac{1}{n!}
\sum_{\pi \in S_n} \mu_{\vec{k},\pi}^2
\geq \sum_{\vec{k}:|\vec{k}|={\red n}} \left(\frac{1}{n!}
\sum_{\pi \in S_n} \mu_{\vec{k},\pi}\right)^2
=\mathrm{Tr}[\Pi_{Sym^n(\mathbb{C}^d)}{H'_S}^2\Pi_{Sym^n(\mathbb{C}^d)}].        
\end{equation}

Therefore, by (\ref{HL=HS}) and (\ref{HL^2geqHS^2}), we have
\begin{align*}
\frac{1}{4}\underset{\ket{\psi} \leftarrow Sym^n(\mathbb{C}^d)}{E}[F_Q(\ket{\psi},H_L)]
&=\frac{\mathrm{Tr}[\Pi_{Sym^n(\mathbb{C}^d)}H_L^2\Pi_{Sym^n(\mathbb{C}^d)}]}{|D|+1}-\frac{\mathrm{Tr}[\Pi_{Sym^n(\mathbb{C}^d)}H_L\Pi_{Sym^n(\mathbb{C}^d)}]^2}{|D|(|D|+1)}\\
&\geq \frac{\mathrm{Tr}[\Pi_{Sym^n(\mathbb{C}^d)}{H'_S}^2\Pi_{Sym^n(\mathbb{C}^d)}]}{|D|+1}-\frac{\mathrm{Tr}[\Pi_{Sym^n(\mathbb{C}^d)}H'_S\Pi_{Sym^n(\mathbb{C}^d)}]^2}{|D|(|D|+1)}\\
&=\frac{1}{4}\underset{\ket{\psi} \leftarrow Sym^n(\mathbb{C}^d)}{E}[F_Q(\ket{\psi},H'_S)].
\end{align*}

Finally, we show that
$
\frac{\mathrm{Tr}({h_S'}^2)}{d}
  -\frac{\mathrm{Tr}(h_S')^2}{d^2}> 0
${\red.}
Let $\tilde{\lambda_j}={\red(\sum_{i=1}^n \lambda_{i,j})/n}$.
Then, 
\begin{align*}
\mathrm{Tr}({h_S'}^2)
-\frac{\mathrm{Tr}(h_S')^2}{d}
&=({\tilde{\lambda_1}}^2+\cdots+{\tilde{\lambda_d}}^2)
-\frac{({\tilde{\lambda_1}}+\cdots+{\tilde{\lambda_d}})^2}{d}\\
&=\left(1-\frac{1}{d}\right){\tilde{\lambda_1}}^2
+\cdots+
\left(1-\frac{1}{d}\right){\tilde{\lambda_d}}^2
-\frac{1}{d}(2\tilde{\lambda_1}\tilde{\lambda_2}
+2\tilde{\lambda_1}\tilde{\lambda_3}
+\cdots+
2\tilde{\lambda_1}\tilde{\lambda_d}
+\cdots+
2{\red\tilde{\lambda}_{d-1}}\tilde{\lambda_d})\\
&=\frac{1}{d}(\tilde{\lambda_1}-\tilde{\lambda_2})^2
+\frac{1}{d}(\tilde{\lambda_1}-\tilde{\lambda_3})^2
+\cdots+
\frac{1}{d}(\tilde{\lambda_1}-\tilde{\lambda_d})^2
+\cdots+
\frac{1}{d}({\red\tilde{\lambda}_{d-1}}-\tilde{\lambda_d})^2.
\end{align*}
From the assumption,
there exists $j \neq j'$ such that
$\sum_{i=1}^n \lambda_{i,j}- \sum_{i=1}^n \lambda_{i,j'}=\Theta(n)$.
Then, there exists
$j \neq j'$ such that $\tilde{\lambda_j} \neq \tilde{\lambda_{j'}}$.
Thus, we have
\[
\frac{\mathrm{Tr}({h_S'}^2)}{d}
  -\frac{\mathrm{Tr}(h_S')^2}{d^2}> 0.
\]

\subsection{Proof of 
Proposition~\ref{onbchikai}}

First, we prove the following claim:

\begin{Claim}\label{claim_onb}
Let 
$\{\ket{\phi_1}, \ket{\phi_2}, \cdots, \ket{\phi_d}\}$
be an arbitrary orthonormal basis of $\mathbb{C}^d$.
Let $\ket{\Phi_1} \in \mathbb{C}^d$ 
be a vector which satisfies
\[
\|\ket{\phi_1}\bra{\phi_1}-\ket{\Phi_1}\bra{\Phi_1}\|_1 
\leq 2\|\ket{\phi_1}-\ket{\Phi_1}\|_2 
\leq
\epsilon_p
\]
Then, there exists 
$\{\ket{\Phi_1}, \ket{\tilde{\phi_2}},\cdots, \ket{\tilde{\phi_d}}\}$
be an orthonormal basis of $\mathbb{C}^d$
such that for all $j=2,\cdots,d$,
\[
\|\ket{\tilde{\phi_j}}\bra{\tilde{\phi_j}}-\ket{\phi_j}\bra{\phi_j}\|_1 
\leq 2\|\ket{\tilde{\phi_j}}-\ket{\phi_j}\|_2 
\leq C' \epsilon_p,
\]
where $C'$ is a constant, which is independent of $n$.
\end{Claim}
\proofname\ \ 
Let ${\red\ket{{\tilde{\phi}}_2}} \in \mathbb{C}^d$
be an arbitrary vector orthogonal to
$\ket{\Phi_1}$.
Then, 
$\langle \tilde{\phi_2}|\Phi_1 \rangle=0$.
By the assumption,
$\|\ket{\phi_1}-\ket{\Phi_1}\|_2
\leq \frac{\epsilon_p}{2}$.
Then,
$\mathrm{Re} \langle \tilde{\phi_2}| \phi_1\rangle
\leq \frac{\epsilon_p}{2}
$.
Thus, there exists ${\red\ket{\tilde{\phi}_2}} \in \mathbb{C}^d$ such that
$\mathrm{Re}\langle \tilde{\phi_2}| \phi_2\rangle
\geq
\sqrt{1-\frac{\epsilon_p^2}{4}}
\approx
1-\frac{\epsilon_p^2}{2}$.
That is,
$\|\ket{\tilde{\phi_2}}\bra{\tilde{\phi_2}}-\ket{\phi_2}\bra{\phi_2}\|_1 
\leq 2\|\ket{\tilde{\phi_2}}-\ket{\phi_2}\|_2 \leq 2\epsilon_p$.

Let ${\red\ket{\tilde{\phi}_3}} \in \mathbb{C}^d$
be an arbitrary vector orthogonal to
$\ket{\Phi_1}$ and $\ket{\tilde{\phi_2}}$.
Then, 
$
\langle \tilde{\phi_3}|\Phi_1 \rangle=0
\ \  \mathrm{and}\ \ \
\langle \tilde{\phi_3}|\tilde{\phi_2} \rangle=0.
$
By the assumption,
$
\|\ket{\phi_1}-\ket{\Phi_1}\|_2
\leq \epsilon_p/2
$ 
and 
$
\|\ket{\phi_2}-\ket{\tilde{\phi_2}}\|_2
\leq \epsilon_p.
$
Then,
$
\mathrm{Re}\langle \tilde{\phi_3} | \phi_1\rangle
\leq \frac{\epsilon_p}{2}
$
and 
$
\mathrm{Re}\langle \tilde{\phi_3}| \phi_2\rangle
\leq \epsilon_p
$.
Thus, there exists $\ket{\tilde{\phi_3}} \in \mathbb{C}^d$ such that
$\mathrm{Re}\langle \tilde{\phi_3}| \phi_3\rangle
\geq \sqrt{1-\frac{17\epsilon_p^2}{4}}
\approx 1-\frac{\sqrt{17}\epsilon_p^2}{2}$.
That is,
$\|\ket{\tilde{\phi_3}}\bra{\tilde{\phi_3}}-\ket{\phi_3}\bra{\phi_3}\|_1 \leq 
2\|\ket{\tilde{\phi_3}}-\ket{\phi_3}\|_2 \leq 
2 \times 17\epsilon_p$.

By repeating this procedure,
we can prove that
there exists {\red an orthonormal basis}
$\{\ket{\Phi_1}, \ket{\tilde{\phi_2}},\cdots, \ket{\tilde{\phi_d}}\}$
{\red of} $\mathbb{C}^d$
such that for all $j=2,\cdots,d$,
\[
\|\ket{\tilde{\phi_j}}\bra{\tilde{\phi_j}}-\ket{\phi_j}\bra{\phi_j}\|_1 \leq
2\|\ket{\tilde{\phi_j}}\bra{\tilde{\phi_j}}-\ket{\phi_j}\bra{\phi_j}\|_2 \leq C' \epsilon_p.
\]
\hfill $\blacksquare$

Then, we prove
Proposition~\ref{onbchikai}.
Let 
$\{\ket{\phi_1}, \ket{\phi_2}, \cdots, \ket{\phi_d}\}$
be an arbitrary orthonormal basis of $\mathbb{C}^d$.
Then, there exists
$\ket{\Phi_1} \in \mathcal{N}_{T,\mathbb{C}^d}$ 
such that
\begin{equation}\label{phi1_norm}
\|\ket{\phi_1}\bra{\phi_1}-\ket{\Phi_1}\bra{\Phi_1}\|_1 
\leq 2\|\ket{\phi_1}-\ket{\Phi_1}\|_2 
\leq \epsilon_p.
\end{equation}

By Claim~\ref{claim_onb}, there exists {\red an orthonormal basis} 
$\{\ket{\Phi_1}, \ket{\phi_{2,1}},\cdots, \ket{\phi_{d,1}}\}$
{\red of} $\mathbb{C}^d$
such that for all $j=2,\cdots,d$,
\begin{equation}\label{claim2,1}
\|\ket{\phi_{j,1}}\bra{\phi_{j,1}}-\ket{\phi_j}\bra{\phi_j}\|_1 
\leq 2\|\ket{\phi_{j,1}}-\ket{\phi_j}\|_2
\leq C' \epsilon_p
\end{equation}
Since
$\ket{\phi_{2,1}} \in \mathbb{C}^d
\setminus\mathrm{span}\{\ket{\Phi_1}\}$,
there exists $\ket{\Phi_2} \in \mathcal{N}_{T,\mathrm{span}\{\ket{\Phi_1}\}^{\perp}}$
such that
\begin{equation}\label{phi2_norm1}
\|\ket{\phi_{2,1}}\bra{\phi_{2,1}}-\ket{\Phi_2}\bra{\Phi_2}\|_1
\leq 2 \|\ket{\phi_{2,1}}-\ket{\Phi_2}\|_2
\leq \epsilon_p.
\end{equation}
By (\ref{claim2,1}) and (\ref{phi2_norm1}), we have
\begin{align}
\|\ket{\phi_{2}}\bra{\phi_{2}}-\ket{\Phi_2}\bra{\Phi_2}\|_1 
&\leq \|\ket{\phi_{2}}\bra{\phi_{2}}-\ket{\phi_{2,1}}\bra{\phi_{2,1}}\|_1 
+\|\ket{\phi_{2,1}}\bra{\phi_{2,1}}-\ket{\Phi_2}\bra{\Phi_2}\|_1 
\leq (C'+1) \epsilon_p\label{phi2_norm}{\red.}
\end{align}
Similarly, we have
\begin{align*}
\|\ket{\phi_{2}}-\ket{\Phi_2}\|_2 
&\leq \|\ket{\phi_{2}}-\ket{\phi_{2,1}}\|_2 
+\|\ket{\phi_{2,1}}-\ket{\Phi_2}\|_2
\leq \frac{(C'+1) \epsilon_p}{2}.
\end{align*}

$\{\ket{\phi_{2,1}}, \ket{\phi_{3,1}}, \cdots, \ket{\phi_{d,1}}\}$
is an orthonormal basis of $\mathrm{span}\{\ket{\Phi_1}\}^{\perp}$.
By (\ref{phi2_norm1}) and Claim~\ref{claim_onb}, there exists 
$\{\ket{\Phi_2}, \ket{\phi_{3,2}},\cdots, \ket{\phi_{d,2}}\}$
be an orthonormal basis of $\mathrm{span}\{\ket{\Phi_1}\}^{\perp}$
such that for all $j=3,\cdots,d$,
\begin{equation}\label{claim2,2}
\|\ket{\phi_{j,1}}\bra{\phi_{j,1}}-\ket{\phi_{j,2}}\bra{\phi_{j,2}}\|_1 \leq 
2\|\ket{\phi_{j,1}}-\ket{\phi_{j,2}}\|_2 \leq C' \epsilon_p.    
\end{equation}
Since
$\ket{\phi_{3,{\red 2}}} \in \mathrm{span}\{\ket{\Phi_1},\ket{\Phi_2}\}^{\perp}$,
there exists $\ket{\Phi_3} \in \mathcal{N}_{T,\mathrm{span}\{\ket{\Phi_1},\ket{\Phi_2}\}^{\perp}}$
such that
\begin{equation}\label{phi3_norm1}
\|\ket{\phi_{3,2}}\bra{\phi_{3,2}}-\ket{\Phi_3}\bra{\Phi_3}\|_1 
\leq 2\|\ket{\phi_{3,2}}-\ket{\Phi_3}\|_2
\leq \epsilon_p.
\end{equation}
By (\ref{claim2,1}), (\ref{claim2,2})
and (\ref{phi3_norm1}), we have
\begin{align}
\|\ket{\phi_{3}}\bra{\phi_{3}}-\ket{\Phi_3}\bra{\Phi_3}\|_1 
&\leq \|\ket{\phi_{3}}\bra{\phi_{3}}-\ket{\phi_{3,1}}\bra{\phi_{3,1}}\|_1 
+\|\ket{\phi_{3,1}}\bra{\phi_{3,1}}
-\ket{\phi_{3,2}}\bra{\phi_{3,2}}\|_{\red 1}
+\|\ket{\phi_{3,2}}\bra{\phi_{3,2}}-\ket{\Phi_3}\bra{\Phi_3}\|_1 \nonumber\\
&\leq (2C_1+1) \epsilon_p{\red.}\label{phi3_norm}
\end{align}
Similarly, we have
we have
\begin{align*}
\|\ket{\phi_{3}}-\ket{\Phi_3}\|_2
&\leq \|\ket{\phi_{3}}-\ket{\phi_{3,1}}\|_2
+\|\ket{\phi_{3,1}}
-\ket{\phi_{3,2}}\|_2
+\|\ket{\phi_{3,2}}-\ket{\Phi_3}\|_2
\leq \frac{(2C_1+1) \epsilon_p}{2}.
\end{align*}

By repeating the same procedure,
we can get a similar inequality
as (\ref{phi1_norm}), (\ref{phi2_norm})
and (\ref{phi3_norm}).
Finally,
we can prove that
\begin{align*}
\ket{\Phi_1} &\in \mathcal{N}_{T,\mathbb{C}^d},\\
\ket{\Phi_2} &\in \mathcal{N}_{T,\mathrm{span}\{\ket{\Phi_1}\}^{\perp}},\\
&\cdots\\
\ket{\Phi_{d-1}} &\in \mathcal{N}_{T,\mathrm{span}\{\ket{\Phi_1},\cdots,\ket{\Phi_{d-2}}\}^{\perp}},\\
\ket{\Phi_d}&\in \mathrm{span}\{\ket{\Phi_1},\cdots,\ket{\Phi_{d-1}}\}^{\perp}
\end{align*}
such that for all ${\red j}=1,\cdots,d$,
\[
\|\ket{\phi_j}\bra{\phi_j}-\ket{\Phi_j}\bra{\Phi_j}\|_1 \leq ((j-1)C_1+1)\epsilon_p
\]
and
\[
\|\ket{\phi_j}-\ket{\Phi_j}\|_2 \leq \frac{((j-1)C_1+1)\epsilon_p}{2},
\]
where $C_1$ is a constant, which is independent of $n$.

\subsection{Proof of Proposition~\ref{epsilonnet}}

Fix $\epsilon>0$.
By setting $\epsilon_p$ and $\epsilon_c$ 
appropriately,
we prove that
for any $H \in S_{LD}$,
there exists
$H_{\mathrm{rep}} \in \mathcal{N}_S$ such that
\[
\|H- H_{\mathrm{rep}}\|_\infty \leq \epsilon.
\]

Let $H$ be an element of
$S_{LD}^{\mathrm{Result\ 1}},S_{LD}^{\mathrm{Result\ 3}}$:
\begin{align*}
H
=A_0+\sum_{m \in I_{\mathrm{coff}}} \mu_m A_m{\red.}
\end{align*}
Then, there exists 
$H_{\mathrm{rep}} \in \mathcal{N}_S$ such that
\begin{align*}
H_{\mathrm{rep}}
=B_0+\sum_{m \in I_{\mathrm{coff}}} ({\blue {\blue \pm B}} {\blue \mp 2\epsilon_c} k_m) B_m
\end{align*}
and
\[
|\mu_m-({\blue {\blue \pm B}} {\blue \mp 2\epsilon_c} k_m)|\leq \epsilon_c,
\forall m\in I_{\mathrm{coff}}
\]
and
\begin{equation}\tag{\ref{1-norm}}
\|\ket{\phi_k}_j\bra{\phi_k}_j-\ket{\Phi_k}_j\bra{\Phi_k}_j\|_1 \leq C\epsilon_p, \forall j \in I_{\mathrm{basis}},
\forall k=1,2,\cdots,d
\end{equation}
and
\begin{equation}\tag{\ref{2-norm}}
\|\ket{\phi_k}_j-\ket{\Phi_k}_j\|_2 \leq \frac{C\epsilon_p}{\sqrt{2}}, \forall j \in I_{\mathrm{basis}},
\forall k=1,2,\cdots,d{\red.}
\end{equation}
Define $H'$ as follows:
\begin{align*}
H'
=B_0+\sum_{m \in I_{\mathrm{coff}}} \mu_m B_m{\red.}
\end{align*}
Then,
\begin{equation}\label{H-Hrep}
\|H-H_{\mathrm{rep}}\|_\infty
\leq \|H-H'\|_\infty+\|H'-H_{\mathrm{rep}}\|_\infty.
\end{equation}

\subsubsection{Evaluation of 
the first term $\|H-H'\|_\infty$}

First, for an arbitrary
$j \in I_{\mathrm{basis}}$,
define a unitary matrix as follows:
\begin{align*}
U_j&=\sum_{k=1}^{d} \ket{\phi_k}_j\bra{k}_j,\\
U_j^{\mathrm{rep}}&=\sum_{k=1}^{d} \ket{\Phi_k}_j\bra{k}_j.
\end{align*}
Define $H''$ as follows:
\begin{align*}
H''
=C_0+\sum_{m \in I_{\mathrm{coff}}} \mu_m C_m{\red.}
\end{align*}
Here,
\begin{align*}
C_m=\sum_{(i_1,\cdots, i_n)}
a_{(i_1,\cdots, i_n),m}
&\left(
\bigotimes_{j ={\red 1}}^{n}\{\ket{k}_j\bra{k}_j\ \mathrm{if} \ i_j=k\}
\right),
\end{align*}
where $(i_1,\cdots, i_n) \in  \{1,2,\cdots,d\} \times \cdots \times \{1,2,\cdots,d\}$.
Then,
\begin{align*}
&\|H-H'\|_\infty\\
\leq
& \Biggm\Vert\left(
\bigotimes_{j \in I_{\mathrm{basis}}}U_j
\right)
\otimes \left(
\bigotimes_{j \in I_{\mathrm{basis}}^c}I
\right)
H''
\left(
\bigotimes_{j \in I_{\mathrm{basis}}}U_j^{\dag}
\right)
\otimes \left(
\bigotimes_{j \in I_{\mathrm{basis}}^c}I
\right)\\
&\ \ \ \ \ \ \ -\left(
\bigotimes_{j \in I_{\mathrm{basis}}}U_j^{\mathrm{rep}}
\right)
\otimes \left(
\bigotimes_{j \in I_{\mathrm{basis}}^c}I
\right)
H''
\left(
\bigotimes_{j \in I_{\mathrm{basis}}}(U_j^{\mathrm{rep}})^{\dag}
\right)
\otimes \left(
\bigotimes_{j \in I_{\mathrm{basis}}^c}I
\right)
\Biggm\Vert_\infty\\
\leq
& \Biggm\Vert\left(
\bigotimes_{j \in I_{\mathrm{basis}}}U_j
\right)
\otimes \left(
\bigotimes_{j \in I_{\mathrm{basis}}^c}I
\right)
-\left(
\bigotimes_{j \in I_{\mathrm{basis}}}U_j^{\mathrm{rep}}
\right)
\otimes \left(
\bigotimes_{j \in I_{\mathrm{basis}}^c}I
\right)
\Biggm\Vert_\infty
\|H''\|_\infty\\
&\ \ \ \ \ +\|H''\|_\infty \Biggm\Vert\left(
\bigotimes_{j \in I_{\mathrm{basis}}}U_j^\dag
\right)
\otimes \left(
\bigotimes_{j \in I_{\mathrm{basis}}^c}I
\right)
-\left(
\bigotimes_{j \in I_{\mathrm{basis}}}(U_j^{\mathrm{rep}})^\dag
\right)
\otimes \left(
\bigotimes_{j \in I_{\mathrm{basis}}^c}I
\right)
\Biggm\Vert_\infty
\end{align*}
By the same discussion as
\cite[Section~8]{kitaev_textbook},
\begin{align*}
\Biggm\Vert\left(
\bigotimes_{j \in I_{\mathrm{basis}}}U_j
\right)
\otimes \left(
\bigotimes_{j \in I_{\mathrm{basis}}^c}I
\right)
-\left(
\bigotimes_{j \in I_{\mathrm{basis}}}U_j^{\mathrm{rep}}
\right)
\otimes \left(
\bigotimes_{j \in I_{\mathrm{basis}}^c}I
\right)
\Biggm\Vert_\infty
\leq
&\sum_{j \in I_{\mathrm{basis}}}
\|U_j-U_j^{\mathrm{rep}}\|_\infty,\\
\Biggm\Vert\left(
\bigotimes_{j \in I_{\mathrm{basis}}}U_j^\dag
\right)
\otimes \left(
\bigotimes_{j \in I_{\mathrm{basis}}^c}I
\right)
-\left(
\bigotimes_{j \in I_{\mathrm{basis}}}(U_j^{\mathrm{rep}})^\dag
\right)
\otimes \left(
\bigotimes_{j \in I_{\mathrm{basis}}^c}I
\right)
\Biggm\Vert_\infty
\leq
&\sum_{j \in I_{\mathrm{basis}}}
\|U_j^\dag-(U_j^{\mathrm{rep}})^\dag \|_\infty.
\end{align*}
Then, 
\begin{align*}
\|U_j-U_j^{\mathrm{rep}}\|_\infty
=
\|\sum_{k=1}^{d} \ket{\phi_k}_j\bra{k}_j
-
\ket{\Phi_k}_j\bra{k}_j
\|_\infty
\leq
\sum_{k=1}^{d}
\|(\ket{\phi_k}_j-\ket{\Phi_k}_j)\bra{k}_j\|_\infty
=
\sum_{k=1}^{d}
\|\ket{\phi_k}_j-\ket{\Phi_k}_j\|_2 \leq \frac{ C\epsilon_p}{\red\sqrt{2}} \times d{\red,}
\end{align*}
and
\begin{align*}
\|U_j^\dag-(U_j^{\mathrm{rep}})^\dag\|_\infty
=
\|\sum_{k=1}^{d} \ket{k}_j\bra{\phi_k}_j
-
\ket{k}_j\bra{\Phi_k}_j
\|_\infty
\leq
\sum_{k=1}^{d}
\|\ket{k}_j(\bra{\phi_k}_j-\bra{\Phi_k}_j)\|_\infty
=
\sum_{k=1}^{d}
\|\ket{\phi_k}_j-\ket{\Phi_k}_j\|_2 \leq \frac{C\epsilon_p}{\red\sqrt{2}} \times d{\red.}
\end{align*}
Thus,
\begin{align*}
\Biggm\Vert\left(
\bigotimes_{j \in I_{\mathrm{basis}}}U_j
\right)
\otimes \left(
\bigotimes_{j \in I_{\mathrm{basis}}^c}I
\right)
-\left(
\bigotimes_{j \in I_{\mathrm{basis}}}U_j^{\mathrm{rep}}
\right)
\otimes \left(
\bigotimes_{j \in I_{\mathrm{basis}}^c}I
\right)
\Biggm\Vert_\infty
\leq
&\sum_{j \in I_{\mathrm{basis}}} \frac{{\red d}C\epsilon_p}{\red\sqrt{2}}
=\frac{s_{\mathrm{basis}} d C\epsilon_p}{\red\sqrt{2}},\\
\Biggm\Vert\left(
\bigotimes_{j \in I_{\mathrm{basis}}}U_j^\dag
\right)
\otimes \left(
\bigotimes_{j \in I_{\mathrm{basis}}^c}I
\right)
-\left(
\bigotimes_{j \in I_{\mathrm{basis}}}(U_j^{\mathrm{rep}})^\dag
\right)
\otimes \left(
\bigotimes_{j \in I_{\mathrm{basis}}^c}I
\right)
\Biggm\Vert_\infty
\leq
&\sum_{j \in I_{\mathrm{basis}}}
\frac{d C\epsilon_p}{\red\sqrt{2}}
=\frac{s_{\mathrm{basis}} d C\epsilon_p}{\red\sqrt{2}}.
\end{align*}
Therefore,
\begin{equation}\label{H-H'}
\|H-H'\|_\infty 
\leq {\red\sqrt{2}}s_{\mathrm{basis}} d C\epsilon_p \|H''\|_\infty
\leq {\red\sqrt{2}}d C s_{\mathrm{basis}}
(s_{\mathrm{coff}}Ba
+\|A_0\|_\infty)
\epsilon_p.
\end{equation}

\subsubsection{Evaluation of 
the second term $\|H'-H_{\mathrm{rep}}\|_\infty$}

\begin{align*}
\|H'-H_{\mathrm{rep}}\|_\infty
\leq  
&\sum_{m \in I_{\mathrm{coff}}} |\mu_m-({\blue {\blue \pm B}} {\blue \mp 2\epsilon_c} k_m)|
\|A_m\|_\infty
\\
\leq 
&\sum_{(i_1,\cdots, i_n) \in I_{\mathrm{coff}}}
|\mu_m-({\blue {\blue \pm B}} {\blue \mp 2\epsilon_c} k_m)|
\|A_m\|_\infty\\
\leq & s_{\mathrm{coff}} a \epsilon_c.
\end{align*}
Thus,
\begin{equation}\label{H'-Hrep}
\|H'-H_{\mathrm{rep}}\|_\infty \leq  s_{\mathrm{coff}} a \epsilon_c.
\end{equation}

\subsubsection{
Ecvaluation of (\ref{H-Hrep})
and definition of $\epsilon_c,\epsilon_p$}

By (\ref{H-Hrep}),(\ref{H-H'}),(\ref{H'-Hrep}),
\begin{align*}
\|H-H_{\mathrm{rep}}\|_\infty 
\leq  {\red\sqrt{2}}d C s_{\mathrm{basis}}
(s_{\mathrm{coff}}Ba
+\|A_0\|_\infty)
\epsilon_p
+s_{\mathrm{coff}} a \epsilon_c.
\end{align*}
Here, if we set
\[
\epsilon_p=\frac{\epsilon}{2{\red\sqrt{2}}d C s_{\mathrm{basis}}
(s_{\mathrm{coff}}Ba
+\|A_0\|_\infty)},\ 
\epsilon_c=\frac{\epsilon}{2 s_{\mathrm{coff}} a},
\]
then
\begin{align*}
\|H-H_{\mathrm{rep}}\|_\infty \leq \epsilon.
\end{align*}

\subsection{Proof of Proposition~\ref{property1}}

Fix $\epsilon>0$.
By setting $\epsilon_p$ and $\epsilon_c$ 
appropriately,
we prove that
for any $H \in S_{LD}$,
there exists $H_{\mathrm{rep}} \in \mathcal{N}_S$ such that
\begin{align*}
\forall \psi, 
\left|\left(F_Q(\ket\psi,H)-\underset{\ket{\psi} \leftarrow
Sym^n(\mathbb{C}^d)}{E}[F_Q(\ket{\psi},H_S)]\right)
-
\left(F_Q(\ket\psi,H_{\mathrm{rep}})-\underset{\ket{\psi} \leftarrow
Sym^n(\mathbb{C}^d)}{E}[F_Q(\ket{\psi},H_{\mathrm{rep},S})]\right)
\right|\leq
\epsilon{\red.}
\end{align*}
For any $\ket\psi$, 
we may evaluate the following value:
\begin{align*}
&\left|\left(F_Q(\ket\psi,H)-\underset{\ket{\psi} \leftarrow
Sym^n(\mathbb{C}^d)}{E}[F_Q(\ket{\psi},H_S)]\right)
-
\left(F_Q(\ket\psi,H_{\mathrm{rep}})-\underset{\ket{\psi} \leftarrow
Sym^n(\mathbb{C}^d)}{E}[F_Q(\ket{\psi},H_{\mathrm{rep},S})]\right)
\right|\\
\leq 
&|F_Q(\ket\psi,H)-(F_Q(\ket\psi,H_{\mathrm{rep}}))|
+\left|
\underset{\ket{\psi} \leftarrow
Sym^n(\mathbb{C}^d)}{E}[F_Q(\ket{\psi},H_S)]
-
\underset{\ket{\psi} \leftarrow
Sym^n(\mathbb{C}^d)}{E}[F_Q(\ket{\psi},H_{\mathrm{rep},S})]\right|{\red.}
\end{align*}

Here,
\begin{align*}
&\frac{1}{4}|F_Q(\ket\psi,H)-F_Q(\ket\psi,H_{\mathrm{rep}})|\\
\leq &\|H^2-H_{\mathrm{rep}}^2\|_\infty \|\ket{\psi}\bra{\psi}\|_1
+\|H\otimes H-H_{\mathrm{rep}}\otimes H_{\mathrm{rep}}\|_\infty \|\ket{\psi}\bra{\psi}\|_1\\
\leq &\|H^2-H_{\mathrm{rep}}^2\|_\infty 
+\|H\otimes H-H_{\mathrm{rep}}\otimes H_{\mathrm{rep}}\|_\infty. 
\end{align*}

\subsubsection{
Evaluation of the first term
$|F_Q(\ket\psi,H)-F_Q(\ket\psi,H_{\mathrm{rep}})|$:
evaluation of
$\|H^2-H_{\mathrm{rep}}^2\|_\infty$}

First,
\begin{equation}\label{H2-Hrep2}
\|H^2-H_{\mathrm{rep}}^2\|_\infty
\leq \|H^2-H'^2\|_\infty+\|H'^2-H_{\mathrm{rep}}^2\|_\infty.
\end{equation}
By the same discussion as
the proof of
Proposition~\ref{epsilonnet},
the first term
$\|H^2-H'^2\|_\infty$ is evaluated as
follows:
\begin{equation}\label{H2-H'2}
\|H^2-H'^2\|_\infty 
\leq s_{\mathrm{basis}} d C\epsilon_p \|H''^2\|_\infty
\leq d C s_{\mathrm{basis}}
(s_{\mathrm{coff}}Ba
+\|A_0\|_\infty)^2
\epsilon_p.
\end{equation}
%%%
The second term
$\|H'^2-H_{\mathrm{rep}}^2\|_\infty$
is evaluated as follows:
\begin{align*}
\|H'^2-H_{\mathrm{rep}}^2\|_\infty
\leq  
&\sum_{m \in I_{\mathrm{coff}}} |\mu_m^2-({\blue {\blue \pm B}} {\blue \mp 2\epsilon_c} k_m)^2|
\|A_m\|_\infty\\
\leq  
&\sum_{m \in I_{\mathrm{coff}}} |\mu_m+({\blue {\blue \pm B}} {\blue \mp 2\epsilon_c} k_m)|
|\mu_m-({\blue {\blue \pm B}} {\blue \mp 2\epsilon_c} k_m)|
\|A_m\|_\infty\\
\leq  &2s_{\mathrm{coff}}B a \epsilon_c.
\end{align*}
Thus,
\begin{equation}\label{H'2-Hrep2}
\|H'^2-H_{\mathrm{rep}}^2\|_\infty \leq  2 s_{\mathrm{coff}} B a \epsilon_c.
\end{equation}
Therefore, by
(\ref{H2-Hrep2}),(\ref{H2-H'2})
and (\ref{H'2-Hrep2}),
\begin{equation}\label{H2-Hrep2_matome}
\|H^2-H_{\mathrm{rep}}^2\|_\infty 
\leq  d C s_{\mathrm{basis}}
(s_{\mathrm{coff}}Ba
+\|A_0\|_\infty)^2\epsilon_p
+2s_{\mathrm{coff}} Ba\epsilon_c{\red.}
\end{equation}
\subsubsection{Evaluation of the first term
$|F_Q(\ket\psi,H)-F_Q(\ket\psi,H_{\mathrm{rep}})|$:
evaluation of
$\|H \otimes H
-H_{\mathrm{rep}} \otimes H_{\mathrm{rep}}\|_\infty$}

First,
\begin{equation}\label{H_tensor-Hrep_tensor}
\|H \otimes H
-H_{\mathrm{rep}} \otimes H_{\mathrm{rep}}
\|_\infty
\leq \|H \otimes H-H' \otimes H'\|_\infty
+\|H' \otimes H'-H_{\mathrm{rep}} \otimes H_{\mathrm{rep}}\|_\infty.
\end{equation}
The first term $\|H \otimes H-H' \otimes H'\|_\infty$
is evaluated as follows:
\begin{align*}
&\|H \otimes H -H' \otimes H'\|_\infty\\
\leq
& \Biggm\Vert\left(
\bigotimes_{j \in I_{\mathrm{basis}}}U_j \otimes U_j
\right)
\otimes \left(
\bigotimes_{j \in I_{\mathrm{basis}}^c}I \otimes I
\right)
H''\otimes H''
\left(
\bigotimes_{j \in I_{\mathrm{basis}}}U_j^{\dag} \otimes U_j^{\dag}
\right)
\otimes \left(
\bigotimes_{j \in I_{\mathrm{basis}}^c}I \otimes I
\right)\\
&\ \ \ \ \ \ \ -\left(
\bigotimes_{j \in I_{\mathrm{basis}}}U_j^{\mathrm{rep}}\otimes U_j^{\mathrm{rep}}
\right)
\otimes \left(
\bigotimes_{j \in I_{\mathrm{basis}}^c}I \otimes I
\right)
H'' \otimes H''
\left(
\bigotimes_{j \in I_{\mathrm{basis}}}(U_j^{\mathrm{rep}})^{\dag} \otimes (U_j^{\mathrm{rep}})^{\dag}
\right)
\otimes \left(
\bigotimes_{j \in I_{\mathrm{basis}}^c}I{\red\otimes I}
\right)
\Biggm\Vert_\infty\\
\leq
& \Biggm\Vert\left(
\bigotimes_{j \in I_{\mathrm{basis}}}U_j \otimes U_j
\right)
\otimes \left(
\bigotimes_{j \in I_{\mathrm{basis}}^c}I \otimes I
\right)
-\left(
\bigotimes_{j \in I_{\mathrm{basis}}}U_j^{\mathrm{rep}} \otimes U_j^{\mathrm{rep}}
\right)
\otimes \left(
\bigotimes_{j \in I_{\mathrm{basis}}^c}I \otimes I
\right)
\Biggm\Vert_\infty
\|H'' \otimes H'' \|_\infty\\
&\ \ \ \ \ +\|H'' \otimes H'' \|_\infty \Biggm\Vert\left(
\bigotimes_{j \in I_{\mathrm{basis}}}U_j^\dag \otimes U_j^\dag
\right)
\otimes \left(
\bigotimes_{j \in I_{\mathrm{basis}}^c}I \otimes I
\right)
-\left(
\bigotimes_{j \in I_{\mathrm{basis}}}(U_j^{\mathrm{rep}})^\dag \otimes (U_j^{\mathrm{rep}})^\dag
\right)
\otimes \left(
\bigotimes_{j \in I_{\mathrm{basis}}^c}I \otimes I
\right)
\Biggm\Vert_\infty{\red.}
\end{align*}
Then, 
\begin{align*}
&\Biggm\Vert\left(
\bigotimes_{j \in I_{\mathrm{basis}}}U_j \otimes U_j
\right)
\otimes \left(
\bigotimes_{j \in I_{\mathrm{basis}}^c}I \otimes I
\right)
-\left(
\bigotimes_{j \in I_{\mathrm{basis}}}U_j^{\mathrm{rep}} \otimes U_j^{\mathrm{rep}}
\right)
\otimes \left(
\bigotimes_{j \in I_{\mathrm{basis}}^c}I \otimes I
\right)
\Biggm\Vert_\infty\\
\leq
&2\Biggm\Vert\left(
\bigotimes_{j \in I_{\mathrm{basis}}}U_j
\right)
\otimes \left(
\bigotimes_{j \in I_{\mathrm{basis}}^c}I
\right)
-\left(
\bigotimes_{j \in I_{\mathrm{basis}}}U_j^{\mathrm{rep}}
\right)
\otimes \left(
\bigotimes_{j \in I_{\mathrm{basis}}^c}I
\right)
\Biggm\Vert_\infty\\
\leq
&2\sum_{j \in I_{\mathrm{basis}}}
\|U_j-U_j^{\mathrm{rep}}\|_{\red\infty}\\
\leq
&2\sum_{j \in I_{\mathrm{basis}}}
\frac{d C\epsilon_p}{\red\sqrt{2}}
={\red\sqrt{2}} d C s_{\mathrm{basis}}\epsilon_p.
\end{align*}
The first and second {\red inequalities
follow} from the same discussion
as
\cite[Section~8]{kitaev_textbook}.
The third inequality follows from
the same discussion
as a proof of Proposition~\ref{epsilonnet}.
Similarly,
\begin{align*}
\Biggm\Vert\left(
\bigotimes_{j \in I_{\mathrm{basis}}}U_j^\dag \otimes U_j^\dag
\right)
\otimes \left(
\bigotimes_{j \in I_{\mathrm{basis}}^c}I \otimes I
\right)
-\left(
\bigotimes_{j \in I_{\mathrm{basis}}}(U_j^{\mathrm{rep}})^\dag \otimes (U_j^{\mathrm{rep}})^\dag
\right)
\otimes \left(
\bigotimes_{j \in I_{\mathrm{basis}}^c}I \otimes I
\right)
\Biggm\Vert_\infty
\leq
{\red\sqrt{2}} d C s_{\mathrm{basis}}\epsilon_p.
\end{align*}
Thus,
\begin{equation}\label{H_tensor-H'_tensor}
\|H \otimes H-H' \otimes H'\|_\infty 
\leq {\red2\sqrt{2}}s_{\mathrm{basis}} d C\epsilon_p  \|H'' \otimes H''\|_\infty
\leq 2{\red\sqrt{2}}d C s_{\mathrm{basis}}
(s_{\mathrm{coff}}Ba
+\|A_0\|_\infty)^2
\epsilon_p.
\end{equation}
Then, we evaluate
the second term
$\|H' \otimes H'-H_{\mathrm{rep}} \otimes H_{\mathrm{rep}}\|_\infty$.
Here,
\begin{align*}
H' \otimes H'
=&\sum_{m \in I_{\mathrm{coff}}} 
\sum_{m' \in I_{\mathrm{coff}}}
\mu_m \mu_{m'}
B_m \otimes B_{m'}
+\sum_{m \in I_{\mathrm{coff}}} 
\mu_m B_m \otimes B_0
+\sum_{m' \in I_{\mathrm{coff}}}
{\red\mu_{m'} B_0 \otimes B_{m'}}
+B_0 \otimes B_0{\red.}
\end{align*}
Also,
\begin{align*}
H_{\mathrm{rep}} \otimes H_{\mathrm{rep}}
=&\sum_{m \in I_{\mathrm{coff}}} 
\sum_{m' \in I_{\mathrm{coff}}}
({\blue {\blue \pm B}} {\blue \mp 2\epsilon_c} k_m)
({\blue {\blue \pm B}} {\blue \mp 2\epsilon_c} k_{m'})
B_m \otimes B_{m'}
+\sum_{m \in I_{\mathrm{coff}}} 
({\blue {\blue \pm B}} {\blue \mp 2\epsilon_c} k_m)
B_m \otimes B_0\\
&+\sum_{m' \in I_{\mathrm{coff}}}
({\blue {\blue \pm B}} {\blue \mp 2\epsilon_c} k_{m'})
{\red B_0 \otimes B_{m'}}
+B_0 \otimes B_0{\red.}
\end{align*}

Then,
\begin{align*}
&\|H' \otimes H'-H_{\mathrm{rep}} \otimes H_{\mathrm{rep}}
\|_\infty\\
\leq
&\Biggm\Vert 
\sum_{m \in I_{\mathrm{coff}}} 
\sum_{m' \in I_{\mathrm{coff}}}
\mu_m \mu_{m'}
B_m \otimes B_{m'}
-
\sum_{m \in I_{\mathrm{coff}}} 
\sum_{m' \in I_{\mathrm{coff}}}
({\blue {\blue \pm B}} {\blue \mp 2\epsilon_c} k_m)
({\blue {\blue \pm B}} {\blue \mp 2\epsilon_c} k_{m'})
B_m \otimes B_{m'}
\Biggm\Vert_\infty\\   
&+\Biggm\Vert 
\sum_{m \in I_{\mathrm{coff}}} 
\mu_m B_m \otimes B_0
-
\sum_{m \in I_{\mathrm{coff}}}
({\blue {\blue \pm B}} {\blue \mp 2\epsilon_c} k_m)
B_m \otimes B_0
\Biggm\Vert_\infty\\
&+\Biggm\Vert 
\sum_{m' \in I_{\mathrm{coff}}}
\mu_{m'} B_0 \otimes B_{m'}
-
\sum_{m' \in I_{\mathrm{coff}}}
({\blue {\blue \pm B}} {\blue \mp 2\epsilon_c} k_{m'})
B_0 \otimes B_{m'}
\Biggm\Vert_\infty{\red.}  
\end{align*}

The first term
is evaluated as follows:
\begin{align*}
&\sum_{m \in I_{\mathrm{coff}}} 
\sum_{m' \in I_{\mathrm{coff}}}
|\mu_m\mu_{m'}-({\blue {\blue \pm B}} \pm 2 \epsilon_c k_m)({\blue {\blue \pm B}} \pm 2 \epsilon_c k_{m'})|\|B_m \otimes B_{m'}\|_\infty
\leq s_{\mathrm{coff}}^2
(2B\epsilon_c+{\red 2}\epsilon_c^2)a^2
\leq s_{\mathrm{coff}}^2(2B+{\red 2})a^2
\epsilon_c{\red.}
\end{align*}
The second term
is evaluated as follows:
\begin{align*}
&\sum_{m \in I_{\mathrm{coff}}} 
|(\mu_m-({\blue {\blue \pm B}} \pm 2 \epsilon_c k_m))|\|B_m \otimes B_0\|_\infty
\leq s_{\mathrm{coff}}\|A_0\|_\infty
{\red a} \epsilon_c{\red.}
\end{align*}
The third term is evaluated
as that of the second term.
To {\red sum} up,
\begin{equation}\label{H'_tensor-Hrep_tensor}
\|H' \otimes H'-H_{\mathrm{rep}} \otimes H_{\mathrm{rep}}\|_\infty
\leq s_{\mathrm{coff}}^2(2B+{\red 2})a^2
\epsilon_c+
2s_{\mathrm{coff}}
\|A_0\|_\infty
{\red a} \epsilon_c{\red.}
\end{equation}
Thus, by
(\ref{H_tensor-Hrep_tensor}),
(\ref{H_tensor-H'_tensor}) and
(\ref{H'_tensor-Hrep_tensor}),
\begin{align*}
&\|H \otimes H
-H_{\mathrm{rep}} \otimes H_{\mathrm{rep}}
\|_\infty\\
\leq &\|H \otimes H-H' \otimes H'\|_\infty
+\|H' \otimes H'-H_{\mathrm{rep}} \otimes H_{\mathrm{rep}}\|_\infty\\
\leq &
 s_{\mathrm{coff}}^2(2B+{\red 2})a^2
\epsilon_c+
2s_{\mathrm{coff}}
\|A_0\|_\infty
{\red a} \epsilon_c
+2{\red\sqrt{2}}d C s_{\mathrm{basis}}
(s_{\mathrm{coff}}Ba
+\|A_0\|_\infty)^2
\epsilon_p.
\end{align*}

\subsubsection{Evaluation of the 
second term
$\left|
\underset{\ket{\psi} \leftarrow
Sym^n(\mathbb{C}^d)}{E}[F_Q(\ket{\psi},H_S)]
-
\underset{\ket{\psi} \leftarrow
Sym^n(\mathbb{C}^d)}{E}[F_Q(\ket{\psi},H_{\mathrm{rep},S})]\right|$}

Note that the Hamiltonians $H_S$
and  $H_{\mathrm{rep},S}$
are represented respectively as follows:
\begin{align*}
H_S&=h_S \otimes I \otimes \cdots \otimes I
+ I \otimes h_S \otimes I \otimes \cdots \otimes I
+\cdots 
+ I \otimes \cdots \otimes I \otimes h_S,\\
H_{\mathrm{rep},S}&=h_{\mathrm{rep},S} \otimes I \otimes \cdots \otimes I
+ I \otimes h_{\mathrm{rep},S} \otimes I \otimes \cdots \otimes I
+\cdots 
+ I \otimes \cdots \otimes I \otimes h_{\mathrm{rep},S},
\end{align*}
for the following single-qudit Hermitian operators
\begin{align*}
h_S&=\sum_{j=1}^{d} 
\frac{\sum_{i=1}^n \mu_{i,j}}{n} 
\ket{\phi_j}\bra{\phi_j},\\
h_{\mathrm{rep},S}&=\sum_{j=1}^{d} 
\frac{\sum_{i=1}^n ({\blue {\blue \pm B}} {\blue \mp 2\epsilon_c} k_{i,j})}{n} 
\ket{\Phi_j}\bra{\Phi_j}.
\end{align*}
By Lemma~\ref{gutaitekinaatai},
we have
\begin{align*}
\underset{\ket{\psi} \leftarrow  Sym^n(\mathbb{C}^d)}{E}[F_Q(\ket{\psi},H_S)]
&=\frac{4n(n+d)}{d+1}\frac{_{n+d-1}C_n}{_{n+d-1}C_n +1}
\left(\frac{\mathrm{Tr}(h_S^2)}{d}
  -\frac{\mathrm{Tr}(h_S)^2}{d^2}
  \right),\\
 \underset{\ket{\psi} \leftarrow  Sym^n(\mathbb{C}^d)}{E}[F_Q(\ket{\psi},H_{\mathrm{rep},S})]
&=\frac{4n(n+d)}{d+1}\frac{_{n+d-1}C_n}{_{n+d-1}C_n +1}
\left(\frac{\mathrm{Tr}(h_{\mathrm{rep},S}^2)}{d}
  -\frac{\mathrm{Tr}(h_{\mathrm{rep},S})^2}{d^2}
  \right).
\end{align*}
Here,
\begin{align*}
|\mathrm{Tr}(h_S^2)-
\mathrm{Tr}(h_{\mathrm{rep},S}^2)|
&\leq
\sum_{j=1}^{d}
\left|
\left(
\frac{\sum_{i=1}^n \mu_{i,j}}{n}
\right)^2
-
\left(
\frac{\sum_{i=1}^n ({\blue {\blue \pm B}} {\blue \mp 2\epsilon_c} k_{i,j})}{n} 
\right)^2
\right|\\
&\leq
\sum_{j=1}^{d}
\left|
\frac{\sum_{i=1}^n \mu_{i,j}}{n}
+
\frac{\sum_{i=1}^n ({\blue {\blue \pm B}} {\blue \mp 2\epsilon_c} k_{i,j})}{n}
\right|
\left|
\frac{\sum_{i=1}^n \mu_{i,j}}{n}
-
\frac{\sum_{i=1}^n ({\blue {\blue \pm B}} {\blue \mp 2\epsilon_c} k_{i,j})}{n}
\right|
\\
&\leq 2dB\epsilon_c.
\end{align*}
Furthermore,
\begin{align*}
|\mathrm{Tr}(h_S)^2-
\mathrm{Tr}(h_{\mathrm{rep},S})^2|
&\leq
|\mathrm{Tr}(h_S)+
\mathrm{Tr}(h_{\mathrm{rep},S})|
|\mathrm{Tr}(h_S)-
\mathrm{Tr}(h_{\mathrm{rep},S})|\\
&\leq 2{\red d}B
\sum_{j=1}^{d}
\left|
\frac{\sum_{i=1}^n \mu_{i,j}}{n}
-
\frac{\sum_{i=1}^n ({\blue {\blue \pm B}} {\blue \mp 2\epsilon_c} k_{i,j})}{n} 
\right|\\
&\leq 2d^{\red 2}B\epsilon_c.
\end{align*}

Therefore,
\begin{align*}
&\left|
\underset{\ket{\psi} \leftarrow
Sym^n(\mathbb{C}^d)}{E}[F_Q(\ket{\psi},H_S)]
-
\underset{\ket{\psi} \leftarrow
Sym^n(\mathbb{C}^d)}{E}[F_Q(\ket{\psi},H_{\mathrm{rep},S})]\right|\\
{\red =} &
\frac{4n(n+d)}{d+1}\frac{_{n+d-1}C_n}{_{n+d-1}C_n +1}
\left|
\left(\frac{\mathrm{Tr}(h_S^2)}{d}
-\frac{\mathrm{Tr}(h_S)^2}{d^2}
\right)
  -
\left(\frac{\mathrm{Tr}(h_{\mathrm{rep},S}^2)}{d}
  -\frac{\mathrm{Tr}(h_{\mathrm{rep},S})^2}{d^2}
  \right)
\right|\\
\leq &
\frac{4n(n+d)}{d+1}\frac{_{n+d-1}C_n}{_{n+d-1}C_n +1}
\left(
\frac{1}{d}
|\mathrm{Tr}(h_S^2)-
\mathrm{Tr}(h_{\mathrm{rep},S}^2)|
+
\frac{1}{d^2}
|\mathrm{Tr}(h_S)^2-
\mathrm{Tr}(h_{\mathrm{rep},S})^2|
\right)\\
\leq &
\frac{4n(n+d)}{d+1}\frac{_{n+d-1}C_n}{_{n+d-1}C_n +1}
{\red 4B\epsilon_c}\\
\leq &
\frac{{\red 16}Bn(n+d)}{d}\epsilon_c{\red.}
\end{align*}

\subsubsection{Evaluation of $\left|\left(F_Q(\ket\psi,H)-\underset{\ket{\psi} \leftarrow
Sym^n(\mathbb{C}^d)}{E}[F_Q(\ket{\psi},H_S)]\right)
-
\left(F_Q(\ket\psi,H_{\mathrm{rep}})-\underset{\ket{\psi} \leftarrow
Sym^n(\mathbb{C}^d)}{E}[F_Q(\ket{\psi},H_{\mathrm{rep},S})]\right)
\right|$ and 
definition of
$\epsilon_p,\epsilon_c$}

\begin{align*}
&\frac{1}{4}\left|\left(F_Q(\ket\psi,H)-\underset{\ket{\psi} \leftarrow
Sym^n(\mathbb{C}^d)}{E}[F_Q(\ket{\psi},H_S)]\right)
-
\left(F_Q(\ket\psi,H_{\mathrm{rep}})-\underset{\ket{\psi} \leftarrow
Sym^n(\mathbb{C}^d)}{E}[F_Q(\ket{\psi},H_{\mathrm{rep},S})]\right)
\right|\\
\leq 
&\frac{1}{4}|F_Q(\ket\psi,H)-(F_Q(\ket\psi,H_{\mathrm{rep}}))|
+\frac{1}{4}\left|
\underset{\ket{\psi} \leftarrow
Sym^n(\mathbb{C}^d)}{E}[F_Q(\ket{\psi},H_S)]
-
\underset{\ket{\psi} \leftarrow
Sym^n(\mathbb{C}^d)}{E}[F_Q(\ket{\psi},H_{\mathrm{rep},S})]\right|
%&\frac{1}{4}|F_Q(\ket\psi,H)-(F_Q(\ket\psi,H_{\mathrm{rep}}))|
%+\frac{1}{4}\left|
%\max_{\ket{\Phi}: \mathrm{separable}}F_Q(\ket{\Phi},H)
%-
%\max_{\ket{\Phi}: \mathrm{separable}}F_Q(\ket{\Phi},H_{\mathrm{rep}}) ))\right|
\\
{\red\leq} &\|H^2-H_{\mathrm{rep}}^2\|_\infty 
+\|H\otimes H-H_{\mathrm{rep}}\otimes H_{\mathrm{rep}}\|_\infty 
+\frac{1}{4}\left|
\underset{\ket{\psi} \leftarrow
Sym^n(\mathbb{C}^d)}{E}[F_Q(\ket{\psi},H_S)]
-
\underset{\ket{\psi} \leftarrow
Sym^n(\mathbb{C}^d)}{E}[F_Q(\ket{\psi},H_{\mathrm{rep},S})]\right|\\
\leq &
 d C s_{\mathrm{basis}}
(s_{\mathrm{coff}}Ba
+\|A_0\|_\infty)^2\epsilon_p
+2s_{\mathrm{coff}} Ba\epsilon_c\\
&+s_{\mathrm{coff}}^2(2B+{\red 2})a^2
\epsilon_c+
2s_{\mathrm{coff}}
\|A_0\|_\infty
{\red a} \epsilon_c
+2{\red\sqrt{2}}d C s_{\mathrm{basis}}
(s_{\mathrm{coff}}Ba
+\|A_0\|_\infty)^2
\epsilon_p
+\frac{{\red 4}Bn(n+d)}{\red d}\epsilon_c\\
\leq &
{\red(1+2\sqrt{2})} d C s_{\mathrm{basis}}
(s_{\mathrm{coff}}Ba
+\|A_0\|_\infty)^2{\red\epsilon_p}
+(2 s_{\mathrm{coff}} Ba
+ s_{\mathrm{coff}}^2(2B+{\red 2})a^2
+
2s_{\mathrm{coff}}
\|A_0\|_\infty
{\red a}
+{\red 4Bn(n+d)/d}) \epsilon_c{\red.}
\end{align*}
If we set
\begin{align*}
    \epsilon_p&=\frac{\epsilon}{8 \ {\red(1+2\sqrt{2})} d C s_{\mathrm{basis}}
(s_{\mathrm{coff}}Ba
+\|A_0\|_\infty)^2},\\
\epsilon_c&=\frac{\epsilon}{8
(2 s_{\mathrm{coff}} Ba
+ s_{\mathrm{coff}}^2(2B+{\red 2})a^2
+
2s_{\mathrm{coff}}
\|A_0\|_\infty
{\red a}
+{\red 4Bn(n+d)/d}) }
\end{align*}
then
\begin{align*}
\left|\left(F_Q(\ket\psi,H)-\underset{\ket{\psi} \leftarrow
Sym^n(\mathbb{C}^d)}{E}[F_Q(\ket{\psi},H_S)]\right)
-
\left(F_Q(\ket\psi,H_{\mathrm{rep}})-\underset{\ket{\psi} \leftarrow
Sym^n(\mathbb{C}^d)}{E}[F_Q(\ket{\psi},H_{\mathrm{rep},S})]\right)
\right|
 \leq \epsilon.
\end{align*}

\subsection{Proof of Proposition~\ref{property3}}

Fix $\epsilon>0$.
By setting $\epsilon_p$ and $\epsilon_c$ 
appropriately,
we prove that
for any $H \in S_{LD}$,
there exists
$H_{\mathrm{rep}} \in \mathcal{N}_S$ such that
\[
\left|\Big(F_Q(\ket\psi,H)-\max_{\ket{\Phi}: \mathrm{separable}}F_Q(\ket{\Phi},H)\Big)
-
\Big(F_Q(\ket\psi,H_{\mathrm{rep}})-\max_{\ket{\Phi}: \mathrm{separable}}F_Q(\ket{\Phi},H_{\mathrm{rep}}) )\Big)
\right|\leq
\epsilon 
\]
For any $\ket\psi$, 
we may evaluate the following value:
\begin{align*}
&\left|\Big(F_Q(\ket\psi,H)-\max_{\ket{\Phi}: \mathrm{separable}}F_Q(\ket{\Phi},H)\Big)
-
\Big(F_Q(\ket\psi,H_{\mathrm{rep}})-\max_{\ket{\Phi}: \mathrm{separable}}F_Q(\ket{\Phi},H_{\mathrm{rep}}) )\Big)
\right|\\
\leq 
&|F_Q(\ket\psi,H)-(F_Q(\ket\psi,H_{\mathrm{rep}}))|
+\left|
\max_{\ket{\Phi}: \mathrm{separable}}F_Q(\ket{\Phi},H)
-
\max_{\ket{\Phi}: \mathrm{separable}}F_Q(\ket{\Phi},H_{\mathrm{rep}}) ))\right|{\red.}
\end{align*}

\subsubsection{Evaluation of the first term
$|F_Q(\ket\psi,H)-F_Q(\ket\psi,H_{\mathrm{rep}})|$}

By the same discussion as 
the proof of Proposition~\ref{property1},
\begin{align*}
&{\red\cfrac{1}{4}}|F_Q(\ket\psi,H)-F_Q(\ket\psi,H_{\mathrm{rep}})|\\
\leq &\|H^2-H_{\mathrm{rep}}\|_\infty \|\ket{\psi}\bra{\psi}\|_1
+\|H\otimes H-H_{\mathrm{rep}}\otimes H_{\mathrm{rep}}\|_\infty \|\ket{\psi}\bra{\psi}\|_1\\
\leq &\|H^2-H_{\mathrm{rep}}^2\|_\infty 
+\|H\otimes H-H_{\mathrm{rep}}\otimes H_{\mathrm{rep}}\|_\infty\\
\leq &  d C s_{\mathrm{basis}}
(s_{\mathrm{coff}}Ba
+\|A_0\|_\infty)^2\epsilon_p
+2s_{\mathrm{coff}} Ba\epsilon_c\\
&+s_{\mathrm{coff}}^2(2B+{\red 2})a^2
\epsilon_c+
2s_{\mathrm{coff}}
\|A_0\|_\infty
{\red a} \epsilon_c
+2{\red\sqrt{2}}d C s_{\mathrm{basis}}
(s_{\mathrm{coff}}Ba
+\|A_0\|_\infty)^2
\epsilon_p{\red.}
\end{align*}

\subsubsection{Evaluation of the 
second term
$\left|
\max_{\ket{\Phi}: \mathrm{separable}}F_Q(\ket{\Phi},H)
-
\max_{\ket{\Phi}: \mathrm{separable}}F_Q(\ket{\Phi},H_{\mathrm{rep}}) )\right|$}

First,
we prove that 
for an arbitrary separable state
$\ket{\Phi_{\mathrm{sep}}}$,
there exists a separable state
$\ket{\Psi_{\mathrm{sep}}}$
such that
\[
F_Q(\ket{\Psi_{\mathrm{sep}}},H_{\mathrm{rep}})-\epsilon_{\mathrm{optimized}\ \mathrm{sep.}} \leq F_Q(\ket{\Phi_{\mathrm{sep}}},H) \leq F_Q(\ket{\Psi_{\mathrm{sep}}},H_{\mathrm{rep}})+\epsilon_{\mathrm{optimized}\ \mathrm{sep.}},
\]
where
\[
\epsilon_{\mathrm{optimized}\ \mathrm{sep.}}=
16 (s_{\mathrm{coff}}Ba+
\|A_0\|_\infty)
s_{\mathrm{coff}} a \epsilon_c
.
\]
First,
we denote
\begin{align*}
H=\sum_{(i_1,\cdots, i_n)}
x_{(i_1,\cdots, i_n)}
&\left(
\bigotimes_{j \in I_{\mathrm{basis}}}\{\ket{\phi_k}_j\bra{\phi_k}_j\ \mathrm{if} \ i_j=k\}
\right) \otimes \left(
\bigotimes_{j \in I_{\mathrm{basis}}^c}\{\ket{k}_j\bra{k}_j\ \mathrm{if} \ i_j=k\}
\right),\\
H_{\mathrm{rep}}
=\sum_{(i_1,\cdots, i_n)}
y_{(i_1,\cdots, i_n)}
&\left(
\bigotimes_{j \in I_{\mathrm{basis}}}\{\ket{\Phi_k}_j\bra{\Phi_k}_j\ \mathrm{if} \ i_j=k\}
\right) \otimes \left(
\bigotimes_{j \in I_{\mathrm{basis}}^c}\{\ket{k}_j\bra{k}_j\ \mathrm{if} \ i_j=k\}
\right).
\end{align*}

For an arbitrary separable state
$\ket{\Phi_{\mathrm{sep}}}$,
there exists 
$c_{j,k} \in [0,1]$ such that
all $c_{j,k}$s are the same
for $(j,k)\in\{1,\cdots,n\} \times 
\{1,\cdots,d\}$,
$\sum_{k=1}^{d}c_{j,k}^2=1$ and
\[
\ket{\Phi_{\mathrm{sep}}}=
\left(
\bigotimes_{j \in I_{\mathrm{basis}}}
\sum_{k=1}^{d}
c_{j,k}\ket{\phi_k}_j
\right)\\
\otimes \left({\red\bigotimes_{j\in I_{\mathrm{basis}}^c}}
\sum_{k=1}^{d}
c_{j,k}\ket{k}_j
\right),
\]
{\red and} we define a separable state
$\ket{\Psi_{\mathrm{sep}}}$ as follows:
\[
\ket{\Psi_{\mathrm{sep}}}=
\left(
\bigotimes_{j \in I_{\mathrm{basis}}}
\sum_{k=1}^{d}
c_{j,k}\ket{\Phi_k}_j
\right)
\otimes \left(
\bigotimes_{j \in I_{\mathrm{basis}}^c}
\sum_{k=1}^{d}
c_{j,k}\ket{k}_j
\right){\red.}
\]
Then,
\begin{align*}
\frac{1}{4}F_Q(\ket{\Phi_{\mathrm{sep}}},H)
=
\sum_{(i_1,\cdots,i_n)} x_{(i_1,\cdots,i_n)}^2
|c_{j i_1}|^2\cdots|c_{j i_n}|^2
-\left(
\sum_{(i_1,\cdots,i_n)} x_{(i_1,\cdots,i_n)}
|c_{j i_1}|^2\cdots|c_{j i_n}|^2
\right)^2{\red.}
\end{align*}
Also,
\begin{align*}
\frac{1}{4}F_Q(\ket{\Psi_{\mathrm{sep}}},H_{\mathrm{rep}})
=
\sum_{(i_1,\cdots,i_n)} y_{(i_1,\cdots,i_n)}^2
|c_{j i_1}|^2\cdots|c_{j i_n}|^2
-\left(
\sum_{(i_1,\cdots,i_n)} y_{(i_1,\cdots,i_n)}
|c_{j i_1}|^2\cdots|c_{j i_n}|^2
\right)^2{\red.}
\end{align*}
Then,
\begin{align*}
&\frac{1}{4}\left|
F_Q(\ket{\Phi_{\mathrm{sep}}},H)-
F_Q(\ket{\Psi_{\mathrm{sep}}},H_{\mathrm{rep}})\right|\\
\leq &\left|
\sum_{(i_1,\cdots,i_n)} (x_{(i_1,\cdots,i_n)}^2-
y_{(i_1,\cdots,i_n)}^2)
|c_{j i_1}|^2\cdots|c_{j i_n}|^2
\right|\\
&+\left|
\left(
\sum_{(i_1,\cdots,i_n)} x_{(i_1,\cdots,i_n)}
|c_{j i_1}|^2\cdots|c_{j i_n}|^2
\right)^2
-
\left(
\sum_{(i_1,\cdots,i_n)} y_{(i_1,\cdots,i_n)}
|c_{j i_1}|^2\cdots|c_{j i_n}|^2
\right)^2
\right|{\red.}
\end{align*}
The first term is evaluated as follows:
\begin{align*}
\left|
\sum_{(i_1,\cdots,i_n)} (x_{(i_1,\cdots,i_n)}^2-
y_{(i_1,\cdots,i_n)}^2)
|c_{j i_1}|^2\cdots|c_{j i_n}|^2
\right|
\leq &\max_{(i_1,\cdots,i_n)} |x_{(i_1,\cdots,i_n)}^2-
y_{(i_1,\cdots,i_n)}^2| \\
\leq &\max_{(i_1,\cdots,i_n)} |x_{(i_1,\cdots,i_n)}+
y_{(i_1,\cdots,i_n)}|
|x_{(i_1,\cdots,i_n)}-
y_{(i_1,\cdots,i_n)}|
\\
\leq &2 (s_{\mathrm{coff}}Ba
+\|A_0\|_\infty)
\times
s_{\mathrm{coff}} a \epsilon_c
\\
{\red =} &2 (s_{\mathrm{coff}}Ba+
\|A_0\|_\infty)
s_{\mathrm{coff}} a \epsilon_c{\red.}
\end{align*}
The second term 
is evaluated as follows:
\begin{align*}
&\left|
\left(
\sum_{(i_1,\cdots,i_n)} x_{(i_1,\cdots,i_n)}
|c_{j i_1}|^2\cdots|c_{j i_n}|^2
\right)^2
-
\left(
\sum_{(i_1,\cdots,i_n)} y_{(i_1,\cdots,i_n)}
|c_{j i_1}|^2\cdots|c_{j i_n}|^2
\right)^2
\right|\\
\leq & 
\left|
\sum_{(i_1,\cdots,i_n)} (x_{(i_1,\cdots,i_n)}-
y_{(i_1,\cdots,i_n)})
|c_{j i_1}|^2\cdots|c_{j i_n}|^2
\right|
\left|
\sum_{(i_1,\cdots,i_n)} (x_{(i_1,\cdots,i_n)}+
y_{(i_1,\cdots,i_n)})
|c_{j i_1}|^2\cdots|c_{j i_n}|^2
\right|\\
\leq & 
\max_{(i_1,\cdots,i_n)} |x_{(i_1,\cdots,i_n)}-
y_{(i_1,\cdots,i_n)}|
\max_{(i_1,\cdots,i_n)} |x_{(i_1,\cdots,i_n)}+
y_{(i_1,\cdots,i_n)}|\\
\leq &2 (s_{\mathrm{coff}}Ba+
\|A_0\|_\infty)
s_{\mathrm{coff}} a \epsilon_c{\red.}
\end{align*}
Thus, 
for an arbitrary separable state
$\phi_{\mathrm{sep}}$,
there exists a separable state
$\psi_{\mathrm{sep}}$
such that
\[
F_Q(\ket{\Psi_{\mathrm{sep}}},H_{\mathrm{rep}})-\epsilon_{\mathrm{optimized}\ \mathrm{sep.}} \leq F_Q(\ket{\Phi_{\mathrm{sep}}},H) \leq F_Q(\ket{\Psi_{\mathrm{sep}}},H_{\mathrm{rep}})+\epsilon_{\mathrm{optimized}\ \mathrm{sep.}},
\]
where
\[
\epsilon_{\mathrm{optimized}\ \mathrm{sep.}}=
16 (s_{\mathrm{coff}}Ba+
\|A_0\|_\infty)
s_{\mathrm{coff}} a \epsilon_c
.
\]
Similarly,
for a separable state $\psi_{\mathrm{sep}}$,
there exists a separable state
$\phi_{\mathrm{sep}}$ such that
\[
F_Q(\ket{\Phi_{\mathrm{sep}}},H)-\epsilon_{\mathrm{optimized}\ \mathrm{sep.}} \leq F_Q(\ket{\Psi_{\mathrm{sep}}},H_{\mathrm{rep}}) \leq F_Q(\ket{\Phi_{\mathrm{sep}}},H)+\epsilon_{\mathrm{optimized}\ \mathrm{sep.}}
\]
Thus,
\[
\max_{\ket{\Phi}: \mathrm{separable}}F_Q(\ket{\Phi},H) \leq \max_{\ket{\Phi}: \mathrm{separable}}F_Q(\ket{\Phi},H_{\mathrm{rep}}) +\epsilon_{\mathrm{optimized}\ \mathrm{sep.}}
\]
and
\[
\max_{\ket{\Phi}: \mathrm{separable}}F_Q(\ket{\Phi},H_{\mathrm{rep}}) \leq \max_{\ket{\Phi}: \mathrm{separable}}F_Q(\ket{\Phi},H) +\epsilon_{\mathrm{optimized}\ \mathrm{sep.}}{\red.}
\]
Therefore
\[
\frac{1}{4}
\left|
\max_{\ket{\Phi}: \mathrm{separable}}F_Q(\ket{\Phi},H)
-
\max_{\ket{\Phi}: \mathrm{separable}}F_Q(\ket{\Phi},H_{\mathrm{rep}}) )\right|
\leq 4 (s_{\mathrm{coff}}Ba+
\|A_0\|_\infty)
s_{\mathrm{coff}} a \epsilon_c{\red.}
\]

\subsubsection{Evaluation of $\left|(F_Q(\ket\psi,H)-\max_{\ket{\Phi}: \mathrm{separable}}F_Q(\ket{\Phi},H))
-
(F_Q(\ket\psi,H_{\mathrm{rep}})-\max_{\ket{\Phi}: \mathrm{separable}}F_Q(\ket{\Phi},H_{\mathrm{rep}}) ))
\right|$ and 
definition of
$\epsilon_p,\epsilon_c$}

\begin{align*}
&\frac{1}{4}\left|\Big(F_Q(\ket\psi,H)-\max_{\ket{\Phi}: \mathrm{separable}}F_Q(\ket{\Phi},H)\Big)
-
\Big(F_Q(\ket\psi,H_{\mathrm{rep}})-\max_{\ket{\Phi}: \mathrm{separable}}F_Q(\ket{\Phi},H_{\mathrm{rep}}) )\Big)
\right|\\
\leq 
&\frac{1}{4}|F_Q(\ket\psi,H)-(F_Q(\ket\psi,H_{\mathrm{rep}}))|
+\frac{1}{4}\left|
\max_{\ket{\Phi}: \mathrm{separable}}F_Q(\ket{\Phi},H)
-
\max_{\ket{\Phi}: \mathrm{separable}}F_Q(\ket{\Phi},H_{\mathrm{rep}}) ))\right|\\
\leq &\|H^2-H_{\mathrm{rep}}^2\|_\infty 
+\|H\otimes H-H_{\mathrm{rep}}\otimes H_{\mathrm{rep}}\|_\infty 
+\frac{1}{4}\left|
\max_{\ket{\Phi}: \mathrm{separable}}F_Q(\ket{\Phi},H)
-
\max_{\ket{\Phi}: \mathrm{separable}}F_Q(\ket{\Phi},H_{\mathrm{rep}}) ))\right|\\
\leq &
 d C s_{\mathrm{basis}}
(s_{\mathrm{coff}}Ba
+\|A_0\|_\infty)^2\epsilon_p
+2s_{\mathrm{coff}} Ba\epsilon_c\\
&+s_{\mathrm{coff}}^2(2B+{\red 2})a^2
\epsilon_c+
2s_{\mathrm{coff}}
\|A_0\|_\infty
{\red a} \epsilon_c
+2{\red\sqrt{2}}d C s_{\mathrm{basis}}
(s_{\mathrm{coff}}Ba
+\|A_0\|_\infty)^2
\epsilon_p\\
&+4 (s_{\mathrm{coff}}Ba+
\|A_0\|_\infty)
s_{\mathrm{coff}} a \epsilon_c\\
\leq &
{\red(1+2\sqrt{2})} d C s_{\mathrm{basis}}
(s_{\mathrm{coff}}Ba
+\|A_0\|_\infty)^2{\red\epsilon_p}
+(2 s_{\mathrm{coff}} Ba
+ s_{\mathrm{coff}}^2(2B+{\red 2})a^2
+
2s_{\mathrm{coff}}
\|A_0\|_\infty{\red a}
+4 (s_{\mathrm{coff}}Ba+
\|A_0\|_\infty)
s_{\mathrm{coff}} a) \epsilon_c{\red.}
\end{align*}
If we set
\begin{align*}
    \epsilon_p&=\frac{\epsilon}{8 \ {\red(1+2\sqrt{2})} d C s_{\mathrm{basis}}
(s_{\mathrm{coff}}Ba
+\|A_0\|_\infty)^2},\\
\epsilon_c&=\frac{\epsilon}{8
(2 s_{\mathrm{coff}} Ba
+ s_{\mathrm{coff}}^2(2B+{\red 2})a^2
+
2s_{\mathrm{coff}}
\|A_0\|_\infty
{\red a}
+4(s_{\mathrm{coff}}Ba+
\|A_0\|_\infty)
s_{\mathrm{coff}} a) }
\end{align*}
then
\begin{align*}
\left|\Big(F_Q(\ket\psi,H)-\max_{\ket{\Phi}: \mathrm{separable}}F_Q(\ket{\Phi},H)\Big)
-
\Big(F_Q(\ket\psi,H_{\mathrm{rep}})-\max_{\ket{\Phi}: \mathrm{separable}}F_Q(\ket{\Phi},H_{\mathrm{rep}}) )\Big)
\right|
 \leq \epsilon.
\end{align*}

\subsection{Proof of Theorem~\ref{upperbound_symmetric}}

We set $d>13$.
The upper bound is {\red evaluated} as follows:
\begin{align*}
&\left(\frac{B-A}{\epsilon_c}+4\right)^{dn}
\left(\frac{5}{\epsilon_p}\right)^{{\red d}(d+1)}
2 \mathrm{exp}
\left(-
\frac{2 _{n+d-1}C_n
\left(c{\red -}\epsilon+\underset{H_{\mathrm{rep}} \in \mathcal{N}_S}{\min} D_{\mathrm{mean-lower}}^{H_{\mathrm{rep}}}
\right)^2}
{{\red 144} \pi^3 \log_e 2(2+ 2\sqrt{2})^2
\Theta(n)^4
}
\right)\\
\leq
&
2 \mathrm{exp}
\left(-
C_1\frac{n^{d-1}}
{n^4}
 +dn\log_2 
\left(\frac{B-A}{\epsilon_c}+1\right)
+{\red d}(d+1)\log_2
\left(\frac{5}{\epsilon_p}\right)
\right)\\
\leq
&
2 \mathrm{exp}
\left(
-C_1\frac{n^{d-1}}
{n^4}
+C_2 dn \times n^5
+C_3 n^8
\right)
\end{align*}
where $C_1,C_2,C_2',C_3$ {\red are constants}.
Here,
\begin{align*}
& -C_1\frac{n^{d-1}}
{n^4}
+C_2 dn \times n^5
+C_3 n^8\\
=& -C_1 n^{d-5}
+C_2' n^6
+C_3 n^8\\
\rightarrow &- \infty\ \ 
(n\rightarrow\infty){\red.}
\end{align*}
Thus,
\begin{align*}
\left(\frac{B-A}{\epsilon_c}+4\right)^{\red dn}
\left(\frac{5}{\epsilon_p}\right)^{\red d(d+1)}
2 \mathrm{exp}
\left(-
\frac{{\red 2 _{n+d-1}C_n} \left(c{\red -}\epsilon+\underset{H_{\mathrm{rep}} \in \mathcal{N}_S}{\min} D_{{\red\mathrm{mean-lower}}}^{H_{\mathrm{rep}}}
\right)^2}
{{\red 144} \pi^3 \log_e 2(2+ 2\sqrt{2})^2
\Theta(n)^4
}
\right)\rightarrow 0\ \ (n \rightarrow \infty){\red.}
\end{align*}

\subsection{Proof of Theorem~\ref{upperbound_random}}
Let 
$s_{\mathrm{basis}}=n$,
$s_\mathrm{coff}=d^{o(n)}$,
$\|A_0\|_\infty=d^{o(n)}$ and
$a=\max_{m \in I_{\mathrm{coff}}} \|A_m\|_\infty=\Theta(n)$.
The upper bound is evalueted as follows:
\begin{align*}
&\left(\frac{B-A}{\epsilon_c}+4\right)^{s_{\mathrm{coff}}}
\left(\frac{5}{\epsilon_p}\right)^{{\red d}(d+1)s_{\mathrm{basis}}}
2 \mathrm{exp}
\left(-
\frac{2d^{n} \left(c{\red -}\epsilon+\underset{H_{\mathrm{rep}} \in \mathcal{N}_S}{\min} D_{\mathrm{optimized}\ \mathrm{sep.-mean}}^{H_{\mathrm{rep}}}
\right)^2}
{{\red 144} \pi^3 \log_e 2(2+ 2\sqrt{2})^2
(s_{\mathrm{coff}}Ba+\|A_0\|_\infty)^4
}
\right)\\
\leq
&
2 \mathrm{exp}
\left(-
C_2\frac{ d^{n}}
{(s_{\mathrm{coff}}Ba+\|A_0\|_\infty)^4}
 +s_{\mathrm{coff}}\log_2 
\left(\frac{B-A}{\epsilon_c}+1\right)
+{\red d(d+1)} s_{\mathrm{basis}}\log_2
\left(\frac{5}{\epsilon_p}\right)
\right)\\
\leq
&
2 \mathrm{exp}
\left(
-C_2\frac{d^{n}}
{(s_{\mathrm{coff}}Ba+\|A_0\|_\infty)^4}
+(C_3 s_{\mathrm{coff}} 
(s_{\mathrm{coff}}^2 a^2+C_3's_{\mathrm{coff}} a\|A_0\|_\infty)
+C_4 s_{\mathrm{basis}}^2
(s_{\mathrm{coff}}a+C_4'\|A_0\|_\infty)^2
\right)
\end{align*}
where $C_1,C_2,{\red C_2',}C_3{\red,C_3'}$ {\red are constants}.
Here,
\begin{align*}
& C_1\frac{ -d^{n}}
{(s_{\mathrm{coff}}Ba+\|A_0\|_\infty)^4}
+(C_2 s_{\mathrm{coff}}
(s_{\mathrm{coff}}^2 a^2+C_2's_{\mathrm{coff}} a\|A_0\|_\infty)
+C_3 s_{\mathrm{basis}}^2
(s_{\mathrm{coff}}a+C_3'\|A_0\|_\infty)^2)\\
=&-d^{n-o(n)}
+
(C_2 d^{o(n)}+C_3 n^2 d^{o(n)})\\
\rightarrow &- \infty\ \ 
(n\rightarrow\infty){\red .}
\end{align*}
Thus,
\begin{align*}
\left(\frac{B-A}{\epsilon_c}+4\right)^{s_{\mathrm{coff}}}
\left(\frac{5}{\epsilon_p}\right)^{{\red d}(d+1)s_{\mathrm{basis}}}
2 \mathrm{exp}
\left(-
\frac{2d^{n} \left(c{\red-}\epsilon+\underset{H_{\mathrm{rep}} \in \mathcal{N}_S}{\min} D_{\mathrm{optimized}\ \mathrm{sep.-mean}}^{H_{\mathrm{rep}}}
\right)^2}
{{\red 144} \pi^3 \log_e 2(2+ 2\sqrt{2})^2
(s_{\mathrm{coff}}Ba+\|A_0\|_\infty)^4
}
\right)\rightarrow 0\ \ (n \rightarrow \infty){\red .}
\end{align*}

\subsection{Proof of Theorem~\ref{globalunitary}}

Let $\psi$ be an arbitrary quantum state.
Let $H$ be
an arbitrary
locally diagonalizable Hamiltonian.
Let $\ket{\lambda_M}$ and 
$\ket{\lambda_m}$ be 
an eigenvector corresponding
to the maximal eigenvalue 
and that of the minimal eigenvalue,
respectively.
Here,
there exists a unitary matrix $U$
such that
\[
U^{\red\dag} \ket{{\red\psi}}
=\frac{\ket{\lambda_M}+
\ket{\lambda_m}}{\sqrt{2}}.
\]
Then,
\[
F_Q(U^{\red\dag} \ket{\psi}, H)
=\max_{\phi: \mathrm{arbitrary}}F_Q(\ket{\phi},H){\red.}
\]
Here,
\begin{align*}
F_Q(U^{\red\dag} \ket{\psi}, H)=F_Q(\ket{\psi}, UHU^\dag){\red.}
\end{align*}
By the two {\red equalities} above,
\begin{equation}\label{arbitrary}
F_Q(\ket{\psi}, UHU^\dag)
=\max_{\phi: \mathrm{arbitrary}}F_Q(\ket{\phi},H){\red.}
\end{equation}
Since the maximal eigenvalues of
$H$ and $UHU^\dag$ are the same
and {\red their} minimal {\red eigenvalues}
are the same,
\begin{equation}\label{unitary}
\max_{\phi: \mathrm{arbitrary}}F_Q(\ket{\phi},H)
=\max_{\phi: \mathrm{arbitrary}}F_Q(\ket{\phi},UHU^\dag){\red.}
\end{equation}
Thus, by (\ref{arbitrary}) and (\ref{unitary})
\[
F_Q(\ket{\psi}, UHU^\dag)
=\max_{\phi: \mathrm{arbitrary}}F_Q(\ket{\phi},UHU^\dag).
\]

\section{Appendix~B : Proofs of Section~5}

\subsection{Proof of Proposition~\ref{symmetric_takai}}

Let 
$\ket{\Psi}$ 
be an arbitrary $n$-qubit quantum state
denoted by
\[
\ket{\Psi}=\sum_{i_1,\cdots,i_n \in \{0,1\}} 
c_{i_1,\cdots,i_n} \ket{i_1,i_2,\cdots,i_n}.
\]

We define an $n$-qubit quantum state $\ket{\Psi_{\mathrm{intermediate}}}$
as follows:
\[
\ket{\Psi_{\mathrm{intermidiate}}}
=\sum_{k=0}^{n}
\sum_{i_1+\cdots+i_n=k}
a_k \ket{i_1,i_2,\cdots,i_n}{\red.}
\]
Then, a linear Hamitonian such as (\ref{H_S})
can be represented as follows:
\begin{align*}
H_S&=h_S \otimes I \otimes \cdots \otimes I
+I \otimes h_S \otimes I \otimes \cdots \otimes I
+\cdots
+I \otimes \cdots \otimes I \otimes h_S\\
&=\sum_{k=0}^{n}
\sum_{i_1+\cdots+i_n=k}((n-k)\lambda_0+k\lambda_1)
\ket{i_1,i_2,\cdots,i_n}\bra{i_1,i_2,\cdots,i_n}.
\end{align*}
Thus, we have
\begin{equation}\label{Psi_intermediate}
F(\Psi)=F(\Psi_{\mathrm{intermediate}}).   
\end{equation}
Therefore, we only need to show that
\begin{equation}\label{intermediate_symmetric}
F(\Psi_{\mathrm{intermediate}})\leq F(\Psi_{\mathrm{symmetric}}).
\end{equation}

Let $X,Y$ be a random variable 
which has the following distribution:
\begin{align*}
g(X=(n-k)\lambda_0+k\lambda_1)&={}_n C_k \ a_k^2\\
h(Y=(n-k)\lambda_0+k\lambda_1)&={}_n C_k \ \frac{a_k^2+a_{n-k}^2}{2}{\red.}
\end{align*}
Then,
\begin{align*}
{\red\cfrac{1}{4}}F(\Psi_{\mathrm{intermediate}})&=\mathrm{Var}[X]\\
{\red\cfrac{1}{4}}F(\Psi_{\mathrm{symmetric}})&=\mathrm{Var}[Y]{\red.}
\end{align*}
For convenience,
we assume that $n$ is even and 
$n=2m$($\exists m \in \mathbb{Z}$).
Let $Z,W$ be a random variable 
which has the following distribution:
\begin{align*}
g(Z=i)&={}_n C_{m+i} \ a_{m+i}^2\\
h(W=i)&={}_n C_{m+i} \ \frac{a_{m+i}^2+a_{n-(m+i)}^2}{2}
\end{align*}
where
$i=-m,-m+1,\cdots,-1,0,1,\cdots,m-1,m$.
Then,
\begin{align*}
X&=n\lambda_0+(Z+m)(\lambda_1-\lambda_0)\\
Y&=n\lambda_0+(W+m)(\lambda_1-\lambda_0){\red.}
\end{align*}
Furthermore,
\begin{align*}
\mathrm{E}[Z^2]&=\mathrm{E}[W^2]\\
\mathrm{E}[Z]^2&\geq 0\\
\mathrm{E}[W]^2&= 0{\red.}
\end{align*}
Therefore,
\[
\mathrm{Var}[Z]=\mathrm{E}[Z^2]-\mathrm{E}[Z]^2
\leq \mathrm{E}[W^2]-\mathrm{E}[W]^2
=\mathrm{Var}[W]{\red.}
\]
Here,
\begin{align*}
\mathrm{Var}[X]
&=\mathrm{Var}[n\lambda_0+(Z+m)(\lambda_1-\lambda_0)]\\
&=\mathrm{Var}[(Z+m)(\lambda_1-\lambda_0)]\\
&=(\lambda_1-\lambda_0)^2 \mathrm{Var}[Z+m]\\
&=(\lambda_1-\lambda_0)^2 \mathrm{Var}[Z].
\end{align*}
Similarly,
\begin{align*}
\mathrm{Var}[Y]
&=(\lambda_1-\lambda_0)^2 \mathrm{Var}[W]{\red.}
\end{align*}
Therefore,
\begin{equation}\tag{\ref{intermediate_symmetric}}
F(\Psi_{\mathrm{intermediate}})={\red 4}\mathrm{Var}[X]\leq {\red 4}\mathrm{Var}[Y]=F(\Psi_{\mathrm{symmetric}}). 
\end{equation}
By
(\ref{Psi_intermediate}) {\red and} (\ref{intermediate_symmetric}),
\[
F(\Psi)\leq F(\Psi_{\mathrm{symmetric}}){\red.}
\]

\subsection{Proof of Proposition~\ref{symmetric_bound}}

Let $\ket{\Psi_{\mathrm{symmetric}}}$ 
be an arbitrary
$n$-qubit quantum state
such that 
\[
\ket{\Psi_{\mathrm{symmetric}}}
=\sum_{k=0}^{n}
\sum_{i_1+\cdots+i_n=k} b_k
\ket{i_1,i_2,\cdots,i_n}
\]
and for all $k=0,1,\cdots,n$,
\[
b_k^2
\leq 2^{-E_g(\ket{\Psi})}
\leq 2^{-n+\frac{2n^{c-1}}{\log_e 2}-\frac{(2-c)\log_e n}{\log_e 2}}.
\]
Then,
\begin{align*}
F(\Psi_{\mathrm{symmetric}})&={\red 4}\mathrm{Var}[Y]\\
&={\red 4}(\lambda_1-\lambda_0)^2 \mathrm{Var}[W]\\
&={\red 4}(\lambda_1-\lambda_0)^2 \mathrm{E}[W^2]{\red.}
\end{align*}
In the proof of {\red Proposition~14},
we define a random variable $W$
which has the following distribution:
\begin{align*}
h(W=i)&={}_n C_{m+i} \ b_{m+i}^2
\end{align*}
where $i=-m,-m+1,\cdots,-1,0,1,\cdots,m-1,m$.
Hereafter,
we evaluate
an upper bound on $\mathrm{E}[W^2]$.
Here,
\begin{align*}
\mathrm{E}[W^2]
&=\sum_{i=-m}^{m} i^2 h(W=i)\\
&=\sum_{i=-m}^{m} i^2 {}_n C_{m+i} \ b_{m+i}^2\\
&=\sum_{j=0}^{n} \left(j-\frac{n}{2}\right)^2 {}_n C_{j} \ b_j^2\\
&=\sum_{j=0,\cdots,n/2-k} \left(j-\frac{n}{2}\right)^2 {}_n C_{j} \ b_j^2
+\sum_{j=n/2-k+1,\cdots, n/2+k-1} \left(j-\frac{n}{2}\right)^2 {}_n C_{j} \ b_j^2
+\sum_{j=n/2+k,\cdots, n} \left(j-\frac{n}{2}\right)^2 {}_n C_{j} \ b_j^2{\red.}
\end{align*}
We set $k=\sqrt{n^c}$.
The second term can be evaluated as follows:
\begin{align*}
\sum_{j=n/2-k+1,\cdots, n/2+k-1} \left(j-\frac{n}{2}\right)^2 {}_n C_{j} \ b_j^2
&\leq \frac{1}{2}\left(\frac{n}{2}+k-1-\frac{n}{2}\right)^2
+\frac{1}{2}\left(\frac{n}{2}-k+1-\frac{n}{2}\right)^2\\
&\leq \frac{1}{2}(k-1)^2 \times 2\\
&\leq n^c{\red.}
\end{align*}
By {\red Hoeffding} bound,
the tail {\red probability} of bimonial distribution
can be evaluated as follows:
\begin{align*}
\sum_{j=0,\cdots,n/2-k} {}_n C_{j}
=\sum_{j=n/2+k,\cdots, n}  {}_n C_{j}
&\leq 2^n e^{-2k^2/n}{\red.}
\end{align*}
Thus, the first and third {\red terms} can be evaluated as follows:
\begin{align*}
&\sum_{j=0,\cdots,n/2-k} \left(j-\frac{n}{2}\right)^2 {}_n C_{j} \ b_j^2
+\sum_{j=n/2+k,\cdots, n} \left(j-\frac{n}{2}\right)^2 {}_n C_{j} \ b_j^2\\
& \leq \left(0-\frac{n}{2}\right)^2
2^n e^{-2k^2/n} 2^{-n+\frac{2n^{c-1}}{\log_e 2}-\frac{(2-c)\log_e n}{\log_e 2}}
+\left(n-\frac{n}{2}\right)^2
2^n e^{-2k^2/n} 2^{-n+\frac{2n^{c-1}}{\log_e 2}-\frac{(2-c)\log_e n}{\log_e 2}}\\
& \leq \frac{n^2}{2} e^{-2n^{c-1}} \ 2^{\frac{2n^{c-1}}{\log_e 2}-\frac{(2-c)\log_e n}{\log_e 2}}\\
& \leq \frac{n^c}{2}{\red.}
\end{align*}
Therefore,
\begin{align*}
F(\Psi_{\mathrm{symmetric}})
&={\red 4}(\lambda_1-\lambda_0)^2 \mathrm{E}[W^2]\\
&\leq {\red 4}(\lambda_1-\lambda_0)^2 \left(n^c+\frac{n^c}{2}\right)\\
&\leq {\red 6} (\lambda_1-\lambda_0)^2 n^c{\red.}
\end{align*}


\begin{thebibliography}{10}

\bibitem{nielsenchung}
Michael~A. Nielsen and Isaac~L. Chuang.
\newblock {\em Quantum Computation and Quantum Information: 10th Anniversary
  Edition}.
\newblock Cambridge University Press, USA, 10th edition, 2011.

\bibitem{raussendorf2001one}
Robert Raussendorf and Hans~J Briegel.
\newblock A one-way quantum computer.
\newblock {\em Physical review letters}, 86(22):5188, 2001.

\bibitem{raussendorf2001computational}
Robert Raussendorf and Hans Briegel.
\newblock Computational model underlying the one-way quantum computer.
\newblock {\em arXiv preprint quant-ph/0108067}, 2001.

\bibitem{walther2005experimental}
Philip Walther, Kevin~J Resch, Terry Rudolph, Emmanuel Schenck, Harald
  Weinfurter, Vlatko Vedral, Markus Aspelmeyer, and Anton Zeilinger.
\newblock Experimental one-way quantum computing.
\newblock {\em Nature}, 434(7030):169--176, 2005.

\bibitem{anders2009computational}
Janet Anders and Dan~E Browne.
\newblock Computational power of correlations.
\newblock {\em Physical Review Letters}, 102(5):050502, 2009.

\bibitem{briegel2001persistent}
Hans~J Briegel and Robert Raussendorf.
\newblock Persistent entanglement in arrays of interacting particles.
\newblock {\em Physical Review Letters}, 86(5):910, 2001.

\bibitem{gross09}
David Gross, Steve~T Flammia, and Jens Eisert.
\newblock Most quantum states are too entangled to be useful as computational
  resources.
\newblock {\em Physical review letters}, 102(19):190501, 2009.

\bibitem{bremner2009random}
Michael~J Bremner, Caterina Mora, and Andreas Winter.
\newblock Are random pure states useful for quantum computation?
\newblock {\em Physical review letters}, 102(19):190502, 2009.

\bibitem{bennett93}
Charles~H. Bennett, Gilles Brassard, Claude Cr\'epeau, Richard Jozsa, Asher
  Peres, and William~K. Wootters.
\newblock Teleporting an unknown quantum state via dual classical and
  einstein-podolsky-rosen channels.
\newblock {\em Phys. Rev. Lett.}, 70:1895--1899, Mar 1993.

\bibitem{boschi98}
D.~Boschi, S.~Branca, F.~De~Martini, L.~Hardy, and S.~Popescu.
\newblock Experimental realization of teleporting an unknown pure quantum state
  via dual classical and einstein-podolsky-rosen channels.
\newblock {\em Phys. Rev. Lett.}, 80:1121--1125, Feb 1998.

\bibitem{aragon2025universality}
Luis Arag{\'o}n-Mu{\~n}oz, Chryssomalis Chryssomalakos, Ana~Gabriela
  Flores-Delgado, John Martin, and Eduardo Serrano-Ens{\'a}stiga.
\newblock Universality in fidelity-based quantum metrology.
\newblock {\em arXiv preprint arXiv:2509.18533}, 2025.

\bibitem{epsilon_net}
Patrick~M. Hayden, Debbie~W. Leung, Peter~W. Shor, and Andreas~J. Winter.
\newblock Randomizing quantum states: Constructions and applications.
\newblock {\em Communications in Mathematical Physics}, 250:371--391, 2003.

\bibitem{aspects}
P~Hayden, D~Leung, and AJ~Winter.
\newblock Aspects of generic entanglement.
\newblock {\em Communications in Mathematical Physics}, 265 (1):95 -- 117, July
  2006.
\newblock Publisher: Springer Other identifier: IDS number 042YZ.

\bibitem{lubkin1978entropy}
Elihu Lubkin.
\newblock Entropy of an n-system from its correlation with a k-reservoir.
\newblock {\em Journal of Mathematical Physics}, 19(5):1028--1031, 1978.

\bibitem{page1993average}
Don~N Page.
\newblock Average entropy of a subsystem.
\newblock {\em Physical review letters}, 71(9):1291, 1993.

\bibitem{foong1994proof}
SK~Foong and S~Kanno.
\newblock Proof of page’s conjecture on the average entropy of a subsystem.
\newblock {\em Physical review letters}, 72(8):1148, 1994.

\bibitem{degen17}
C.~L. Degen, F.~Reinhard, and P.~Cappellaro.
\newblock Quantum sensing.
\newblock {\em Rev. Mod. Phys.}, 89:035002, Jul 2017.

\bibitem{paris2009quantum}
Matteo~GA Paris.
\newblock Quantum estimation for quantum technology.
\newblock {\em International Journal of Quantum Information},
  7(supp01):125--137, 2009.

\bibitem{metrology_rev}
Nathan Shettell.
\newblock Quantum information techniques for quantum metrology.
\newblock https://arxiv.org{\slash}pdf{\slash}2201.01523, 2022.

\bibitem{ligo2011gravitational}
A gravitational wave observatory operating beyond the quantum shot-noise limit.
\newblock {\em Nature Physics}, 7(12):962--965, 2011.

\bibitem{aasi2013enhanced}
Junaid Aasi, Joan Abadie, BP~Abbott, Richard Abbott, TD~Abbott, MR~Abernathy,
  Carl Adams, Thomas Adams, Paolo Addesso, RX~Adhikari, et~al.
\newblock Enhanced sensitivity of the ligo gravitational wave detector by using
  squeezed states of light.
\newblock {\em Nature Photonics}, 7(8):613--619, 2013.

\bibitem{abbott16}
B.~P. et~al. Abbott.
\newblock Observation of gravitational waves from a binary black hole merger.
\newblock {\em Phys. Rev. Lett.}, 116:061102, Feb 2016.

\bibitem{wasi10}
W.~Wasilewski, K.~Jensen, H.~Krauter, J.~J. Renema, M.~V. Balabas, and E.~S.
  Polzik.
\newblock Quantum noise limited and entanglement-assisted magnetometry.
\newblock {\em Phys. Rev. Lett.}, 104:133601, Mar 2010.

\bibitem{sewell12}
R.~J. Sewell, M.~Koschorreck, M.~Napolitano, B.~Dubost, N.~Behbood, and M.~W.
  Mitchell.
\newblock Magnetic sensitivity beyond the projection noise limit by spin
  squeezing.
\newblock {\em Phys. Rev. Lett.}, 109:253605, Dec 2012.

\bibitem{correa15}
Luis~A. Correa, Mohammad Mehboudi, Gerardo Adesso, and Anna Sanpera.
\newblock Individual quantum probes for optimal thermometry.
\newblock {\em Phys. Rev. Lett.}, 114:220405, Jun 2015.

\bibitem{de2019quantum}
Antonella De~Pasquale and Thomas~M Stace.
\newblock Quantum thermometry.
\newblock In {\em Thermodynamics in the Quantum Regime: Fundamental Aspects and
  New Directions}, pages 503--527. Springer, 2019.

\bibitem{mehboudi2019thermometry}
Mohammad Mehboudi, Anna Sanpera, and Luis~A Correa.
\newblock Thermometry in the quantum regime: recent theoretical progress.
\newblock {\em Journal of Physics A: Mathematical and Theoretical},
  52(30):303001, 2019.

\bibitem{glm04}
Vittorio Giovannetti, Seth Lloyd, and Lorenzo Maccone.
\newblock Quantum-enhanced measurements: Beating the standard quantum limit.
\newblock {\em Science}, 306(5700):1330--1336, 2004.

\bibitem{glm06}
Vittorio Giovannetti, Seth Lloyd, and Lorenzo Maccone.
\newblock Quantum metrology.
\newblock {\em Phys. Rev. Lett.}, 96:010401, Jan 2006.

\bibitem{glm11}
Vittorio Giovannetti, Seth Lloyd, and Lorenzo Maccone.
\newblock Advances in quantum metrology.
\newblock {\em Nature Photonics}, 5:222--229, 2011.

\bibitem{Toth14}
Géza Tóth and Iagoba Apellaniz.
\newblock Quantum metrology from a quantum information science perspective.
\newblock {\em Journal of Physics A: Mathematical and Theoretical},
  47(42):424006, oct 2014.

\bibitem{hb93}
M.~J. Holland and K.~Burnett.
\newblock Interferometric detection of optical phase shifts at the heisenberg
  limit.
\newblock {\em Phys. Rev. Lett.}, 71:1355--1358, Aug 1993.

\bibitem{braunstein94}
Samuel~L Braunstein and Carlton~M Caves.
\newblock Statistical distance and the geometry of quantum states.
\newblock {\em Physical Review Letters}, 72(22):3439, 1994.

\bibitem{hayashi05}
Masahito Hayashi.
\newblock {\em Asymptotic theory of quantum statistical inference: selected
  papers}.
\newblock World Scientific, 2005.

\bibitem{shettell2022quantum}
Nathan Shettell and Damian Markham.
\newblock Quantum metrology with delegated tasks.
\newblock {\em Physical Review A}, 106(5):052427, 2022.

\bibitem{morimae2013blind}
Tomoyuki Morimae and Keisuke Fujii.
\newblock Blind quantum computation protocol in which alice only makes
  measurements.
\newblock {\em Physical Review A}, 87(5):050301, 2013.

\bibitem{takeuchi2019quantum}
Yuki Takeuchi, Yuichiro Matsuzaki, Koichiro Miyanishi, Takanori Sugiyama, and
  William~J. Munro.
\newblock Quantum remote sensing with asymmetric information gain.
\newblock {\em Physical Review A}, 99(2):022325, 2019.

\bibitem{shimony1995degree}
Abner Shimony.
\newblock Degree of entanglement a.
\newblock {\em Annals of the New York Academy of Sciences}, 755(1):675--679,
  1995.

\bibitem{wei2003geometric}
Tzu-Chieh Wei and Paul~M Goldbart.
\newblock Geometric measure of entanglement and applications to bipartite and
  multipartite quantum states.
\newblock {\em Physical Review A}, 68(4):042307, 2003.

\bibitem{boixo2007generalized}
Sergio Boixo, Steven~T Flammia, Carlton~M Caves, and John~M Geremia.
\newblock Generalized limits for single-parameter quantum estimation.
\newblock {\em Physical review letters}, 98(9):090401, 2007.

\bibitem{boixo2008quantum}
Sergio Boixo, Animesh Datta, Steven~T Flammia, Anil Shaji, Emilio Bagan, and
  Carlton~M Caves.
\newblock Quantum-limited metrology with product states.
\newblock {\em Physical Review A—Atomic, Molecular, and Optical Physics},
  77(1):012317, 2008.

\bibitem{zwierz10}
Marcin Zwierz, Carlos~A. P\'erez-Delgado, and Pieter Kok.
\newblock General optimality of the heisenberg limit for quantum metrology.
\newblock {\em Phys. Rev. Lett.}, 105:180402, Oct 2010.

\bibitem{zwierzx12}
Marcin Zwierz, Carlos~A. P\'erez-Delgado, and Pieter Kok.
\newblock Ultimate limits to quantum metrology and the meaning of the
  heisenberg limit.
\newblock {\em Phys. Rev. A}, 85:042112, Apr 2012.

\bibitem{random_qfi}
M.~Oszmaniec, R.~Augusiak, C.~Gogolin, J.~Ko\l{}ody\ifmmode~\acute{n}\else
  \'{n}\fi{}ski, A.~Ac\'{\i}n, and M.~Lewenstein.
\newblock Random bosonic states for robust quantum metrology.
\newblock {\em Phys. Rev. X}, 6:041044, Dec 2016.

\bibitem{haar_random}
Antonio~Anna Mele.
\newblock Introduction to {H}aar {M}easure {T}ools in {Q}uantum {I}nformation:
  {A} {B}eginner's {T}utorial.
\newblock {\em {Quantum}}, 8:1340, May 2024.

\bibitem{kitaev_textbook}
A.~Yu. Kitaev, A.~H. Shen, and M.~N. Vyalyi.
\newblock {\em Classical and Quantum Computation}.
\newblock American Mathematical Society, USA, 2002.

\end{thebibliography}
\end{document}